\documentclass[a4paper,11pt]{article}
\pdfoutput=1 

\usepackage[dvipsnames]{xcolor}
\usepackage{graphicx,color}
\usepackage{jcappub} 
\usepackage{enumitem}
\usepackage{braket}
\usepackage{slashed}
\usepackage[utf8]{inputenc}
\usepackage{hyperref}
\usepackage[most]{tcolorbox}
\usepackage{float}
\usepackage{caption}
\usepackage{subfigure}
\usepackage{makecell}
\usepackage{hyperref}
\usepackage{epsfig}
\usepackage{multirow}
\usepackage{amsmath,amssymb}
\usepackage{tikz}
\usetikzlibrary{shapes.geometric,arrows}
\usepackage{dsfont}

\newcommand{\tb}{\Tilde{t}}
\newcommand{\bv}{\Bar{v}}
\newcommand{\lr}[1]{\left(#1\right)}

\title{Cosmic superstrings in large volume compactifications:\\ \textbf{\noindent  \Large{\it PTAs, LISA and time-varying tension}}}

\author{Anish Ghoshal$^{a}$, Filippo Revello$^{b,c,d}$ and Gonzalo Villa$^{e}$}

\affiliation[a]{Institute of Theoretical Physics, Faculty of Physics,
University of Warsaw, \\ ul. Pasteura 5, 02-093 Warsaw, Poland}
\affiliation[b]{Instituut voor Theoretische Fysica, K.U. Leuven, \\
Celestijnenlaan 200D, B-3001 Leuven, Belgium}
\affiliation[c]{Leuven Gravity Institute, KU Leuven, \\ Celestijnenlaan 200D, B-3001 Leuven, Belgium}
\affiliation[d]{Institute for Theoretical Physics,
Utrecht University, \\ Princetonplein 5, 3584 CC Utrecht, The Netherlands}
\affiliation[e]{DAMTP, Centre for Mathematical Sciences, University of Cambridge, \\ Wilberforce Road, Cambridge, CB3 0WA, UK}

\emailAdd{anish.ghoshal@fuw.edu.pl}
\emailAdd{filippo.revello@kuleuven.be}
\emailAdd{gv297@cam.ac.uk}

\abstract{The Stochastic Gravitational Wave Background (SGWB) from cosmic superstrings offers one of the few known possibilities to test String Theory within current experimental reach. However, in order to be compatible with the existing constraints, the tension of a cosmic superstring network is required to lie several orders of magnitude below the Planck scale.  This is naturally realized in string compactifications where the volume of the extra dimensions is parametrically large (in string units). We estimate the GW spectrum arising from cosmic superstrings in such scenarios, providing analytical formulae as well as numerical results. Crucially, we do so within a semi-realistic string cosmology scenario, taking into account various modified cosmological epochs (such as kination or early matter domination) induced by the presence of moduli and a \textit{time-dependent} string tension. 
We show that part of the spectrum generically lies within reach of LISA and ET, with a large class of models predicting a characteristic drop in the amplitude
which may be robustly probed by LISA.
The corresponding signal would encode information on the dynamics of moduli and reheating. On the other hand, the ultra-high frequency part of the spectrum can be significantly enhanced by a long, early phase of kination with time-varying tension, yielding a spectral index unique to this set-up.}

\begin{document}

\maketitle

\section{Introduction}
The observation of gravitational waves (GWs) has started a new era in observational cosmology \cite{LIGOScientific:2016aoc}. So far, most efforts have concentrated on the detection of events that are astrophysical in nature, such as black holes or neutron star mergers \cite{Bailes:2021tot}.
It is however possible that the discovery of (putative) stochastic GW backgrounds of cosmological origin might ignite a similar revolution in fundamental physics. Within this context, cosmic strings are a particularly interesting target, being one of the very few universal predictions\footnote{Their existence as objects in the spectrum of the theory is guaranteed.
Whether they are produced in the early Universe is model-dependent.} of String Theory which might be accessible to current experimental methods. 
The eventual detection of the GW background from a network of cosmic (super)strings could give a spectacular confirmation of String Theory, which is very likely beyond the reach of planned collider experiments. In order to be consistent with observations, the tension $\mu$ of superstrings in our universe would have to be parametrically smaller than the Planck scale, $ \mu \lesssim 10^{-11}-10^{-12} \, M_P^2$, where the strongest bounds come from GW searches with Pulsar Timing Arrays (PTAs) \cite{NANOGrav:2023gor,EPTA:2023fyk,Reardon:2023gzh,Xu:2023wog} at nHz frequencies and throughout the paper we use the reduced Planck mass $M_P=1/\sqrt{8\pi G}\simeq 2.4\times 10^{18}$ GeV.\footnote{Large scale interferometers such as LVK also give significant constraints, although these may be relaxed by early-time modifications to standard cosmology \cite{Ellis:2023tsl,Datta:2024bqp}. This will indeed be the case in the scenarios we study.}

In String Theory, a very natural and model-independent way to realize this is in large volume compactifications, where the size of the extra dimensions is very large in units of $M_P$ \cite{Balasubramanian:2005zx,Conlon:2005ki}. These scenarios and their phenomenology have been studied intensely, as the large volumes also realize the limit of control where computations can be reliably performed within the supergravity approximation. It is interesting to notice that realistic values for the extra-dimensional volume in such constructions fall exactly in the right ballpark to reproduce the recently observed signal from PTAs \cite{NANOGrav:2023gor,EPTA:2023fyk,Reardon:2023gzh,Xu:2023wog} in terms of GWs from cosmic superstrings.\footnote{See~\cite{Conzinu:2024cwl} for another possible explanation of the PTA signal with a string cosmology origin, in the Pre-Big-Bang scenario~\cite{Gasperini:1992em}.} While an astrophysical explanation of the signal (super massive black hole binaries) is the most conservative \cite{NANOGrav:2023gor,Ellis:2023dgf,EPTA:2023xxk},\footnote{See \cite{NANOGrav:2023hvm,Figueroa:2023zhu,Ellis:2023oxs,Ratzinger:2020koh} for an in-depth comparison between many cosmological possibilities.} cosmic superstrings also provide a good fit to the data \cite{Ellis:2023tsl,Avgoustidis:2025svu}.
This is to be contrasted with the situation for gauge strings, which cannot provide a good fit to the amplitude and the slope of the signal \cite{NANOGrav:2023hvm,Ellis:2023tsl,Ellis:2020ena, Blasi:2020mfx,Blanco-Pillado:2021ygr}. \footnote{In particular, the predicted slope is not steep enough to accommodate the data.}

In addition to revealing the existence of (super)strings, the detection of \emph{any} cosmic string GW background (i.e. not limited to string theory ones) could at the same time serve as a powerful probe of early universe cosmology. Indeed, one particularly appealing feature of cosmic strings is that they evolve in a self-similar way (dubbed as the scaling regime~\cite{Albrecht:1984xv,Bennett:1987vf,Allen:1990tv}) throughout cosmic history, thus emitting GWs throughout their whole evolution, unlike other cosmological sources\footnote{An obvious exception being the stochastic GW background from inflation.} (such as phase transitions) whose emission is localized in time.
Therefore, their signal would be spread across the whole observable spectrum and potentially visible to experiments of very different types, ranging from PTAs to large scale interferometers such as LIGO-Virgo-KAGRA (LVK) \cite{KAGRA:2013rdx}, Einstein Telescope \cite{Sathyaprakash:2012jk} (ET) and LISA \cite{Audley:2017drz,Blanco-Pillado:2024aca}.

It is well known (see e.g.~\cite{Gouttenoire:2019kij} and references therein) that the spectral index of the GW background depends on the equation of state of the Universe at the time that the GWs are sourced.
Thus, if cosmic strings were present in our Universe, these could well be studied through measurement of stochastic backgrounds of gravitational waves in different ranges of frequencies.
More concretely, the GW spectrum sourced by a network of cosmic strings during radiation domination is flat.
Deviations from radiation domination thus induce boosts or dips in the spectrum (see e.g.~\citep{Gouttenoire:2019kij}).
Departures from the ``standard cosmology" between inflation and radiation domination have been recently studied in the context of large-volume string compactifications~\cite{Conlon:2022pnx,Apers:2022cyl,Revello:2023hro,Apers:2024ffe,Conlon:2024hgw,Conlon:2024ene} (see \cite{Cicoli:2023opf} for a review), where the physics of moduli seem to suggest that the early Universe had a long, interesting history with epochs with various equations of state.

A prominent example of such epochs is a period of matter domination caused by moduli that are oscillating around the minimum of their scalar potential. This period typically lasts until the oscillating moduli decay to radiation, giving rise to reheating. An important goal of this work is to characterize its impact on the GW background from cosmic superstrings, together with prospects for observability. We anticipate how, in the context of large volume compactifications, this leads to sharp predictions that depend on a very small number of parameters. The latter are compatible with the signal observed in PTAs being sourced by cosmic superstrings, and are particularly relevant for frequency ranges within the interferometer band (both space and earth based). An eventual detection would encode precious information about properties of the moduli relevant to reheating (masses, couplings, etc), which in turn depend on fundamental, microscopic parameters of the theory (such as the scale of supersymmetry breaking or the size of the extra dimensions).

An advantage unique to GWs is that they are also able to probe epochs before Big Bang Nucleosynthesis, so far unconstrained by observation \cite{Allahverdi:2020bys,Cicoli:2024bwq} and where stringy effects are expected to be even more relevant \cite{Apers:2024ffe}. In trying to access information about such early times, however, an obvious difficulty is posed by the fact the GW signals sourced very early on will peak at very high frequencies. This is an unavoidable consequence of causality; the wavelength of a GW must lie within the Hubble radius at the time it is emitted, resulting in a lower bound on the frequency. Interesting frequency values for early universe cosmology roughly span the $ 10 \, {\rm{kHz}} $$- 100\,  {\rm{GHz}}$ band, in what is known as the Ultra High Frequency (UHF) GW regime. As a limitation/opportunity, these are currently beyond the range of current detectors, which cannot cover frequencies much larger than 10 kHz. However, no astrophysical sources are known to emit GWs in this region, making it a particularly compelling target: any measured signal would have to be of cosmological origin or new physics in the astrophysical regime, such as primordial black holes or exotic objects. UHF gravitational waves are a very active topic of research, and extensive theoretical and experimental efforts are being spent to investigate new detection techniques (See \cite{Aggarwal:2020olq,Roshan:2024qnv,Aggarwal:2025noe} for reviews). 

Finally, let us also make a general point which does not hinge on the specific scenarios discussed in our paper. One of the most generic consequences of String Theory (aside from the existence of strings) is the fact that there exist no free parameters, and all couplings arise from expectation values of scalar fields. It follows that the properties of cosmic strings, whether field theoretic or fundamental, can also depend on the moduli in a stringy setting. To make a simple example, the fundamental string tension $m_s^2$ in a generic compactification will depend on the stabilised values of the volume and the dilaton, as reviewed below. In scenarios where the moduli are evolving on cosmological timescales, such as those mentioned in the first paragraph, this will invariably lead to cosmic string networks with a varying tension. For this reason, a main goal of the paper will be to perform a model-independent analysis of how the usual properties of cosmic string networks are modified in the case of a time-dependent tension, and how it affects their GW emission. We will present explicit examples, both a toy version and a more realistic scenario, where this gives rise to a spectral index that cannot be accommodated by strings with constant tension.

The paper is structured as follows. In Section \ref{sc:superstrings}, we review cosmic strings in String Theory, mostly focusing on the properties of fundamental string networks and the effect of a varying tension. Section \ref{sec:stringy} provides a brief introduction to large volume compactifications and the typical cosmological histories arising in that setting.
In particular, we discuss how moduli dynamics induce periods of time-varying tension and matter domination, where the latter gives rise to a prediction for the LISA band in a large class of models.
The computation of GW spectrum is contained in Section \ref{sc:GW} where, under the assumption of a scaling regime (shown to occur with time-varying tension in~\cite{Revello:2024gwa}), we provide novel analytic tools for the computation of GW spectra with constant and time-dependent tensions, accommodating the different dependences on the intercommutation probability proposed in the literature. Section \ref{sec:scenarios} is devoted to a few selected benchmark scenarios and their observational consequences, most notably including a concrete prediction for a drop of the spectrum in the LISA band. Finally, we conclude in Section \ref{sc:conc} with some future prospects.\\

\noindent \textbf{Note to the reader:} while sections \ref{sc:superstrings} and \ref{sec:stringy} contain (for the most part) a review of existing literature, they draw upon material from very different corners of the subject, and to our knowledge they have never been presented in a unified way. We have also taken care to emphasize the role of certain details which are crucial to the computation of the GW spectrum.
Particularly important is the definition of the parameters $\beta$ and $\chi$ in Section~\ref{sec:network-properties}, which are later used to reproduce the different dependences in the intercommutation probability discussed in the literature. In any case, the sections are written to be relatively self-contained, and a knowledgeable reader may wish to skip directly to Section \ref{sc:GW}.

\section{Cosmic superstrings}\label{sc:superstrings}

Cosmic strings encompass a variety of one-dimensional, topological defects that may have formed in the early universe. In Quantum Field Theory (QFT), they are realized as string-like, solitonic field configurations with a finite energy per unit length, given by the tension $\mu$.\footnote{We are neglecting possible IR divergences, which may be regulated on physical grounds.} Perhaps the simplest example is that of a quartic scalar theory with a spontaneously broken $\rm{U}(1)$ symmetry, where the winding around a string in spacetime corresponds to winding around the valley of the potential. In String Theory, they can rather arise as microscopic, fundamental strings that have been stretched by Hubble expansion, for example during inflation. Of course, the two pictures are compatible, as field theory strings can naturally appear in low-energy limits of string theory corresponding to a QFT. In that case, the field-theoretic singularity at the core of the string is UV completed into either a fundamental string or a wrapped brane \cite{Reece:2023czb}.

The main focus of this work will be on cosmic superstrings, which can consist of either fundamental (F-)strings, or D$p$ branes wrapped around a $(p-1)-$ dimensional cycle. For simplicity, we will only consider the case of fundamental F-strings, as they give the dominant contribution to the GW spectrum in the frequency range we are interested in.
In this section we intend to give a brief overview of their phenomenological properties, relevant to the computation of gravitational waves. The interested reader may also wish to consult Section 2 of \cite{Revello:2024gwa}, as well as some of the original references on the topic \cite{Witten:1985fp,Polchinski:1988cn,Copeland:2003bj,Damour:2004kw,Jackson:2004zg,Copeland:2009ga}.

\subsection{Microscopic properties}

Stable strings in the early universe can form macroscopic networks, which are the source of the gravitational waves considered in this paper. Such networks can consist of fundamental (F-) strings, D1 branes and bound states of the two, known as $(p,q)$-strings.\footnote{A $(p,q)$ string is precisely a bound state of $p$ F-strings and $q$ D-strings.} However, in the region of frequencies we are interested in, the largest contribution to the GW signal comes from the F-strings only (See \cite{Avgoustidis:2025svu} for the most up-to-date analysis of a multi-component $(p,q)$-string network). Therefore, we will assume only F-strings are present.
The two fundamental, microscopic parameters determining the evolution of a fundamental cosmic string network are the string tension $\mu$ and the intercommutation probability $P$. The string tension is determined by the string scale in the case of fundamental strings, or by the scalar vev of some symmetry breaking field in the scalar case. The intercommutation probability $P$, on the other hand, is the probability for two long string segments to form a loop when they collide. In field theory $P \simeq 1$, while in a string theoretic setting it usually suppressed by the string coupling $g_s$. This should already make it clear that cosmic superstrings networks will behave rather differently compared to their field-theoretic counterpart, with a crucial impact for observations mainly due to their modified evolution. Interestingly, it is precisely thanks to these differences that the cosmic superstring interpretation of the recently measured signal from PTAs \cite{NANOGrav:2023gor,EPTA:2023fyk,Reardon:2023gzh,Xu:2023wog} is favored with respect to that of (stable) field theory cosmic strings \cite{Ellis:2023tsl}.

\subsubsection{String tension}\label{ssc:stt}
 
 A distinguishing feature of the String Theory case is that the tension of the strings is not necessarily constant, as it can inherit a dependence on the various moduli of a given compactification \cite{Revello:2024gwa}. This is a specific instance of a more general principle, according to which energy scales in string theory are not fixed \emph{a priori}, but rather determined by dynamical processes. As an example, the fundamental string tension in a type II compactification (without warping) is given by
\begin{equation}\label{eq:tension}
    \mu=\frac{1}{2\pi \alpha'}= m_s^2=\frac{g_s^{1/2}}{2\mathcal{V}}M_P^2
\end{equation}
where $\sqrt{\alpha'}$ is the fundamental unit of length in string theory (the Regge slope), $m_s$ the string scale, $g_s$ the string coupling\footnote{In terms of the string dilaton $\varphi$, $g_s=e^{\varphi}$.} and $\mathcal{V}$ the volume of the extra dimensions (in the Einstein frame and units of $2\pi \sqrt{\alpha'}$).
In the above, the string scale $m_s$ is related to $M_P$ by the expectation values of the dilaton and volume modulus respectively.
If such fields exhibit a non-trivial time dependence during the evolution of the universe, the tension of the strings will vary with time, translating to a significant impact on the evolution of the string network \cite{Revello:2024gwa}. Moreover, this is not unique to String Theory strings - the same phenomenon is expected to be generic for field theory strings \emph{within a stringy UV completion}. Indeed, time-varying moduli can also induce time variations of scalar field vevs (rather than fundamental scales like $m_s$), and thus of QFT string tensions. One of the questions explored in this work is whether such a distinctive phenomenology may leave imprints on the stochastic gravitational wave background. 

From an observational perspective, the tension $\mu$ is constrained to be orders of magnitude below $M_P^2$ from a variety of sources, implying a parametric suppression of the string scale. The analysis of CMB anisotropies gives the robust limit $G \mu < 10^{-7}-10^{-8}$ \cite{Charnock:2016nzm}, while Pulsar Timing Array (PTA) searches for stochastic GW backgrounds \cite{NANOGrav:2023hvm,EPTA:2023fyk,Reardon:2023gzh,Xu:2023wog,NANOGrav:2023gor,EPTA:2023sfo,EPTA:2023fyk} can improve the bound by many orders of magnitude, $G \mu \lesssim 10^{-11}-10^{-12}$ \cite{Blanco-Pillado:2021ygr,Ellis:2023tsl,Marfatia:2023fvh,Avgoustidis:2025svu}. This already rules out cosmic strings in heterotic string compactifications \cite{Witten:1985fp}, where the tension is essentially fixed in terms of the GUT fine structure constant as $G \mu \sim \frac{\alpha_{\rm{GUT}}}{16 \pi} \sim 10^{-3}$. On the other hand, \eqref{eq:tension} shows how low tensions are not difficult to obtain in the type II setting. In particular, it is clear that a parametrically small string tension can most naturally be achieved in the weakly coupled limit of large compactification volumes (and eventually small string couplings). For this reason, it is very natural to focus on scenarios with large extra dimensions, where $\mathcal{V} \gg 1$.

Large Extra Dimensions were initially suggested as a purely phenomenological way to ameliorate or solve the hierarchy problem \cite{Arkani-Hamed:1998jmv,Antoniadis:1998ig}, but later found a string theoretic incarnation in terms of the Large Volume Scenario (LVS) \cite{Balasubramanian:2005zx,Conlon:2005ki}\footnote{It must be remarked, however, that the similarity is only qualitative. While in both cases the extra-dimensional radius is parametrically larger than the string length, in the LVS construction it is well below the $R \sim 1\, \rm{mm}$ advocated in \cite{Arkani-Hamed:1998jmv}.} and its generalizations. The LVS is a construction arising from flux compactifications of type IIB, and is one of the better established scenarios of moduli stabilization, featuring in particular a mechanism to stabilize the volume modulus at exponentially large values. Moreover, it is very appealing from the point of view of phenomenology as the exponentially large volume easily allows to engineer hierarchies between different scales in Nature - such as the one between $m_s$ and $M_P$. Realistic values for the stabilized volume and dilaton in LVS are in the range $\mathcal{V} \sim 10^6-10^{13}$ and $g_s \sim 0.1$, resulting a string tension $G \mu \sim 10^{-9}-10^{-16}$. These values not only allow to evade the constraints mentioned above, but are in the right ballpark to be tested by upcoming GW detectors.

For these reasons, we shall take the LVS and related constructions as an interesting benchmark in the study of gravitational waves from string theory. Another reason, to be explored in detail in Section \ref{sec:stringy}, is that scenarios with steep runaway potentials towards large volume strongly suggest modifications to the $\Lambda$CDM models of cosmology in the very early universe which can further modify the spectrum of GW from cosmic strings. All in all, this paints the compelling and consistent picture of a well-motivated corner of String Theory which allows for observationally viable (but not yet excluded) cosmic superstrings and at the same time can exhibit smoking-gun signatures of its string theoretic origin, in the form of a time-dependent string tension. 

For completeness, let us also mention another possible mechanism to obtain a low, fundamental string tension. This can happen if the string scale is locally suppressed by a significant amount of warping, which can happen if the compactification space (usually taken to be a Calabi-Yau manifold) develops throats. In that case, the metric can locally be described by~\cite{Dasgupta:1999ss,Giddings:2001yu,Giddings:2005ff}:
\begin{equation}\label{eq:amb-metric}
    d s^2=
    \lr{1+\frac{e^{4\mathcal{A}(y)}}{\mathcal{V}^{2/3}}}^{-1/2}
    g_{\mu \nu} d x^{\mu} d x^{\nu} +\lr{1+\frac{e^{4\mathcal{A}(y)}}{\mathcal{V}^{2/3}}}^{1/2}
    g_{m n} d y^n d y^m,
\end{equation}
where $\mathcal{A}(y)$ is a warping factor determined by the backreaction of sources of stress tensor like branes and fluxes.
Then, the tension for a string localized at a region of large warping $y=y_0$, $e^{4\mathcal{A}(y_0)}\gg \mathcal{V}^{2/3}$ can be written as
\begin{equation}
   \mu= 
   \frac{1}{2\pi \alpha'}\lr{1+\frac{e^{4\mathcal{A}(y_0)}}{\mathcal{V}^{2/3}}}^{-1/2}
   \simeq \frac{g_s^{1/2}}{2\mathcal{V}^{2/3}}e^{-2\mathcal{A}(y_0)}M_P^2\, ,
\end{equation}
with a clear exponential suppression.\footnote{This expression corrects a factor of $2\pi$ in~\cite{Revello:2024gwa} and uses the 10D Einstein frame expression for the warp factor~\cite{Giddings:2005ff}.}
While this mechanism is relatively common in string compactifications (for example in the KKLT construction \cite{Kachru:2003aw}), many of the details are model-dependent, and we shall not pursue this case further. 

\subsubsection{Intercommutation probability}

The value of the intercommutation probability can also be affected by the large extra dimensions. In highly symmetric scenarios like those in~\cite{OCallaghan:2010rlo,OCallaghan:2010mtk} $P$ will be suppressed by the ratio of the extra-dimensional volume to the string volume, as the strings would now be able to ``miss" each other in the extra dimensions.
For the values of $\mathcal{V}$ considered here, this would lead to a negligibly small $P$, seriously impacting the ability to even form a cosmic superstring network in the first place.
However, the scenario in the back of our minds considers generic ingredients on model-building such as branes and fluxes.
As remarked in \cite{Jackson:2004zg},\footnote{In a more detailed model for cosmic superstring evolution one may introduce a time dependence on the effective volume parameter $w$ and on $g_s$, which would then determine the resulting time-dependence of the various interaction coefficients.
Other discussions on effective volumes probed by cosmic strings appear in~\cite{Avgoustidis:2007ju,Avgoustidis:2012vc}.} the position of a string along the extra dimensions is a modulus like others, and will presumably be stabilised in a realistic compactification, reducing the effective length along which the strings will be able to propagate.
Taking this into account, $P$ can be estimated as~\cite{Jackson:2004zg} (see Eq. (8.1))
\begin{equation}\label{eq:JJP-prob}
   P \simeq 0.5 g_s^2 \left[\frac{6 \pi}{\log \left( A\mathcal{V}\right)} \right]^3 \, ,
\end{equation}
where $A$ is an order one (for weak warping) and model-dependent numerical constant which we henceforth set to one.
Although this expression is also a function of $\mathcal{V}$, the dependence is very weak, and we will consider it constant throughout the rest of the work. As a benchmark value, we can consider $g_s =0.11$ and $\mathcal{V} = 10^{10}/3$, leading to $P \simeq 3.8 \cdot 10^{-3}$ and $G\mu=2.0 \cdot 10^{-12}$.

\subsection{Properties of the network}\label{sec:network-properties}

In this work, we will make the assumption that a population of cosmic superstrings is present after inflation, in sufficient numbers to be able to form a network. A typical production mechanism for superstrings is brane-antibrane annihilation, whose attractive potential is often advocated as a mechanism for inflation in String Theory \cite{Dvali:1998pa,Burgess:2001fx,Jones:2002cv,Sarangi:2002yt,Jones:2003da,Pogosian:2003mz,Kachru:2003sx} (See also \cite{Burgess:2022nbx,Cicoli:2024bwq} for more recent developments). 
More generally, any scenario achieving a thermalized energy density of order one or more in string units will generically excite fundamental string oscillator modes, rendering large, heavy strings.
Depending on the details of the compact space and whether there are open strings\footnote{This is related to the stability or instability of the strings discussed in~\cite{Copeland:2003bj}.} in the spectrum (i.e: on whether there are branes present where this energy density is reached), the thermodynamics is dominated by either a bath of open strings (rendering a Hagedorn phase, recently discussed in~\cite{Frey:2023khe,Frey:2024jqy}) or by an order one number of long, closed strings.\footnote{This is sometimes called a Hagedorn phase transition in the literature.
This is however different from what was considered in the seminal paper of Atick and Witten~\cite{Atick:1988si} as the latter requires a finite string coupling whilst the computations rendering a long string are performed in the $g_s \to 0$ limit.
There is by now evidence (see e.g~\cite{Barbon:2004dd}) that the Atick-Witten transition leads to black hole nucleation, so both pictures are related by the Horowitz-Polchinski correspondence~\cite{Horowitz:1996nw,Horowitz:1997jc} between black holes and highly excited strings, stating that highly excited strings become black holes as $g_s$ gets large.
In this article we work in the limit where the highly excited strings can be described as such.}
We refer to~\cite{Abel:1999rq,Frey:2023khe} and Appendix A of~\cite{Frey:2024jqy} for further details on string thermodynamics, and simply point out that the latter case, with an order one number of highly excited strings per Hubble patch, that is of interest to us in this article.
More recently, it has also been suggested that, in the presence of a time-dependent string tension, if a small number of strings were nucleated from the vacuum they would be able to percolate and form a network \cite{Conlon:2024uob} (see also~\cite{Brunelli:2025ems}). 

It has been known since the early days of the subject that, whenever a network of cosmic strings is formed, it will quickly reach an attractor known as the scaling regime, where the energy density of the strings constitutes a fixed fraction of the background \cite{Kibble:1984hp,Bennett:1985qt,Austin:1993rg,Vilenkin:2000jqa}. This is a self-similar configuration where correlation length of the network grows proportionally to the Hubble radius. Modulo some assumptions to be reviewed in this section, a similar configuration is reached in the case of varying tension, with the only difference that the relative energy density in strings will now evolve proportionally to the tension. In formulae, 
\begin{equation}
   \Omega_{\rm{strings}} (t) \equiv \frac{\rho_{\rm{strings}}}{3 H^2} = \Omega_{\rm{strings}} (t_0) \frac{\mu(t)}{\mu(t_0)},
\end{equation} 
for any given reference time $t_0$. The scaling regime can be described by a semi-analytical model known as the Velocity One Scale (VOS) model, introduced in \cite{Martins:1996jp,Martins:2000cs}. 
The VOS model was recently extended to case of varying tension in \cite{Revello:2024gwa}, and here we give a brief overview of the results.

\subsubsection{The VOS model (with a time dependent tension)}\label{ssc:tdte}
The VOS model \cite{Martins:1996jp,Martins:2000cs} can be obtained by coarse-graining of the equations of the classical equations of motion for the strings, supplemented by information obtained through numerical simulations. Denoting by $\left\{ X^{\mu},\zeta_a \right\}$ spacetime and worldsheet coordinates respectively, the motion of a single string with a (space)time-dependent tension is described by 
\begin{equation}
    S=-  \int d^2 \zeta\, \mu \left(X^{\rho}(\zeta)\right)   \sqrt{-\rm{det} \left( \gamma_{ab} \right)} \quad \quad \quad \quad \gamma_{ab}=\frac{\partial X^{\mu}(\zeta)}{\partial \zeta^a} \frac{\partial X^{\nu}(\zeta)}{\partial \zeta^b} g_{\mu \nu}.
\end{equation}
 The corresponding equations of motion are \cite{Emond:2021vts,Conlon:2024uob}
\begin{equation}
   \frac{1}{ \sqrt{-\gamma}} \partial_a \left(  \sqrt{-\gamma} \gamma^{ab} g_{\rho \sigma} X^{\sigma}_{\, ,b} \right)-\frac{1}{2} \partial_{\rho} g_{\lambda \sigma} \gamma^{a b} X^{\lambda}_{\, ,a} X^{\sigma}_{\, ,b} =0.
\end{equation}
The background metric is assumed to be FRW, $ds^2 = a^2(\tau) \eta_{\mu \nu}$, and we further split the spacetime coordinates as $X^{\rho}= (X^0,x^i)$.
The EOMs simplify significantly in the gauge where $\gamma_{01}=\gamma_{10}=0$ and $X^0=\tau$.

From the microscopic description, one can extract two fundamental, coarse-grained quantities that will be sufficient to describe the evolution of a network. The first one is the correlation length $L$,
defined by
\begin{equation}\label{eq:rhostrings}
    \rho_{\rm{strings}} = \frac{\mu(t)}{L(t)^2}.
\end{equation}
While $L$ is related to the energy density of the network by definition, it can also be interpreted as the average length between the long strings. During scaling, it typically grows as $L \propto t \propto H^{-1}(t)$. The other one is the RMS velocity $v$ of the string, given by
\begin{equation}
    v^2= \frac{\int d \zeta^1 \,\varepsilon \dot{x}^2 }{\int  d \zeta^1 \, \varepsilon }, \quad \quad \rm{where} \quad \quad \varepsilon \equiv \sqrt{\frac{x'^2}{1-\dot{x}^2}}.
\end{equation}
If the background is dominated by a fluid with $\rho \sim a^{-n}$ and the string tension scales as $t^{-q}$, $v$ and $L$ obey the system of equations \cite{Revello:2024gwa}
\begin{equation}\label{eq:L}
\begin{cases}
2\frac{{\rm d} L}{{\rm d}t} &= 2 H L(1+v^2)+\tilde {c}v+\frac{L}{\mu} \frac{{\rm d} \mu}{{\rm d}t} v^2,\\
\frac{{\rm d} v}{{\rm d}t} &= (1-v^2) \left[\frac{k(v)}{L}-\left(2 H + \frac{1}{\mu} \frac{{\rm d} \mu}{{\rm d}t} \right) v \right],
   \end{cases}
\end{equation}
where $H$ is the Hubble ratio (in terms of $dt = a(t) d \tau$), $\tilde{c}$ the loop chopping efficiency (defined later in \eqref{eq:eloss}) and $k(v)$ a phenomenologically motivated function known as the momentum parameter. For our numerical estimates, we will take $\tilde{c}=0.23$, derived from simulations in \cite{Martins:2000cs}. An analytical expression for $k(v)$ was also suggested in \cite{Martins:2000cs}, based on a combination of analytical arguments and a comparison to simulations:
\begin{equation}\label{eq:k(v)}
    k(v)=\frac{2 \sqrt{2}}{\pi}\left(1-v^2\right)\left(1+2 \sqrt{2} v^3\right) \frac{1-8 v^6}{1+8 v^6}.
\end{equation}
While the original analysis was restricted to the constant tension case, \cite{Revello:2024gwa} provided evidence that the same formula should also describe strings with varying tension. Finally, the loop chopping efficiency $\tilde{c}$ is a phenomenological parameter describing energy loss into loops.

We refer to Section 4 of \cite{Revello:2024gwa} for a full analysis of the dynamical system, which in general can have multiple attractors. However, for reasonable initial conditions, the late time solutions will always approach a scaling solution characterized by $L= \xi t$ and
\begin{equation}\label{eq:fpsc}
  \xi= \sqrt{\frac{n}{8} \frac{k(\bar{v})\left(k(\bar{v}) + \tilde{c} \right)}{\left(1-\frac{q n}{4} \right) \left(1-\frac{2}{n} \right)}} \quad \quad \quad \quad \bar{v} = \sqrt{\frac{n}{2} \frac{k(\bar{v})}{k(\bar{v})+\tilde{c}} \frac{1-\frac{2}{n}}{1-\frac{q n}{4}}},
\end{equation}
for a power-law cosmology $a\sim t^{2/n}$ with varying tension $\mu \sim t^{-q}$ (See \cite{Revello:2024gwa} for more details).\footnote{Notice the notation switch from $q$ in this paper to $m$ in~\cite{Revello:2024gwa}.} In turn, these determine the number density of long strings through \eqref{eq:rhostrings}. This will be a fundamental quantity to compute to compute the GW spectrum in Section \ref{sc:GW}. For the rest of this work, we will assume such a scaling regime has been reached after a short transient.

We are now in a position to discuss the effect of varying the intercommutation probability, which is the main difference between cosmic strings and superstrings at the network level.
The idea is that the loop-chopping parameter is reduced by a power of the intercommutation probability, $\Tilde{c} \to \Tilde{c}P^{1-\beta}$.
Considering $\beta\neq 0$ takes into account the fact that, because the intercommutation probability is reduced (and so the number of interactions), there is an increase in the small scale structure of the string.
This increase leads to further contacts between the points of the string, an effect which competes against the reduction of the probability.
The VOS predicts $\beta=0$ (since the increase of small scale structure generates a new scale that is not present in the model, see~\cite{Avgoustidis:2005nv}), whilst simulations indicate $\beta=1/2$ in flat space~\cite{Sakellariadou:2004wq} and $\beta=2/3$ in an expanding spacetime~\cite{Avgoustidis:2005nv}.
It has been shown~\cite{Avelino:2012qy} that weakly coupled strings reach a scaling regime with $\xi \sim \Tilde{c}$, so for our present purposes we will take the effect of varying the intercommutation probability $P$ as taking $\Tilde{c} \to P^{1-\beta} \Tilde{c}$ and $\xi \to P^{1-\beta} \xi$, in agreement with previous studies~\cite{Avelino:2012qy,Sousa:2016ggw}.
Recent analyses in the context of PTA data take $\beta=1/2$~\cite{NANOGrav:2023hvm,Figueroa:2023zhu,Ellis:2023tsl} and $\beta=2/3$~\cite{Avgoustidis:2025svu}.
In Sec.~\ref{sc:GW} we will study the GW spectrum analytically for all $\beta$, but in Sec.~\ref{sec:scenarios} we will take $\beta=1/2$.
There is another effect that we will not be considering in this article, namely the existence of $D$-strings and the possibility of forming bound states with junctions, for which the VOS has been extended in~\cite{Avgoustidis:2007aa,Avgoustidis:2009ke,Pourtsidou:2010gu}.
This effect only affects the low-frequency part of the spectrum~\cite{Avgoustidis:2009ke,Avgoustidis:2025svu} and will not play a major role in our discussion, although an interesting future direction would be to study it in the context of varying tension.

\subsubsection{The behavior of loops}

When long strings intersect, they can form closed loops with a probability $P$. One can estimate the energy injected into the loop population as \cite{Kibble:1984hp} 
\begin{equation}\label{eq:eloss}
    \frac{d \rho}{dt}=\tilde{c} \bv \frac{\rho}{L}\, 
\end{equation}
which also provides a definition of the loop chopping parameter. Larger loops are the most relevant for GW emission \cite{Siemens:2006yp}, and numerical simulations \cite{Blanco-Pillado:2013qja} show that a fraction $\mathcal{F}_{\alpha} \simeq 0.1$ of the emitted loops have length $\mathcal{\ell} (t_i)=\alpha t_i$, with $\alpha \simeq 0.1$.

It is not known what the appropriate value of $\alpha$ is in the case of cosmic superstrings.
In particular, Ref.~\cite{Sousa:2016ggw} argues that a smaller intercommutation probability should affect the initial size of the loops.
This size is related to the correlation length $L$ via $\ell=\alpha_L L$ (which defines $\alpha_L$).
Since $L=\xi t$ and $\xi$ depends on the intercommutation probability, it follows that $\alpha=\alpha_L \xi$ inherits a $P$-dependence.\footnote{Note that assuming $\alpha_L$ constant implies that $\alpha$ inherits a dependence in the equation of state of the background and the rate of varying tension through $\xi$, c.f. Eq.~\eqref{eq:fpsc}.
In Sec.~\ref{sc:GW} we study the behaviour of the spectrum for arbitrary $\alpha$ but in Sec.~\ref{sec:scenarios} we will take $\alpha=0.1$.}
As discussed in~\cite{Sousa:2016ggw} and Sec.~\ref{sc:GW} explicitly, this results in a shift in the frequency of features in the GW spectrum.
To take into account this effect, we define the parameter $\chi$ so that, for cosmic superstrings,
\begin{equation}\label{eq:def-chi}
    \alpha \to \alpha P^{\chi} \, ,
\end{equation}
reproducing~\cite{Sousa:2016ggw} for $\chi=1/3$.
Recent studies~\cite{NANOGrav:2023hvm,Figueroa:2023zhu,Ellis:2023tsl} take $\chi=0$, while Ref.~\cite{Avgoustidis:2025svu} considers $\chi\neq 0$.\footnote{We cannot provide a value for $\chi$ that reproduces~\cite{Avgoustidis:2025svu} because we work under the simplifying assumption that the scaling regime is achieved throughout cosmic history.}
Our parametrization is therefore useful in unifying and potentially comparing the approaches.
In Sec.~\ref{sc:GW} we will discuss the dependence of the amplitude of the GW spectrum and transition frequencies in $\chi$, while in Sec.~\ref{sec:scenarios} we will take the benchmark of~\cite{Ellis:2023tsl} and allow for different values of $\chi$.

All in all, loops are produced at a rate
\begin{equation}
    \mathcal{F}_\alpha\frac{d \rho}{dt}=\mu \gamma \ell \frac{dn}{dt} 
\end{equation}    
where $\gamma$ is a Lorentz factor, $\ell$ the proper length of the loop.
This implies a loop production rate:
\begin{equation}\label{eq:lprod}
    \frac{dn}{dt}=\frac{\mathcal{F}_\alpha}{\gamma}\frac{\tilde{c} \bv}{\alpha_L}\frac{1}{L^4}
    \equiv \frac{\mathcal{F}_\alpha}{\alpha} \frac{C_{\rm{eff}}}{t^4} \, ,
\end{equation}
where the last equality defines $C_{\rm{eff}}$.
This number has been computed in~\cite{Cui:2018rwi} to be dependent on the equation of state of the Universe, with $C_{\rm{eff}}\in \lbrace 0.39,\, 5.4,\, 30.6\rbrace$ during matter, radiation and kination domination, respectively.
We will assume these values for the case where the tension varies in Sec.~\ref{sec:scenarios}, and we note for later reference that the loop production rate scales with intercommutation probability like
\begin{equation}\label{eq:p-dependence}
    \frac{dn}{dt} \to \frac{dn}{dt} P^{-2(1-\beta)-\chi}\, .
\end{equation}

Once they are produced, the loops will then start oscillating and emitting gravitational radiation~\cite{Vilenkin:1981bx,Vachaspati:1984gt,Turok:1984cn,Burden:1985md,Olum:1999sg,Moore:2001px}.
The evolution of their length is given by the combination of the effects due to the varying tension and the loss of energy due to gravitational radiation.
The former is obtained by integration of the equations of motion for the loop~\cite{Conlon:2024uob,Revello:2024gwa} and the latter is given by
\begin{equation}\label{eq:energy-loss}
    \frac{dE_{\rm{GW}}}{dt}=-\Gamma G\mu^2\, ,
\end{equation}
where $\Gamma$ is a phenomenological parameter describing energy loss to GWs.
From simulations, it can be estimated as $\Gamma \simeq 50$ \cite{Blanco-Pillado:2017oxo}. 
This number might have to be revisited for superstrings.
 We will keep it arbitrary in our analytic estimates and choose $\Gamma=50$ for the plots.
Combined, these effects imply the following evolution of equation for the length of the loops:
\begin{equation}\label{eq:lt}
   \frac{{\rm d} \ell}{{\rm d}t} =  - \ell \frac{1}{2\mu}\frac{d\mu}{dt}- \Gamma G \mu \, .
\end{equation}
As first noticed in \cite{Conlon:2024uob}, a peculiarity of \eqref{eq:lt} is that it allows loops to grow faster than the scale factor if the tension decreases quickly enough. This can lead to novel phenomena, such as percolation, which we will however not consider in this work.\footnote{See Section 4.4 of~\cite{Revello:2024gwa} for estimates of the percolation time in our setting.} With a power-law dependence of the tension on time, $\mu (t) = \mu_0 \left(\frac{t_0}{t}\right)^q$, and an initial condition $\ell(t_i)=\alpha P^\chi t_i$, Eq.\eqref{eq:lt} is solved by
\begin{equation}\label{eq:loop-ev-vary-tension}
    \ell(t)=\lr{\frac{t}{t_i}}^{q/2} \lr{\alpha P^\chi-\frac{\Gamma G \mu_0}{\frac{3q}{2}-1}\lr{\frac{t_0}{t_i}}^q}t_i+\frac{\Gamma G\mu_0}{\frac{3q}{2}-1}\lr{\frac{t_0}{t}}^{q-1}t_0 \, ,
\end{equation}
for $q \neq 2/3$, and
\begin{equation}\label{eq:loop-tracker}
    \ell(t)=\left[\alpha P^\chi  - \Gamma G\mu_0 \lr{\frac{t_0}{t_i}}^{2/3}\log \lr{\frac{t}{t_i}}\right]\lr{\frac{t}{t_i}}^{1/3}t_i \, 
\end{equation}
for $q=2/3$.

\medskip

\section{String Theory and the First Half of the Universe}\label{sec:stringy}

As suggested towards the end of Section \ref{ssc:stt}, fundamental strings with a time-varying tension typically arise when moduli fields acquire a non-trivial time dependence. In this section we give some concrete examples of alternative, early universe cosmologies arising in a String Theory setting, mostly inspired from \cite{Apers:2024dtn} (see also \cite{Cicoli:2023opf} for a review).
In order to make our calculations fully explicit, we specialise to a particular corner of the string theory landscape, type IIB flux compactifications at large volume.
The generic situation where exotic, stringy epochs are expected to appear is that of scalar fields (moduli) displaced from their minimum, which start rolling along (typically steep) potentials. In realistic scenarios, where other forms of energy density are present, this can lead to a cascade of events, resulting in the multi-phase cosmologies reviewed here, comprising kination, tracker and matter domination epochs. Let us also remark that these scenarios also assume inflation took place, although they are agnostic with respect to the actual mechanism. To be compatible with observational constraints, such non-standard epochs have to end before the beginning of BBN, and satisfy the constraints from $\Delta N_{\rm{eff}}$ \cite{Allahverdi:2020bys}. 

\subsection{Compactifications at large volume}

As remarked in Section \ref{ssc:stt}, cosmic superstrings (in absence of warping) require a very large compactification volume in order to be phenomenologically viable. For this reason, we will consider explicit examples where moduli stabilization occurs at very large values of the volumes. The paradigmatic example is the Large Volume Scenario (LVS) \cite{Balasubramanian:2005zx,Conlon:2005ki}, arising from flux compactifications of Type IIB String Theory (See \cite{Conlon:2006gv} for a review). It is one of the few known cases where all moduli can be stabilized, offering a rich and interesting phenomenology \cite{Conlon:2005jm,Conlon:2006tq,Conlon:2007gk,Blumenhagen:2009gk,Cicoli:2012sz,Cicoli:2012aq,Hebecker:2014gka}. While such a specific choice might appear restrictive, the vast majority of what we are going to say does not depend on the fine details of the construction and equally applies to different vacua in the asymptotics of moduli space. All that is needed is a steep (close to exponential) potential in the canonically normalized volume that leads to a vacuum for exponentially large values of $\mathcal{V} \gtrsim 10^{10}$. The LVS is simply one of the best developed scenarios and allows one to make computations very concrete, for example by giving numerical values to $\mathcal{O}(1)$ coefficients. Another advantage is that most physical observables are (for the most part) determined by a single parameter, the volume $\mathcal{V}$. In Section \ref{sc:GW}, we will see how this leads to a unique prediction for the GW spectrum once the overall amplitude of the signal (equivalently, the string tension) are fixed, for example by experimental data.
In particular, as we will discuss, natural volumes in LARGE volume scenarios give string tensions and reconnection probabilities in a range that could explain the PTA signal~\cite{Ellis:2023tsl,Avgoustidis:2025svu}.

In its simplest incarnation, it is characterized by a single light modulus $\Phi$, which controls the overall size of the compactification manifold: their relation can be expressed as
\begin{equation}\label{eq:vol}
    \Phi = \sqrt{\frac{2}{3}} M_P \log \mathcal{V},
\end{equation}
where $\mathcal{V}$ is the volume of the extra-dimensional manifold in units of $(2 \pi \sqrt{\alpha'})^6$. Including the uplift term to de Sitter, its potential is given by
\begin{equation}\label{eq:VLVS}
V_{LVS}(\Phi) = V_0 \left[(1- \epsilon_{LVS} \Phi^{3/2})e^{-\lambda \frac{\Phi}{M_P}}+ \delta e^{-\sqrt{6}  \frac{\Phi}{M_P}} \right],
\end{equation}
where $\lambda = \sqrt{\frac{27}{2}}$ for LVS. It has a minimum for $\Phi \simeq 1/\epsilon_{LVS}^{2/3}$, which can be tuned in terms of microscopic parameters such as $g_s$, allowing the volume at the minimum to take exponentially large values. From a microscopic perspective, this results from the interplay between non-perturbative correction to the superpotential and the well known $\mathcal{N}=2$ BBHL correction \cite{Becker:2002nn,Bonetti:2016dqh} to the K\"{a}hler potential appearing at order $\mathcal{O}(\alpha'^3)$.
Moreover, if one assumes the validity of the anti-D3 brane uplift, the coefficient $\delta$ can be fine-tuned to achieve a small, positive cosmological constant at the minimum. 

As well as the LVS, one can consider more general mechanisms (potentials) relying on additional $\alpha'$ and $g_s$ perturbative corrections to the large volume scalar potential (See \cite{Burgess:2020qsc,Cicoli:2021rub} for a systematic analysis). Examples include logarithmic field redefinitions involving the volume modulus \cite{Weissenbacher:2019mef,Weissenbacher:2020cyf,Klaewer:2020lfg} or $\mathcal{N}=1$ mixed corrections to the K\"{a}hler potential at order $\mathcal{O}(g_s ^2 \alpha'^3)$ \cite{Antoniadis:1997eg,Minasian:2015bxa,Antoniadis:2018hqy,Antoniadis:2019rkh,Leontaris:2022rzj}.
In general, the tree-level K\"{a}hler potential is given by $K=-3 \log \mathcal{V}$, where the volume $\mathcal{V}$ has a power-law dependence on the K\"{a}hler moduli $T_i$. It follows that \emph{any} perturbative correction to the potential, once expressed terms in terms of canonically normalized fields, will consist of exponential terms. Likewise, logarithmic field redefinitions will manifest as power-law correction these exponential terms. For this reason, we can schematically parametrize the form of more general, large volume scalar potentials as
\begin{equation}\label{eq:Vlv}
    V_{\rm{LV},1}(\Phi) = V_0  \left[(1- \epsilon \Phi^{\kappa})e^{-\lambda_1 \frac{\Phi}{M_P}}+ \delta e^{- \lambda_2  \frac{\Phi}{M_P}} \right], 
\end{equation}
which makes use of the logarithmic dependence in brackets to obtain a hierarchy $\Phi \simeq /\epsilon^{1/\kappa}$, or
\begin{equation}
    V_{\rm{LV},2}(\Phi)=V_0
    \left[ Ae^{-\lambda_1 \frac{\Phi}{M_P}}+B e^{-\lambda_2 \frac{\Phi}{M_P}}+e^{-\lambda_3 \frac{\Phi}{M_P}}\right]\, ,
\end{equation}
which only features a minimum at large volume if there is a hierarchy between the coefficients.
We discuss an example in Appendix~\ref{sec:scenario-highfreq}, inspired by a recent analysis of brane-antibrane inflation~\cite{Cicoli:2024bwq}.
Here the constants $V_0, \kappa, \lambda_1,\lambda_2,\lambda_3$, $\delta, A$ and $B$ depend on the case under consideration.\footnote{A fully realistic model would arrange them so that a small and positive contribution to the observed dark energy is reproduced - a task beyond the scope of this article. }
From a qualitative level, the cosmological evolution outlined in the next section will not be affected significantly. However, it is clear that the different scenarios may give rise to quantitatively different predictions for the GW spectrum in Section \ref{sec:scenarios}.

\subsubsection{Kination, trackers and all that}\label{ssc:kin}

It is commonly assumed that, after inflation, the universe underwent a long epoch of radiation domination before finally reheating.
However, this period is almost completely unconstrained by observations \cite{Allahverdi:2020bys}, and alternatives are certainly on the table. In \cite{Conlon:2022pnx,Apers:2024ffe}, it was suggested that string theoretic considerations would indeed paint a very different picture, due to the evolution of the moduli fields. In the following, we will assume the presence of a single, cosmologically active modulus, the volume $\Phi$. This is justified by the existence of the LVS compactifications (and related scenarios) described in the previous section, where all other moduli can stabilized with parametrically higher masses. 

To reproduce the observed spectrum of primordial perturbations, cosmic inflation is expected to have taken place at very high energy scales, of the order $\Lambda_{\rm{inf}} \lesssim 10^{16} \, {\rm{GeV}}$ (corresponding to $H_{\rm{inf}} \lesssim 10^{13} \, {\rm{GeV}}$, with the upper limit set by Planck~\cite{Planck:2018vyg}), although we are agnostic about its origin.\footnote{In particular, we do not necessarily identify $\Phi$ with the inflaton.}
With the potential given by \eqref{eq:VLVS}, if $\Phi$ found itself too close the minimum of the potential after inflation, this would imply a parametric suppression of the string scale (which controls all the scales of the theory) below that of inflation, raising the question of how one might have constructed a successful embedding of inflation into String Theory in the first place. It therefore makes sense to consider the case where, after inflation, the evolution begins with $\Phi \sim \mathcal{O}(1) M_P$, which subsequently starts rolling along the (approximately) exponential part of the potential, giving rise to an epoch of kinetic energy domination (kination). Since kinetic energy redshifts as $\rho_K \sim a(t)^{-6}$, faster than any other known fluid, this will be especially relevant right at the end of inflation, where other sources can be diluted and will not immediately dominate the energy budget of the universe. Explicit examples of inflation terminating with a kination phase for the volume modulus in string compactifications can be found in \cite{Conlon:2008cj,Burgess:2022nbx}, where the latter either acts as the inflaton or the waterfall field terminating inflation \cite{Linde:1993cn} (see also \cite{Chen:2024roo}). Although the example we have presented here is for a specific (albeit well-motivated) construction, we expect similar behavior to be generic within string theoretic constructions, due to the (typically steep) exponential potentials that often appear for the moduli. For instance, scalar field mediated, stringy cosmologies that aim to match observational data on the known evolution of the universe often lead to a kination epoch when extrapolated back into the past \cite{Andriot:2024jsh,Andriot:2024sif}. 

Whenever a scalar field starts rolling along a steep potential, its potential energy will be quickly depleted to the advantage of the kinetic one, until the former becomes almost negligible. In absence of other sources, the only obstacle to its motion will be from Hubble friction, in a regime known as kination. In this limit, the equations of motion for a generic scalar $\phi$,\footnote{We will reserve the symbol $\Phi$ for the volume modulus, and $\phi$ for a generic scalar.} reduce to
\begin{equation}
    \ddot{\phi} + 3 H \dot{\phi} =0 \quad \quad \quad {\rm and} \quad \quad \quad 3 H^2 = \frac{1}{2} \dot{\phi}^2,
\end{equation}
where the dots denotes derivatives with respect to cosmic time. They are easily solved by
\begin{equation}\label{eq:volev}
    \phi=\phi_0+\sqrt{\frac{2}{3}} M_P \ln \left(\frac{t}{t_0}\right) \, ,\quad \quad \quad a(t)=a_0\left(\frac{t}{t_0}\right)^{1 / 3} \, , \quad \quad \quad  \quad H=\frac{1}{3 t},
\end{equation}
in terms of some (for now arbitrary) initial time $t_0$, and with initial values $\phi_0, a_0$. Assuming a previous phase of inflation, characterized by an almost constant potential $V(\phi_0)$, the beginning of kination can be taken at a time $t_0$ satisfying
\begin{equation}
    \frac{M_P^2}{3 t_0^2} \sim V\left(\phi_0\right) \sim \Lambda_{\mathrm{inf}}^4,
\end{equation}
where $\Lambda_{\mathrm{inf}} \sim \sqrt{H_{ {\rm{inf}} }M_P}$. As already noticed, this phase cannot last forever, as any other source that is initially present (even in a tiny amount) will eventually catch-up. Indeed, any another component (denoted by the index $\gamma)$ with 
equation of state parameter
\begin{equation}\label{eq:EoS}
    P_{\gamma}= (\gamma -1)\rho_{\gamma}, \quad \quad \quad 0 < \gamma < 2,
\end{equation}
will redshift as $\rho_{\gamma} \sim a^{-3\gamma}$. If its initial abundance is
\begin{equation}
    \Omega_\gamma\left(t_0\right) \equiv \frac{\rho_\gamma\left(t_0\right)}{3 H_{\mathrm{inf}}^2 M_P^2}=\frac{\rho_\gamma\left(t_0\right)}{\Lambda_{\mathrm{inf}}^4}=\varepsilon_0 \ll 1,
\end{equation}
kination can only last until the time
\begin{equation}
    t_{\gamma} = t_0 \, \varepsilon_0^{\frac{1}{\gamma-2}},
\end{equation}
defined as the time where the kinetic and fluid energy become equal.\footnote{For a full derivation, including factors of $\mathcal{O}(1)$, see \cite{Apers:2024ffe}.} What happens afterwards? To answer this question, let us now specialize to the situation where $\phi$ can be identified with the volume modulus $\Phi$. Sufficiently far from the minimum, the potential \eqref{eq:Vlv} can be approximated as a simple exponential,
\begin{equation}\label{eq:vexp}
    V(\Phi) = V_0 e^{-\frac{\lambda(\Phi-\Phi_0)}{M_P}}.
\end{equation}
We also assume a fluid is also present, with equation of state as in \eqref{eq:EoS}. In this case, the equations of motion can be famously rephrased \cite{Wetterich:1987fm,Copeland:1997et,Ferreira:1997hj} as the autonomous, dynamical system 
\begin{equation}\label{eq:sys}
    \left\{ \begin{aligned} x^{\prime}(N) & =-3 x-\frac{V^{\prime}(\Phi)}{V(\Phi)} \sqrt{\frac{3}{2}} y^2+\frac{3}{2} x\left[2 x^2+\gamma\left(1-x^2-y^2\right)\right] \\ y^{\prime}(N) & =\frac{V^{\prime}(\Phi)}{V(\Phi)} \sqrt{\frac{3}{2}} x y+\frac{3}{2} y\left[2 x^2+\gamma\left(1-x^2-y^2\right)\right] \\ H^{\prime}(N) & =-\frac{3}{2} H\left(2 x^2+\gamma\left(1-x^2-y^2\right)\right) \\ \Phi^{\prime}(N) & =\sqrt{6} x\end{aligned}\right. ,
\end{equation}
where 
\begin{equation}
    x=\frac{\dot{\Phi}}{M_P} \frac{1}{\sqrt{6} H}, \qquad y=\sqrt{\frac{V(\Phi)}{3}} \frac{1}{M_P H},
\end{equation}
and derivatives are with respect to $N= \ln a$. For suitable values of the parameters, the evolution of the system will necessarily converge to an attractor solution where all components redshift in the same way, and whose energy densities have fixed ratios.
Such solutions are known as trackers,\footnote{We follow e.g. \cite{Cicoli:2023opf} in calling this regime a tracker.
This situation is also called a (tracking) scaling solution in the literature - a term we avoid in this context to avoid confusion with the scaling regime of the strings.}
because both the kinetic and potential energy of the scalar field track or mimic the behaviour of the fluid component, and redshift as the latter would if it were dominating the energy density of the universe alone.
In formulae, the scale factor and Hubble rate behave as
\begin{equation}
   a(t)=a (t_t)\left(\frac{t}{t_t}\right)^{2 / 3 \gamma} \, ,\quad \quad \quad \quad H =\frac{2}{3 \gamma}\frac{1}{t},
\end{equation}
with $t_t$ the beginning of the tracker epoch. On a tracker, the scalar field evolves as
\begin{equation}
    \phi=\phi(t_t)+\frac{2 M_P}{\lambda} \ln \left(\frac{t}{t_t}\right),
\end{equation}
which does not depend on the value of $\gamma$. In terms of the abundances $\Omega_i \equiv \frac{\rho_i}{3 H^2 M_P^2}$ of the various components,
\begin{equation}\label{eq:tracker}
    \Omega_k=x^2=\frac{3}{2} \frac{\gamma^2}{\lambda^2} \, ,\quad \Omega_p=y^2=\frac{3(2-\gamma) \gamma}{2 \lambda^2} \, ,\quad \Omega_\gamma=1-x^2-y^2=1-\frac{3 \gamma}{\lambda^2},
\end{equation}
where the subscripts $k$ and $p$ denote kinetic and potential energies respectively. In practice, realistic values of $\gamma$ are $\gamma = 1$ and $\gamma=\frac{4}{3}$, corresponding to a matter or radiation tracker respectively.
Possible sources in the early universe are a population of Primordial Black Holes (PBHs) for the former \cite{Conlon:2022pnx,Apers:2024ffe} (See for example \cite{Bhaumik:2022pil,Bhaumik:2022zdd}) or decay products from the volume (such as the volume axion) in the latter case \cite{Conlon:2022pnx}. The tracker solution will last until the volume reaches the minimum of the potential, and starts oscillating around it.
A point of concern, dubbed the overshoot problem \cite{Brustein:1992nk}, is that the field might not be trapped in the minimum and runaway to the boundary of field space.
As discussed in \cite{Barreiro:1998aj,Huey:2000jx,Brustein:2004jp,Barreiro:2005ua,Conlon:2008cj,Conlon:2022pnx}, the Hubble friction induced by the fluid driving the tracker period is sufficient to ensure that this is the case for the type of potentials at hand.
The tracker phase is thus crucial to ensure the field does not overcome the potential barrier to infinity.
From this moment onwards, as we now discuss, the evolution will proceed with a phase of matter domination, followed by reheating.

\subsubsection{Matter domination from moduli decay(s)}\label{ssc:mdmd}

In the usual picture, moduli can generically be displaced from the minimum of their potential after inflation. This is especially for the volume modulus,\footnote{The same is true for the dilaton. The attentive reader may worry that displaced, oscillating moduli would quickly overtake the energy density of the kinating volume after inflation. However, such moduli do not redshift as matter if their mass is decreasing with time. In LVS, for example, the energy density of the complex structure moduli (including the axio-dilaton) redshifts exactly as $\rho^{-6}$ during kination \cite{Apers:2024dtn}.} which couples to all energy sources that are present (for example, the inflaton potential). If the displacement is small, when the Hubble rate drops down to $H =\mathcal{O}\left(m_{\Phi}\right)$ it will start to oscillate, giving rise to an early epoch of matter domination lasting until the moduli decay. To distinguish it from other scenarios of early matter domination discussed in the literature, we shall denote this epoch as moduli-induced Early Matter Domination (mEMD).

With a non-standard evolution, it is in principle possible for the volume to reach the minimum well \emph{after}\footnote{If it were to reach the minimum with $H \gg m_{\Phi}$, the modulus would simply freeze and then start oscillating again when $H \sim m_{\Phi}$, as in the standard case.} its Hubble ratio has dropped below $m_{\Phi}$. Therefore, the beginning of the matter domination epoch is in general given by
\begin{equation}
   t_m \simeq {\rm Max}  \left\{ \frac{1}{m_{\Phi}}, \frac{1}{H(t_{{\rm min}})} \right\},
\end{equation}
where $t_{{\rm min}}$ is the time it takes to reach the minimum. In the specific case of LVS, however, it turns out that the two coincide, as $t_m \sim \frac{1}{m_{\Phi}} \sim \frac{1}{H(t_{{\rm min}})}$. To see this, let us refer back to Sec. \ref{ssc:kin} and notice that the tracker regime has to be attained before $\Phi$ reaches the minimum, in order to avoid the overshoot problem. One can then use the fact that $y$ is approximately constant (i.e. \eqref{eq:tracker}) to relate the volume to the Hubble scale, resulting in 
\begin{equation}\label{eq:hfin}
    \frac{H(t_{{\rm min}})}{M_P} = \frac{\lambda}{3}\left( \frac{2 V_0}{(2-\gamma)\gamma M_P^4}\right)^{1/2} e^{-\frac{ \lambda(\Phi_{{\rm fin}}-\Phi_0)}{2M_P}}.
\end{equation}
On the other hand, for a potential that is (a sum of) exponential(s) the mass of $\Phi$ close to the minimum is of order\footnote{Here $V(\Phi_{{\rm fin}})$ is the value of the potential at the minimum before uplifting.}
\begin{equation}\label{eq:mmin}
  m_{\Phi} \sim \sqrt{\frac{|V(\Phi_{{\rm fin}})|}{M_P^2}} \sim \sqrt{\frac{V_0}{M_P^2}}  e^{- \frac{\lambda( \Phi_{{\rm fin}}-\Phi_0)}{2M_P}},
\end{equation}
which aside from $\mathcal{O}(1)$ factors has the exact same dependence on the field as Eq. \eqref{eq:hfin}. Therefore, at least in this scenario one can never really disentangle the two scales $H(t_{{\rm min}})$ and $m_{\Phi}$ in a parametric manner.

The mEMD epoch ends when all moduli have decayed to radiation, which happens when $H \lesssim \Gamma_{\Phi}$ for all the moduli that are oscillating. Since moduli only couple gravitationally, their decay rate can generically estimated as
\begin{equation}\label{eq:mdecays}
    \Gamma_{\Phi_i} = \frac{c_{i}}{48 \pi} \frac{m^3_{\Phi}}{M_P^2},
\end{equation}
with $c_i$ a model-dependent, $\mathcal{O}(1)$ factor.
Therefore, the last modulus to decay is always the lightest, which in the scenarios we are considering is typically the volume $\Phi$. In the approximation of instantaneous decay, valid as long as $\Gamma_{\Phi} \gg H $, the universe will be reheated at a time when $3 H \simeq \Gamma_{\Phi}$, corresponding to a reheating temperature of \cite{Cicoli:2012aq,Hebecker:2014gka} 
\begin{equation}\label{eq:reheat}
    T_{\text {reheat }} \sim \sqrt{\Gamma_{\Phi} M_P} \sim \mathcal{O}(1) \mathrm{GeV}\left(\frac{m_{\Phi}}{3 \times 10^6 \mathrm{GeV}}\right)^{3 / 2}.
\end{equation}
To be compatible with cosmological observations, $T_{\text {reheat }}$ must be well above the MeV scale \cite{Coughlan:1983ci,Banks:1993en,deCarlos:1993wie}, resulting in the bound $m_{\Phi} \gtrsim  30 \,  {\rm TeV}$, which is surprisingly saturated by typical values of microscopic parameters for cases of interest in Sec.~\ref{sec:scenarios}. Moreover, volume decays can contribute to the dark radiation problem, and are constrained by bounds on the effective number of neutrino species $N_{eff}$~\cite{Planck:2018vyg}. In LVS this is a significant problem, as there is always one decay channel (the volume axion) with an $\mathcal{O}(1)$ branching ratio \cite{Cicoli:2012aq,Higaki:2012ar,Hebecker:2014gka} (see also \cite{Leedom:2024qgr} for possible enhancements due to parametric resonance). 

A generic solution was recently suggested in \cite{Cicoli:2022fzy}, by realizing that in models with a high supersymmetry-breaking scale the Higgs inherits a parametrically enhanced coupling to the volume modulus due to the fine tuning of its mass. The idea is that the trilinear coupling between the Higgs and volume should not be mediated by the (observed) fine-tuned Higgs mass $m_H$, but rather feel the effect of its un-tuned value, whose magnitude is set by the soft terms (and hence proportional to the gravitino mass $m_{3/2}$). In formulae, the unnaturally small Higgs mass arises from the highly fine-tuned cancellation between the tree-level and one-loop term \emph{at the stabilised value} for the volume:\footnote{In \eqref{eq:mhiggs}, $\alpha_0$ and $\alpha_{\rm{loop}}$ are some generic tree-level and one-loop coefficients, $m_{KK}$ the Kaluza-Klein scale and $W_0$ the flux contribution to the superpotential.} 
\begin{equation}\label{eq:mhiggs}
    m_H^2 \sim m_{3 / 2}^2\left[\alpha_0+\alpha_{\mathrm{loop}} \ln \left(\frac{m_{\mathrm{KK}}}{m_{3 / 2}}\right)\right] \simeq \left(\frac{W_0}{\mathcal{V}}\right)^2\left[\alpha_0+\alpha_{\text {loop }} \ln \left(\frac{\mathcal{V}^{1 / 2}}{W_0}\right)\right] M_P^2.
\end{equation}
However, the volume dependence of the two terms is different, and expanding $\mathcal{V}$ in terms of the canonically normalized volume \eqref{eq:vol} around $\mathcal{V}=e^{\sqrt{\frac{3}{2}}\frac{\Phi_0+\delta \Phi}{M_P}}$ leads to a trilinear coupling of the form 
\begin{equation}
    \mathcal{L} \supset \frac{1}{2} \sqrt{\frac{3}{2}} \alpha_{\rm {loop}}    m_{3 / 2}^2 h^2 \frac{\delta \Phi}{M_P} \sim m_{3 / 2}^2 \alpha_{\text {loop }} h^2 \frac{\delta \Phi}{M_P},
\end{equation}
where $\alpha_{\rm{loop}} \sim 1/16 \pi$. This leads to a decay rate 
\begin{equation}\label{eq:fasdtdecay}
    \Gamma_{\Phi \rightarrow h h} \simeq \frac{1}{16 \pi} \frac{m_{3 / 2}^4 \alpha_{\rm{loop }}^2}{m_{\Phi} M_P^2} \sim\left(\alpha_{\rm{loop}} \mathcal{V}\right)^2 \frac{m_{\Phi}^3}{M_P^2},
\end{equation}
which is parametrically enhanced with respect to \eqref{eq:mdecays}, and thus dominates over all other decay channels. If the lifetime of the volume receives such a large suppression, it is however possible for heavier moduli, whose decay rate is still of the form \eqref{eq:mdecays}, to outlive it. This will depend on the fine details of the model, see \cite{Cicoli:2022fzy} for some examples. Another caveat is that, in the above, we have implicitly assumed that the soft terms are of the same order as the gravitino mass,  $m_{\rm{soft}} \sim m_{3/2}$, the naive expectation for gravity-mediated SUSY breaking. This is not true for \emph{sequestered} scenarios where the SM brane is localized on a small cycle away from the bulk of the Calabi-Yau, where soft terms can be hierarchically smaller thanks to certain cancellations \cite{Blumenhagen:2009gk,Aparicio:2014wxa} (see also \cite{Reece:2015qbf} for an EFT analysis). In particular, depending on the precise model building details (for example the uplifting mechanism) one can have soft scalar masses with a parametric scaling 
\begin{equation}
    m_H \sim m_{3/2} \sqrt{\frac{m_{3/2}}{M_P}} \sim M_P \left(\frac{M_P}{\mathcal{V}}\right)^{3/2}\quad \quad {\rm{or}} \quad \quad m_H \sim  \frac{m_{3/2}^2}{M_P} \sim M_P \left(\frac{M_P}{\mathcal{V}}\right)^{2}.
\end{equation}
In either case, the decay rates $\Gamma_{\phi_b \rightarrow h h}$  would have no parametric enhancement with respect to the decay rate \eqref{eq:mdecays} into volume axions.

In conclusion, the end of matter domination epoch is determined by the decay rate of the longest-lived, cosmologically active modulus. While this may ultimately depend on the details of the post-inflationary dynamics and of moduli stabilization, we can distinguish two well-motivated scenarios involving decays of the (universal) volume modulus. The first one corresponds to the case where the volume decays promptly through an enhanced coupling to the Higgs, and it is responsible for reheating (\emph{i.e} no other moduli outlive it). Then, there will be a relatively much shorter matter domination epoch, with a duration given by
\begin{equation}
    \frac{H_{\rm{m}}}{H_{\rm{reh}}} \simeq \frac{m_{\Phi}}{\Gamma_{\Phi\rightarrow h h}} \simeq \frac{M_P^2}{m_{\Phi}^2} \frac{1}{\left(\alpha_{\rm{loop}} \mathcal{V} \right)^2}.
\end{equation}
The other case is when the last modulus to decay (which could be the volume in the case of suppressed soft terms, but possibly something else) does so with the rate \eqref{eq:mdecays}, so that
\begin{equation}
    \frac{H_{\rm{m}}}{H_{\rm{reh}}} \simeq \frac{m_{\Phi}}{\Gamma_{\Phi\rightarrow h h}} \simeq \frac{M_P^2}{m_{\Phi}^2}.
\end{equation}
Depending on the model-building details, all intermediate cases may also be allowed.

As will be remarked extensively in the next section, the length of the matter domination epoch is actually a crucial detail for GW phenomenology. Matter epochs give rise to a GW spectrum with negative slope (see Sec. \ref{ssc:fdct}), which can strongly impact the shape of a cosmic string GW background. We will see in Section \ref{sc:GW} how this will dramatically impact the signal in the interferometer band, and is thus very relevant to observations. As a result, there can also be a suppression of the whole high-frequency spectrum to a lower amplitude by many orders of magnitude.
In what follows we will distinguish two scenarios:
\begin{itemize}
    \item \textbf{Vanilla LVS.}
    We assume that the volume modulus decays gravitationally as in Eq.~\eqref{eq:mdecays} and ignore dark radiation issues.
    Fitting to the PTA signal, we will see that the vanilla LVS predicts a change of slope in the LISA band, with the particular frequency depending on the mass of the modulus.
    Due to the gravitational decay, the period of early matter domination is large and dilutes any high-frequency signal.
    \item \textbf{Short Volume Lifetime (SVL-LVS).}
    We consider instead decay as in Eq.~\eqref{eq:fasdtdecay}.
    The prediction at the LISA band is then as in standard cosmology.
    However, we will observe that for a large hierarchy between the inflation and late-time scales, the high-frequency signal can be very large with a spectral index that cannot be reproduced in scenarios with constant tension.
\end{itemize}

Finally, it should be noted that non-standard cosmologies have the potential to boost GW signals other than the ones coming from cosmic superstrings. This is most relevant for sources that are active at very early times, such as the primordial GW background from inflation. In \cite{Co:2021lkc,Gouttenoire:2021wzu,Gouttenoire:2021jhk} (see also~\cite{Chowdhury:2022gdc,Chowdhury:2023opo}), it was found that a long, early kination epoch can enhance the primordial GW background to observable levels, and one might worry this could dominate the signal we describe. For the scenarios we consider, however, the spectrum from cosmic superstrings will always be larger by many orders of magnitude.\footnote{Best-case scenarios for the GW background from inflation estimate $h^2 \Omega_{\rm{inf}} \sim 10^{-16}$, around 8 orders of magnitude below the radiation plateau in our benchmark in Section~\ref{sec:scenarios}.
It is a well known fact (and a simple conclusion from Section~\ref{sc:GW}) that the dependence on $G\mu$ of the amplitude of the radiation plateau is $h^2 \Omega_{\rm{GW}}\sim \sqrt{G\mu}$.
For the amplitudes to be equal we would thus require (with probabilities as in our benchmark in Section~\ref{sec:scenarios}) $G\mu \sim 10^{-28}$, or an internal volume $\mathcal{V}\sim 10^{26}$, which yields a TeV string scale.}
At low frequencies, where the effect of such modified epochs cannot be felt, the magnitude of $h^2 \Omega_{\rm{GW}}$ from cosmic superstrings will lie close to the region probed by PTA observations for realistic values of the parameters (such as the string tension), far above the prediction for the inflationary background. As the frequency increases, the amplitude for both signals can either decrease during matter domination (with the same rate) or increase during kination (with a larger rate for superstrings).\footnote{Only in the case of varying tension.
Also, the turning point frequencies may be shifted.}
In particular, the effect of the modified expansion rate translates into a peak for the inflationary background, given by
\begin{eqnarray}
    h^2\Omega_{\rm{inf}}^{(\rm{peak})} \simeq  h^2 \Omega_{\rm{inf}} (f_{\Delta}) e^{2 N_{kin}-N_m}.
\end{eqnarray}
In the above, $f_{\Delta}$ is the frequency corresponding to the end of the matter domination at $t_{\Delta}$, and $N_{kin},N_m$ are the e-folds of kination and matter domination respectively. This must compared with \eqref{eq:kination-peak}, which is always larger by a factor $\lr{\frac{\mu(t_0)}{\mu{(t_{\Delta})}}}^2$, the ratio of the string tensions at the initial and final times.

\subsection{Effect on the string tension}

To close of this section, let us examine how the fundamental string tension depends on time in the examples we have considered. 
Let us remind the reader that this can be obtained by simply plugging in the time-dependence of the compactification volume\footnote{And also of the dilaton, which here is constant.} in \eqref{eq:tension}. In the case of volume kination, one can combine Eqs \eqref{eq:vol} and \eqref{eq:volev} to obtain
\begin{equation}\label{eq:kin-tension}
    \mu_{\rm{kin}} (t) = \frac{g_s^2 M_P^2}{4 \pi \mathcal{V}_0^{\rm{kin}}} \frac{t_0^{\rm{kin}}}{t},
\end{equation}
where $t_0^{\rm{kin}}$ and $\mathcal{V}_0^{\rm{kin}}$ are the time and volume corresponding to the beginning of kination. This result is universal, and does not depend on the details of the compactification, such as the potential for the volume. Moreover, during kination $\rho \sim a^{-6}$, so that in the notation of Section \ref{ssc:tdte} $n=6$ and $q=1$. Since $ q <2$ and $ n q > 4$, the analysis of \ref{ssc:tdte} applies.

Similarly, during a radiation or a matter tracker with a potential $V=V_0 e^{-\lambda \frac{\phi}{M_P}}$,\footnote{The expression would change for a tracker obtained from an arbitrary fluid, with equation of state $P=(\gamma-1) \rho$.} one obtains
\begin{equation}
    \mu_{\rm{tr}} (t) = \frac{g_s^2 M_P^2}{4 \pi \mathcal{V}_0^{\rm{tr}}}  \left(\frac{t_0^{\rm{tr}}}{t} \right)^{\frac{2}{\lambda}\sqrt{\frac{3}{2}}}.
\end{equation} 
The result now depends on the slope of the scalar potential, making it in principle an observable quantity. For the value of in LVS, $\lambda = \sqrt{\frac{27}{2}}$, the expression becomes
\begin{equation}
     \mu^{LVS}_{\rm{tr}} (t)=  \frac{g_s^2 M_P^2}{4 \pi \mathcal{V}_0^{\rm{tr}}}  \left(\frac{t_0^{\rm{tr}}}{t} \right)^{\frac{2}{3}}.
\end{equation}
In this case, $n=4$ and $q=\frac{2}{3}$.

\subsection{Schematic timeline}

\begin{figure}[h]
    \centering
    \includegraphics[width=0.8\linewidth]{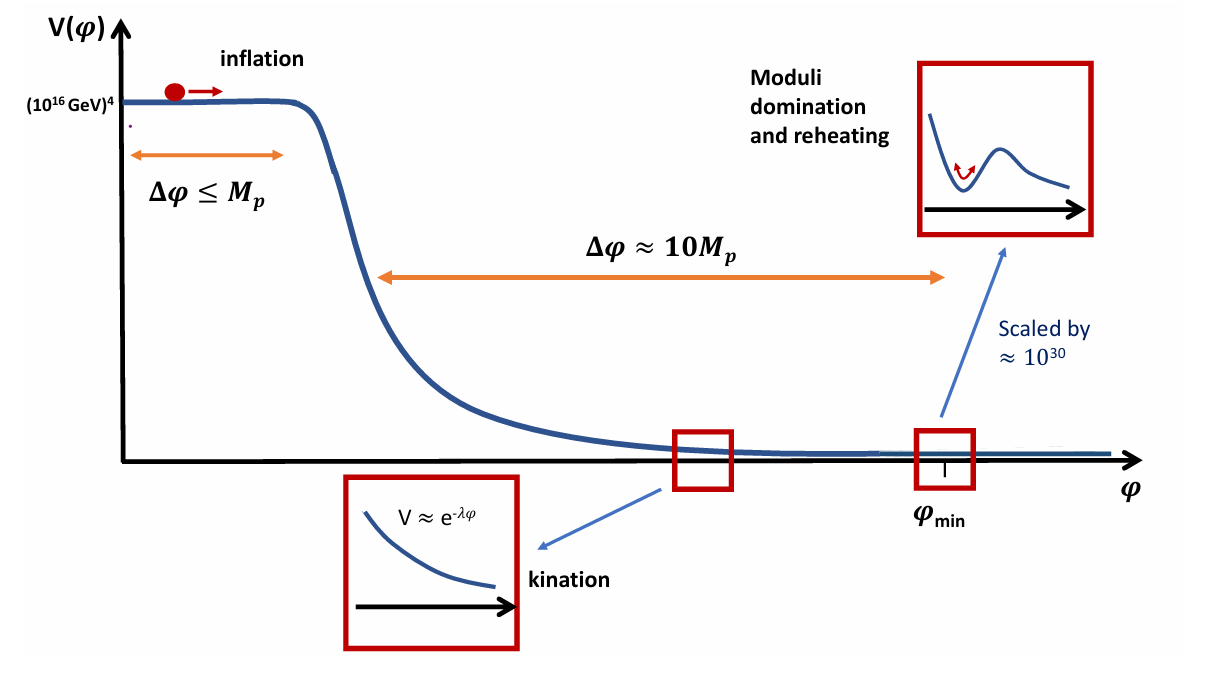}
    \caption{\it Schematic timeline of the scenario. Adapted from~\cite{Cicoli:2023opf}. }
    \label{fig:timeline}
\end{figure}

We have seen how a typical stringy, post-inflationary cosmology can be approximated by a sequence of distinct epochs, each dominated by different components.
Each epoch is characterized by a constant equation of state, with relatively short transients (lasting around one e-fold) in between. A rather general case, studied in this paper, is when the evolution is driven by the rolling of the universal volume modulus. Let us now summarize for later convenience the phases arising in this case, in the context of the Large Volume Scenario or any other decaying exponential potential for the volume modulus, sketched in Fig.~\ref{fig:timeline} (see also Table 1 of \cite{Apers:2024ffe}).
\footnote{In this article we are not concerned about the mechanism driving inflation.
If the volume modulus was not the inflaton, we neglect the physics that stabilises it during inflation and assume that the inflaton is stabilised after inflation.
If it was the inflaton, we neglect the physics that provides a slow-roll potential in the inflationary regime.
These approximations are well justified sufficiently far (in field space) from the inflationary expectation value of the modulus.}

After inflation, there will be a kination phase, followed by a brief epoch of radiation/matter domination and then a radiation/matter tracker. Once the volume reaches the vacuum, it will give rise to early matter domination by oscillating around the minimum of the potential, and finally reheat the Universe by decaying to radiation. While the different transients between these epochs can partly be described analytically, we will not give all of the details here, and the interested reader is referred to Section 2 of \cite{Apers:2024ffe}. In any case, we will also make us of numerical solutions to \eqref{eq:sys} to calculate GW spectra. With these assumptions in mind, a schematic timeline proceeds as follows.
Kination begins at a time
\begin{equation}
    t_0= \frac{M_P}{\sqrt{3} \Lambda^2_{{\rm inf}}},
\end{equation}
where $\Lambda_{{\rm inf}}$ denotes the scale of inflation, \emph{i.e.} $ \Lambda_{{\rm inf}}= V_0^{1/4} e^{-\lambda \frac{\Phi_0}{4 M_P}}$.
During this phase, as discussed around Eq.~\eqref{eq:kin-tension}, the string tension will evolve as $G \mu(t) \sim 1/t$.
Let us now assume the existence of another fluid, with equation of state $P= (\gamma-1) \rho$ and abundance $\Omega_{\gamma} \equiv \varepsilon_0 \ll 1$. Neglecting the potential, one finds \cite{Apers:2024ffe} 
\begin{equation}
    x(a)= \frac{1}{\sqrt{1+ \varepsilon_0 \left(\frac{a}{a} \right)^{3(2-\gamma)}}},
\end{equation}
so that kination-radiation/matter equality is reached at
\begin{equation}
    t_{\gamma} \simeq t_0\varepsilon_0^{\frac{1}{\gamma-2}}.
\end{equation}
During this radiation/matter dominated epoch, one can further estimate that kinetic-potential energy equality is reached at 
\begin{equation}
    t \simeq t_0 \varepsilon_0^{\frac{4+\gamma \left(\lambda \sqrt{\frac{2}{3}} -2\right) }{4(\gamma-2)}},
\end{equation}
at which point $\dot{\Phi}$ (and so the time dependence on the tension) becomes negligible. The tracker is reached when the potential catches up with the matter/radiation, corresponding to a time
\begin{equation}
    t_t \simeq t_0 \varepsilon_0^{\frac{1+\frac{\lambda}{2}+
    \sqrt{\frac{2}{3}} - \frac{2}{3} }{\gamma-2}}.
\end{equation} 
In the tracker regime, the string tension will scale as $G \mu(t) \sim 1/t^{2/3}$, both in the matter and radiation cases. This epoch will last until the onset of matter domination, when the volume will start oscillating around its minimum at $t \simeq m_{\Phi}^{-1}$. As the last step, all of the energy will be converted into radiation during reheating, at a time
\begin{equation}
 t_{{\rm reh}} \simeq \frac{1}{\Gamma_{\Phi}}, \quad \quad {\rm{with}} \quad \quad
\frac{1}{48 \pi} \frac{m_{\Phi}^3}{M_P^2} \lesssim \Gamma_{\Phi} \lesssim \frac{(\alpha_{{\rm{loop}}} \mathcal{V})^2}{16 \pi} \frac{m_{\Phi}^3}{M_P^2}.
\end{equation}
The lower and upper bound correspond to the two scenarios discussed towards the end of section \ref{ssc:mdmd}, Vanilla LVS and SVL-LVS respectively, where the decay is either gravitational in nature or Higgs-mass mediated. In principle, any intermediate value could also be allowed, but this will not be considered explicitly in the rest of the paper.

To close off, let us remark that the above picture is indeed schematic, with the equations being correct up to $\mathcal{O}(1)$ factor and intended to provide simple, analytical approximations to the evolution of the background useful for an intuitive understanding of the physics at play. For the actual computations of GW spectra presented in the next sections, the values of the various background cosmological parameters (scale factor, Hubble rate, string tension...) are computed numerically. In particular, we numerically solve the system \eqref{eq:sys} from the end of inflation to the beginning of the matter-dominated epoch, when the field starts oscillating around the minimum. From there onward, it is simpler to assume a standard radiation-matter universe, with a matter component that decays to radiation as described in \ref{ssc:mdmd} and with initial conditions determined by the previous evolution. Finally, this can be patched to the standard cosmological history before the onset of BBN.

\section{The GW background}\label{sc:GW}

In this Section we follow~\cite{Cui:2018rwi} and provide analytic estimates for the amplitude and spectral index of the GW spectrum sourced by a network of cosmic superstrings, as a function of the equation of state of the background when the loops are created and when they source the dominant contribution to the spectrum.
In Section~\ref{ssc:fdct} we reproduce the results of~\cite{Cui:2018rwi}, generalizing them to include expressions for the amplitude as a function of $G\mu$, $\alpha$, $P$ and $\chi$ (the latter is defined in Eq.~\eqref{eq:def-chi}).
In Section~\ref{sec:fdvt}, we generalize the computation to the case where the tension varies as a power law in time, as motivated by moduli dynamics in string theory constructions.
We observe that this variation can render qualitatively different changes in the GW spectrum.
In particular, it induces spectral indices that cannot be reproduced in scenarios with constant tension, and the maximum contribution to the amplitude at a given frequency is sourced at the time of formation of the loop (while with constant tension this occurs at half of its lifetime).

Let us thus compute the spectral shape of the GW background released by the loops (see~\cite{Sousa:2024ytl} for a recent review).
First, we assume that the loops with length $\ell$ radiate in discrete harmonics given by
\begin{equation}\label{eq:emission-freq}
    \Tilde{f}_k=\frac{2k}{\ell}\, , \qquad k \in \mathbb{N}^+
\end{equation}
and let us for the moment focus on the first mode $k=1$.

The spectrum of GWs with frequency $f$ today created by loops emitting in their fundamental mode is given by summing the contribution over all emission times $\tb$ of loops formed at a time $t_i$ and which have length $\ell=2/\Tilde{f}=2 a(\tb)/f a_0$ at the time $\tb$.
Noting that GW radiation redshifts like $a^{-4}$, while nonrelativistic loops redshift like $a^{-3}$, we have
\begin{equation}
    \Omega_{\rm{GW}}^{(1)} (f) \equiv \frac{1}{\rho_c} \frac{d\rho_{\rm{GW,0}}}{d \log f}=\frac{f}{\rho_c}
    \int_{t_0}^{t_i}{d\tilde{t}\, 
    \lr{\frac{a(\Tilde{t})}{a_0}}^4 \frac{dE^{(1)}_{\rm{GW}}}{d\Tilde{t}}
    \lr{\frac{a(t_i)}{a (\Tilde{t})}}^3 \frac{d \ell (\Tilde{t})}{df} \frac{dt_i}{d\ell (\Tilde{t})} \frac{dn}{dt_i} }\, . 
\end{equation}
This equation sums over all times $\Tilde{t}$ the (redshifted) energy in GWs sourced from loops with length $\ell(\Tilde{t})=\frac{2 a(\Tilde{t})}{f a_0}$.
The density of the number of loops per unit length is written in terms of the (redshifted) initial loop distribution, taking special care of the variation of the size of the interval from $(t_i,t_i +dt_i)$ to $(\ell(\Tilde{t}),\ell(\Tilde{t})+d\ell(\Tilde{t}))$.
In addition, $E^{(1)}_{\rm{GW}}$ denotes the fraction of energy that is released into the first mode.
In general, one writes
\begin{equation}
    \frac{dE_{\rm{GW}}}{d\Tilde{t}}=\sum_{k=1}^{\infty} \frac{dE^{(k)}_{\rm{GW}}}{d\Tilde{t}} \, ,
\end{equation}
which in the language of Eq.~\eqref{eq:energy-loss} corresponds to
\begin{equation}\label{eq:def-gamma-k}
    \Gamma=\sum_{k=1}^{\infty}\Gamma^{(k)} \equiv 
    \sum_{k=1}^{\infty} \frac{\Gamma}{k^{4/3} \zeta (4/3)}\, ,
\end{equation}
where the last identity is written in terms of the Riemann zeta function, $\zeta(q)=\sum_{k=1}^{\infty}k^{-q}$ and defines $\Gamma^{(k)}$.
The power of $4/3$ assumes that the contribution to the GW spectrum is sourced by cusps in the loop (for reference, $\zeta(4/3)\simeq 3.60$).

Throughout the rest of the paper, we use $\rho_c^{-1}=\frac{8\pi}{3} \frac{G}{H_0^2}$ and Eqs.~\eqref{eq:lprod} and~\eqref{eq:def-gamma-k} to write the contribution from the mode $k$ to the gravitational wave spectrum as:
\begin{equation}\label{eq:spectrum}
\begin{split}
   & \Omega_{\rm{GW}}^{(k)}(f)=
    \frac{16\pi}{3}\frac{\Gamma^{(k)} k}{H_0^2f}
    \frac{\mathcal{F}_{\alpha}}{\alpha P^{2(1-\beta)+\chi}} \times \\ 
    \int_{t_i}^{t_0} d\tb \, 
    (G\mu (\Tilde{t}))^2 \, &
     \frac{C_{\rm{eff}}(t_i)}{t_i^4}  
    \frac{dt_i}{dl(\Tilde{t})}
    \left(\frac{a(t_i)}{a(\tb)}\right)^3\left(\frac{a(\tb)}{a_0}\right)^5  
    \Theta \left[ t_i-t_{\rm{osc}}\right]
    \Theta \left[ \alpha t_i-\ell_*\right]\,.
    \end{split}
\end{equation}
The Heaviside theta functions  ensure that the loop formation time cannot be extrapolated before the network is formed, and impose a high frequency cutoff in terms of a minimum length $\ell_*$ that loops can reach before other channels become more efficient (see~\cite{Ghoshal:2023sfa} and references therein).
Of course, the total spectral shape is given by the sum of these contributions, namely
\begin{equation}
    \Omega_{\rm{GW}}(f)=\sum_{k=1}^{\infty}\Omega_{\rm{GW}}^{(k)}\, ,
\end{equation}
and the plots sum up to $k=10^4$ modes.
\subsection{Frequency dependence: constant tension}\label{ssc:fdct}

The GW spectrum in standard cosmology can be broadly classified into a growing contribution, whose peak ought to explain the PTA signal, and a flat plateau at higher frequencies corresponding to a period of radiation domination.
Higher frequencies correspond to earlier time scales, rendering the high-frequency regime as an interesting parameter space to test pre-BBN cosmology. 

As such, one of the most interesting features of cosmic strings is that, if they are present, because of the properties of the scaling regime, their associated GW background carries information about the cosmology following their formation.
In particular, the frequency dependence of the spectrum will vary depending on the equation of state of the background.
In addition, a time-dependent string tension - as would be expected if the volume modulus was rolling down its potential - leads to different spectral dependence and thus provides a smoking-gun signature for scenarios as in~\cite{Apers:2022cyl,Conlon:2022pnx,Apers:2024ffe}.

The background behavior of the spectral index is well understood~\cite{Cui:2018rwi,Gouttenoire:2019kij}, but a similar analysis with a time-dependent tension has only been carried out in a special case~\cite{Emond:2021vts}.
Let us first review the case where the string tension is constant and solely focus on the behavior of the first mode; higher modes yield a qualitatively different behavior only in the case of matter domination (the effect of higher modes has been studied analytically in~\cite{Gouttenoire:2019kij}).
To do so, we follow the logic in~\cite{Cui:2018rwi} and generalize their approach to include the dependence on the intercommutation probability and the parameter $\chi$.
Let us consider two cosmological phases with 
\begin{equation}
    \rho (t)=  \begin{cases} 
      \rho_\Delta \lr{\frac{a_\Delta}{a(t)}}^m\, , \qquad & t<t_\Delta \\
      \rho_\Delta \lr{\frac{a_\Delta}{a(t)}}^n \, , \qquad & t\geq t_\Delta 
   \end{cases} \, .
\end{equation}
In what follows we assume that the scaling regime is maintained throughout an immediate transition between the two phases in favor of an analytic understanding of the situation, but note that these issues have been studied in the past~\cite{Sousa:2013aaa,Sousa:2016ggw,Avgoustidis:2025svu}.

There are two possibilities to take into account: either the loops are formed and emit GWs during $t<t_\Delta$ or $t>t_\Delta$, or they are formed at $t<t_\Delta$ but source their dominant contribution to the spectrum during $t>t_\Delta$.
In the latter case (which simply reproduces the others by setting $n=m$), we have
\begin{equation}
    \lr{\frac{a(t_i)}{a(\Tilde{t})}}^3=\lr{\frac{a(t_i)}{a_\Delta}}^3\lr{\frac{a_\Delta}{a(\Tilde{t})}}^3=\lr{\frac{t_i}{t_\Delta}}^{6/m}\lr{\frac{t_\Delta}{\Tilde{t}}}^{6/n}\, .
\end{equation}
The behaviour of the spectrum with respect to $f$ is understood by studying, at a fixed $f$, which is the time $\Tilde{t}$ where the loop population sources the largest contribution to the GW amplitude.
Notice that the formation time carries information about the length of the loop at time $\Tilde{t}$, from~\eqref{eq:lt} we have:
\begin{equation}
    t_i=\frac{1}{\alpha P^{\chi}+\Gamma G\mu}\lr{\frac{2k}{f}\frac{a(\Tilde{t})}{a_0}+\Gamma G\mu \Tilde{t}} \, , \qquad \frac{dt_i}{d\ell(\Tilde{t})}=\frac{1}{\alpha P^{\chi}+\Gamma G\mu}.
\end{equation}
It is useful to define the quantity $f_\Delta$:
\begin{equation}\label{eq:ti-fdelta}
    f_\Delta \equiv \frac{2}{\alpha P^\chi}\frac{1}{t_\Delta z_\Delta} \, , \qquad \frac{t_i}{t_\Delta}\simeq \frac{f_\Delta}{f}\lr{\frac{\Tilde{t}}{t_\Delta}}^{2/n}+\frac{\Gamma G\mu}{\alpha P^{\chi}} \frac{\Tilde{t}}{t_\Delta}\, , 
\end{equation}
where here and in what follows we neglect $\Gamma G\mu$ over $\alpha P^\chi$, and $z_{\Delta}$ is the redshift at the time $t_{\Delta}$.
Notice that keeping the second term in Eq.~\eqref{eq:ti-fdelta} does not contradict this statement since the relative terms could be of the same order for a choice of times and frequencies.
Indeed, it is precisely this regime that turns out to be interesting for loops in cosmologies with constant string tension (and we will see how this need not be the case otherwise).

We are now in a position to estimate the behavior of the gravitational wave spectrum.
At fixed $f$, noting that for any $n$ of interest $t_i$ grows monotonically with $t$, we can decompose the integral into two pieces:
\begin{equation}\label{eq:gw-std-lf}
    h^2 \Omega_{\rm{GW},1}^{(1)}=A_\Delta \lr{\frac{f}{f_\Delta}}^{3-6/m}
    \int_{t_F}^{t_M (f)} \frac{d\Tilde{t}}{t_\Delta}
     \, \lr{\frac{\Tilde{t}}{t_\Delta}}^{2(6/m-4)/n+4/n } \, .
\end{equation}
\begin{equation}\label{eq:gw-std-hf}
    h^2 \Omega_{\rm{GW},2}^{(1)}=A_\Delta \lr{\frac{f}{f_\Delta}}^{-1} \lr{\frac{\Gamma G\mu}{\alpha P^{\chi}}}^{6/m-4}
    \int_{t_M (f)}^{t_0} \frac{d\Tilde{t}}{t_\Delta}
     \, \lr{\frac{\Tilde{t}}{t_\Delta}}^{6/m-4+4/n} \, .
\end{equation}
Here we have defined the amplitude
\begin{equation}\label{eq:univ-amplitude}
    A_\Delta=\frac{8\pi}{3} \frac{\Gamma^{(1)}}{(H_0t_0)^2} \frac{\mathcal{F}_\alpha}{\alpha P^{2(1-\beta)+\chi}} 
    \lr{G\mu_\Delta}^2 \, C_{\rm{eff}} \lr{\frac{t_0}{t_\Delta z_\Delta^2}}^2
    \equiv 
    A \lr{G\mu_\Delta}^2\,  C_{\rm{eff}}
    \lr{\frac{t_0}{t_\Delta z_\Delta^2}}^2\, , 
\end{equation}
where $A$ is universal and we define $G\mu_\Delta \equiv G\mu (t_\Delta)$ anticipating the case where the tension varies.
We note here that the effect of varying the intercommutation probability is a uniform boost in the amplitude for $\chi=0$ (as in~\cite{Ellis:2023tsl}), and an additional contribution to the amplitude and a shift in the characteristic frequency $f_\Delta$ for $\chi \neq 0$.

In addition, we observe that the frequency dependence of the spectrum depends on the exponents in Eqs.~\eqref{eq:gw-std-lf} and~\eqref{eq:gw-std-hf}, and therefore in the equation of state of the fluid governing the expansion of the Universe.
For values of interest $(m,n)\in \lbrace 3,4,6 \rbrace$ the first exponent of $\Tilde{t}/t_\Delta$ ($p_1$) is always larger than $-1$ while the second ($p_2$) is larger than $-1$ except in a matter-radiation transition (\emph{i.e.} at the end of a period of early matter domination) and during matter domination.
We will have more to say about such periods in Sec.~\ref{sec:scenarios}, since the spectrum is sensitive to higher modes, but let us for the moment assume that $p_1 >-1$ and $p_2<-1$.
Then, the integral is dominated at a time $t_M$ defined as the time at which the two contributions to $t_i$ are equal, namely
\begin{equation}\label{eq:t-max-std}
    \frac{t_M}{t_\Delta} \equiv \lr{\frac{f_\Delta}{f} \frac{\alpha P^{\chi}}{\Gamma G\mu}}^{n/(n-2)} \, .
\end{equation}
It is also easy to see that at $t_M$ the loops are half of their initial length.
All in all, we find that in the regime $t_0>t_M(f)>t_F$, the dominant contribution to the spectrum reads (up to an order one constant and corrections in powers of $t_M(f)/t_0$ and $t_F/t_M(f)$)
\begin{equation}\label{eq:gw-standard}
    h^2 \Omega_{\rm{GW}}^{(1)}\simeq  A_\Delta \lr{\frac{f}{f_\Delta}}^{B}\, \lr{\frac{\alpha P^{\chi}}{\Gamma G\mu}}^{C}\, . 
\end{equation}

Where $B=2(mn-m-3n)/(m(n-2))$ and $C=(12+mn-4m)/(m(n-2))$.
There are subtleties to the analysis above that will be important later.
Firstly, we must note that we have assumed that $t_M(f)<t_0$, where $t_0$ indicates the onset of a different phase of the Universe (eg: radiation-matter transition), or the age of the Universe today.
Otherwise the dominant contribution to the spectrum at this frequency must be computed in the later phase with times $t>t_0$ or, in the case of the latest epoch of matter domination, it is just evaluated today.
In the latter case:
\begin{equation}
    h^2\Omega_{\rm{GW}}^{(1)} \simeq A_0 \lr{\frac{f}{f_0}}^{3-6/m} \, .
\end{equation}
This reproduces the standard growth of the spectrum at very low frequencies.

The second subtlety has to do with the effect of theta functions.
The effect of these is to change the limit of integration by another quantity (which may or may not depend on the frequency).
There are two possible cases of this: by assumption, we need to require  $t_i (f,t_M)>t_F$ (the loops are formed after network formation) and this gives an upper frequency
\begin{equation}
    f<f_F \lr{\frac{\alpha P^{\chi}}{\Gamma G\mu}}^{2/n}2^{1-2/n}\, ,
\end{equation}
where $f_F$ evaluates $f_\Delta$ at the formation time of the network.

At larger frequencies one simply evaluates~\eqref{eq:gw-std-hf} at $t_F$, finding
\begin{equation}
    h^2\Omega_{\rm{GW}}^{(1)} \simeq A_F \lr{\frac{f}{f_K}}^{-1} \lr{\frac{\alpha P^{\chi}}{\Gamma G\mu }}^{6/m-4}\, ,
\end{equation}
which is particularly sensitive to the addition of higher modes.
The other case is the imposition that $t>t_i(f,t)$.
Because the maximum contribution to these integrands occurs at $t_M(f)$, which is by construction larger than $t_i$, this constraint does not give any additional features.
We will see this ceases to be the case in the more complicated case where the string tension is allowed to vary.

Table~\ref{tab:standard-fluid} summarizes the possible slopes that typical fluid equation of states considered in the literature provide.
We will see that varying the string tension results in different predictions, providing a possible smoking gun signature for scenarios with varying moduli.
\begin{table}[h!] 
\centering
\begin{tabular}{ |c||c|c|c|  }
 \hline
$ m $ vs $n$  & $m=3$ & $m=4$ & $m=6$\\
 \hline\hline
 $n=3$ & $-1 $   & $-1$ &   $1$\\
 \hline
 $n=4$ &  $-1$  & $0$   & $1$ \\
 \hline
 $n=6$ & $-1/2$ & $1/4$ &  $1$\\
 \hline
\end{tabular}
\caption{Expected spectral index of the GW spectrum sourced by the first mode of loops that are created when the background energy density redshifts like $\rho \sim a^{-m}$ and radiate GWs with $\rho \sim a^{-n}$, according to Eq.~\eqref{eq:gw-standard}.
Negative spectral indices are altered by the effect of higher modes to $-1/3$.}
\label{tab:standard-fluid}
\end{table}

\subsection{Frequency dependence: varying tension}\label{sec:fdvt}
Let us now consider what happens if the tension of the strings varies for a period of time.
The strategy is the same as before: allow the background equation of state to vary, but this time accompanied by a change in the variation of the tension, namely:
\begin{gather}
\rho (t)=
    \begin{cases}
        \rho_\Delta \lr{\frac{a_\Delta}{a(t)}}^m \, , \qquad t<t_\Delta \\
        \rho_\Delta \lr{\frac{a_\Delta}{a(t)}}^n \, ,\qquad t \geq t_\Delta     \end{cases}
    \, , \qquad 
    G\mu (t) =
    \begin{cases}
        G\mu_\Delta \lr{\frac{t_\Delta}{t}}^p\, , \qquad t<t_\Delta \\
        G\mu_\Delta \lr{\frac{t_\Delta}{t}}^q \, ,\qquad t\geq t_\Delta
    \end{cases}
    \, .
\end{gather}
Particularly interesting are the cases $p=1$, $q=2/3$, $m=6$, $n\in \lbrace 3,4 \rbrace$ and $p=2/3$, $q=0$, $m\in \lbrace 3,4 \rbrace$ and $n=3$ since these correspond, respectively, to a kination period followed by a tracker, and the end of a (matter or radiation) tracker followed by a period of matter domination given by oscillations of the modulus responsible for varying the string tension (the volume modulus being a good candidate for these choices of $p$ and $q$).

In this case the results are different due to the evolution equation of the loops \eqref{eq:loop-ev-vary-tension}-\eqref{eq:loop-tracker}, which we re-write here for convenience:
\begin{gather}
    \ell(t)=\lr{\frac{t}{t_i}}^{p/2} \lr{\alpha P^{\chi}-\frac{\Gamma G \mu_0}{\frac{3p}{2}-1}\lr{\frac{t_0}{t_i}}^p}t_i+\frac{\Gamma G\mu_0}{\frac{3p}{2}-1}\lr{\frac{t_0}{t}}^{p-1}t_0  , \qquad (p\neq 2/3)
    , \\
    \ell(t)=\left[\alpha P^{\chi} - \Gamma G\mu_0 \lr{\frac{t_0}{t_i}}^{2/3}\log \lr{\frac{t}{t_i}}\right]\lr{\frac{t}{t_i}}^{1/3}t_i  \, , \qquad (p=2/3) \, ,
\end{gather}
(note that $p=0$ reproduces the standard case with constant tension).
To make progress, let us assume that the periods where the string tension varies are short enough\footnote{This is immediate for $p\geq 2/3$ which are the scenarios with varying tension that we consider in this article.} so that 
\begin{equation}
    \lr{\frac{t}{t_i}}^{1-3p/2} \ll \frac{\alpha P^{\chi}}{\Gamma G\mu (t_i)}\, .
\end{equation}
If this is the case, one can neglect the effects of energy (length) loss through GW emission in the evolution of the length of the loop, which is then only stretched due to its tension varying.
One can thus approximate
\begin{equation}\label{eq:ti-vary-same}
    \frac{\ell(t)}{t_\Delta}\simeq \alpha P^{\chi} \lr{\frac{t}{t_\Delta}}^{p/2} \lr{\frac{t_i}{t_\Delta}}^{1-p/2}\, \rightarrow \frac{t_i}{t_\Delta} \simeq \lr{\frac{f}{f_\Delta}}^{2/(p-2)} \lr{\frac{t}{t_\Delta}}^{(4/m-p)/(2-p)}\, ,
\end{equation}
\begin{equation}
    \frac{dt_i}{d\ell(t)}\simeq \frac{2}{(2-p)\alpha P^{\chi}} \lr{\frac{t_i}{t}}^{p/2} \, .
\end{equation}

This equation applies for loops that are created and source GWs within the same period.
Otherwise we need to match the evolution equations for different $p$ and $q$, resulting in:
\begin{equation}\label{eq:ti-vary-diff}
    \frac{\ell(t)}{t_\Delta}\simeq \alpha P^{\chi}\lr{\frac{t}{t_\Delta}}^{q/2}\lr{\frac{t_i}{t_\Delta}}^{1-p/2}\, \rightarrow \frac{t_i}{t_\Delta} = \lr{\frac{f}{f_\Delta}}^{2/(p-2)} \lr{\frac{t}{t_\Delta}}^{(4/n-q)/(2-p)}\, .
\end{equation}
\begin{equation}
    \frac{dt_i}{d \ell(t)} \simeq \frac{2}{(2-p)\alpha P^{\chi}} \lr{\frac{t_i}{t_\Delta}}^{p/2} \lr{\frac{t_\Delta}{t}}^{q/2}\, .
\end{equation}
These quantities are all we need to compute the GW spectrum sourced by these loops (noting, however, that depending in the details of the cosmic evolution this spectrum might always be subdominant).
Let us begin by computing the contribution at times $t<t_\Delta$.
Introducing the above expressions into Eq.~\eqref{eq:spectrum}, we find: 
\begin{gather}
    h^2\Omega_{\rm{GW}}^{(1)} \simeq \frac{2A_\Delta}{2-p}  \lr{\frac{f}{f_\Delta}}^D\int_{t_F}^{t_\Delta}{\frac{dt}{t_\Delta} \, \lr{\frac{t}{t_\Delta}}^{E}} \label{eq:gw-spec-vary}\\
    D=\frac{6(m-2)}{m(2-p)}\, , \qquad 
    E=\frac{4}{m}-\frac{5p}{2}+\frac{4-p\, m}{m(2-p)}\lr{\frac{6}{m}+\frac{p}{2}-4}\, ,
\end{gather}
which reproduces the low-frequency results of the section above (where the length-loss due to GW emission is negligible) for $p=0$.
The behaviour is now qualitatively different because in all cases of interest $m=6,\, p=1$ and $m\in \lbrace 3,4 \rbrace\, , p=2/3$ the quantity $E$ is smaller or equal than $-1$ and the spectrum is dominated by emission at \textit{early} times.
The integral is thus dominated at the earliest times, where the loops of size $\ell(t_{\rm{min}}(f))$ are formed.
This $t_{\rm{min}}(f)$ can be computed by requiring $t_i(t_{\rm{min}}(f),f)=t_{\rm{min}}$, finding\footnote{This is true provided $f<f_F$.
For $f>f_F$ the condition $t_i(t_{\rm{min}}'(f),f)>t_F$ is the relevant one, which provides a new $t_{min}(f)'$ given by
\begin{equation*}
    t_{\rm{min}}(f)'/t_F \geq (f/f_F)^{-2m/(4-p m)}\, ,
\end{equation*}
For $p \, m>4$, there is no spectrum at frequencies larger than $f<f_F$ (at least sourced by the first mode).
Otherwise the condition can be satisfied and gives the cutoff at higher frequencies, resulting in a different spectral index.
We will be interested in $p\,m>4$ in what follows.}
\begin{equation}
    \frac{t_{\rm{min}}}{t_F}=\lr{\frac{f}{f_F}}^{-m/(m-2)}\, .
\end{equation}
Thus, neglecting order one factors (and a logarithmic dependence for kination), one finds
\begin{equation}\label{eq:gw-varying}
    h^2 \Omega_{\rm{GW}}^{(1)} \simeq  A_\Delta \lr{\frac{f}{f_\Delta}}^F \, , \qquad F=\frac{2(m(p+1)-4)}{m-2} \, .
\end{equation}
This behaviour lasts for a range of frequencies whose limits can be obtained as follows.
First, impose $t_i(f,t_\Delta)<t_\Delta$ to find a minimum frequency where these loops radiate, namely $f>f_\Delta$.
Second, requiring $t_{\rm{min}}>t_F$ gives
\begin{equation}\label{eq:peak-f}
    f_\Delta <f <f_\Delta \lr{\frac{t_\Delta}{t_F}}^{(m-2)/m}=f_F\, .
\end{equation}
This in turn allows us to estimate the amplitude of the peak of the spectrum in terms only of $t_{kin}$ and $t_F$:
\begin{equation}\label{eq:gw-kin-peak}
    h^2 \Omega_{\rm{GW}}^{\rm{(1,peak)}} \simeq A_\Delta \lr{\frac{t_{kin}}{t_F}}^{2(1+p-4/m)}\, .
\end{equation}
It is important to point out that extrapolating this logic, gravitational waves arising from even earlier times (e.g: while the string network is forming) could dominate the spectrum, possibly invalidating our predictions.
We believe this is a feature rather than a bug, since it illustrates the possibility of studying the early Universe with gravitational waves.
This claim should therefore be checked against simulations that take into account the formation of the network and its approach to the scaling regime.

The case where the loops are created at $t<t_\Delta$ but source at $t>t_\Delta$ can be similarly studied, resulting in 
\begin{equation}
    D=\frac{6(m-2)}{m(2-p)}\, , \qquad 
    E=\frac{4}{n}-\frac{5q}{2}+\frac{4-q\, n}{n(2-p)}\lr{\frac{6}{m}+\frac{p}{2}-4}\, ,
\end{equation}
The integral is again dominated by \textit{early} times (namely $E<-1$) for kination (regardless of whether it is followed by a matter or radiation period, this being a tracker or not), while for the tracker periods it is dominated by late times.
Thus, in the case where only the tracker is present one would need to compare the early and late time contribution to compute the frequency behaviour, etc.
We postpone this analysis for future work and focus in the case of volume modulus kination for concreteness, where the earliest times dominate.
Let us also point out, however, that in these scenarios the spectral index will be generically different than in scenarios with constant tension.
In the case of volume modulus kination, for instance, $\Omega_{\rm{GW}}\sim f^4$, resulting in a very large boost in the signal.

\section{Benchmark scenarios}\label{sec:scenarios}
We now turn to illustrating the GW spectra arising from cosmic superstrings in the two scenarios described in Sec.~\ref{ssc:mdmd}.
We operate under the assumption that the PTA signal is explained by cosmic superstrings, and search for further consequences of this assumption\footnote{One may instead consider, given the scenarios, what are the values of microscopic parameters that are allowed by current data. We intend to explore this in more generality in future work.}.
Following Ref.~\cite{Ellis:2023tsl}, where $\beta=1/2$ and $\chi=0$, the preferred string tension and intercommutation probability are\footnote{We note here that in~\cite{Avgoustidis:2025svu} the choice is $\beta=2/3$ and $\chi \neq 0$, which is why they find a different best-fit than in~\cite{Ellis:2023tsl}.}
\begin{equation}
    G\mu= 2 \times 10^{-12}\, , \qquad P=4\times 10^{-3}\, .
\end{equation}
Using Eqs.~\eqref{eq:tension} and~\eqref{eq:JJP-prob}, this fixes the volume and the string coupling to
\begin{equation}\label{eq:benchmark}
    \mathcal{V}=3.3 \times 10^{9}\, , \qquad g_s=0.11\, ,
\end{equation}
which are completely natural values for LARGE volume scenarios.

The vanilla LVS makes a concrete prediction in the LISA band in terms of the mass and decay rate of the volume mode.
The transition frequency depends on model-dependent order one numbers, with the most likely scenario being that the LISA mission would be able to measure a deviation from standard cosmology.
Due to the large suppression of the signal from a long period of early matter domination, there is no meaningful enhancement in the high-frequency portion of the spectrum (despite an epoch of early kination). The SVL-LVS has an interesting behaviour: the high-frequency signal grows for larger hierarchies between the inflation and late-time vacuum, with a long period of kination in between.
As we will discuss, it is not easy to explain the PTA signal whilst being consistent with a large signal at high-frequencies.
We nevertheless provide an example where such a large, high-frequency signal can occur as an illustration of the interesting physics behind the fact that the signals grow with large hierarchies.
The basic physics can be applied to other well-motivated scenarios which can potentially lead to large signals (see for example \cite{Brunelli:2025ems}).

\subsection{Vanilla LVS and LISA}\label{sec:vanilla-lvs}

Consider the mass of the volume modulus in LVS~\cite{Conlon:2005ki}:\footnote{Following~\cite{Ebelt:2023clh}, we define $W_0=\sqrt{\frac{2}{\pi}}|e^{K_{\rm{cs}}/2}W_{\rm{GVW}}|$ in terms of the K\"{a}hler potential of the complex structure moduli and the Gukov-Vafa-Witten~\cite{Gukov:1999ya} superpotential.
This is because~\cite{Ebelt:2023clh} finds a $W_0$ following a Gaussian distribution around 0.}
\begin{equation}
    m_\Phi= \mathcal{O}(1) \times\frac{W_0}{\mathcal{V}^{3/2}g_s^{1/4}}
    M_P 
\end{equation}
which results in a gravitational decay rate, using Eq.~\eqref{eq:mdecays}:
\begin{equation}
    \Gamma_\Phi=\frac{\mathcal{O}(1)}{48\pi}\cdot \frac{W_0^3}{\mathcal{V}^{9/2} g_s^{3/4}}M_P\, .
\end{equation}
This decay rate then translates into a time for decay of the modulus, which in turn predicts a deviation from the flat spectrum at a frequency that we compute as follows.
Because, as discussed in Sec.~\ref{ssc:fdct}, the largest contribution to the amplitude at a frequency $f$ from loops that are formed during matter but source during radiation arises from the latest times, the spectrum at $f$ is dominated by the loops that are formed and source during radiation domination.
Out of these, the largest contribution arises at $t_M(f)$ as defined in Eq.~\eqref{eq:t-max-std}.
The transition from the flat spectrum sourced by loops that are created and emit during radiation domination is therefore predicted to occur when $t_i(t_M(f),f)>1/\Gamma_\Phi$, meaning that no loop created during radiation domination sources at its point of maximum emission $t_M(f)$ in this frequency bin.
In formulae, using Eqs.~\eqref{eq:ti-fdelta} and~\eqref{eq:t-max-std}, this implies a transition $f_{\rm{T}}$ given by:
\begin{equation}
    f_{\rm{T}}= f_\Gamma \lr{\frac{\alpha P^{\chi}}{\Gamma G\mu}}^{1/2}=
    \frac{2}{\lr{\alpha P^{\chi} \times \Gamma G\mu}^{1/2}}\frac{\Gamma_\Phi}{z_\Gamma}\, ,
\end{equation}
where the second equality defines $f_\Gamma$ as a particular case of $f_\Delta$ (cf: Eq.~\eqref{eq:ti-fdelta}) in terms of the time of reheating $t_\Gamma=1/\Gamma_\Phi$ and its redshift, $z_\Gamma$.
Assuming instantaneous reheating and standard cosmology after volume decay:
\begin{equation}\label{eq:LISA-freq}
    f_{\rm{T}} =8.6 \times 10^{-4} \text{ Hz }\,  \times \, \lr{\frac{0.1}{\alpha P^{\chi}} \times \frac{2 \times 10^{-12}}{G\mu}
    \times \frac{c_i}{48\pi}}^{1/2}
    \times  \lr{\frac{m_\Phi}{30 \text{ TeV}}}^{3/2}\, ,
\end{equation}
where we have used $t_\Delta \simeq 1/\Gamma_\Phi$.
The number is somewhat sensitive to the addition of higher modes so it should be taken as an order of magnitude approximation.
The dependence on the mass and probability, however, remains the same upon addition of higher modes.
It is worth noting that the prediction for the transition frequency is largely insensitive to $\chi$, since the dependence is $P^{-\chi/2}$, which in our other benchmarks is $\chi=0$ or $\chi=1/3$.
This renders the prediction for the frequency robust, although the amplitude does change upon varying $\chi$.

\begin{figure}[t]
    \centering
    \includegraphics[width=1.0\linewidth]{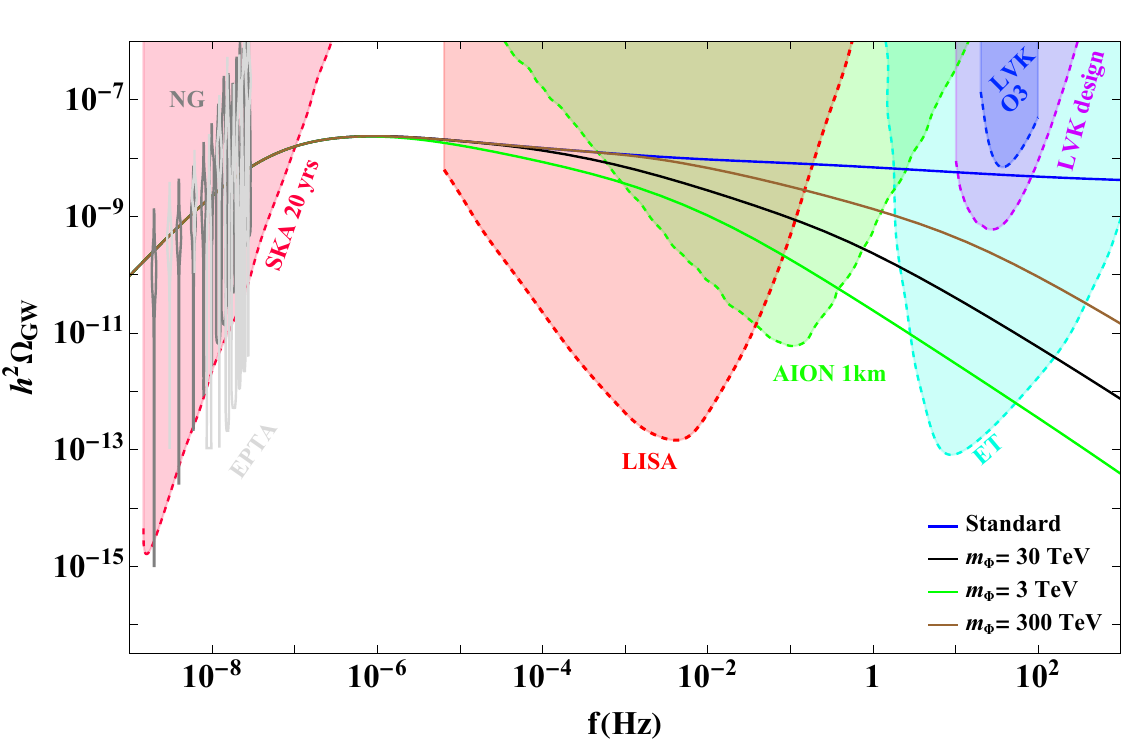}
    \caption{\it GW spectrum sourced by cosmic superstrings with $G\mu=2 \times 10^{-12}$ and $P=4 \times 10^{-3}$, in the
    vanilla LVS scenario, discussed in Section~\ref{sec:vanilla-lvs}.
    A period of early matter domination induced by the volume modulus ends by gravitational decay.
    This induces a dip in the spectrum that will be robustly probed by the LISA mission.
    A reasonable estimate of the volume mass is 30 TeV (solid black) and is related to the dip frequency by Eq.~\eqref{eq:LISA-freq}.
    We also show the spectrum in ``Standard" cosmology, i.e: radiation domination with constant tension. }
    \label{fig:vanilla-LVS}
\end{figure}

\begin{figure}[t]
    \centering
    \includegraphics[width=1.0\linewidth]{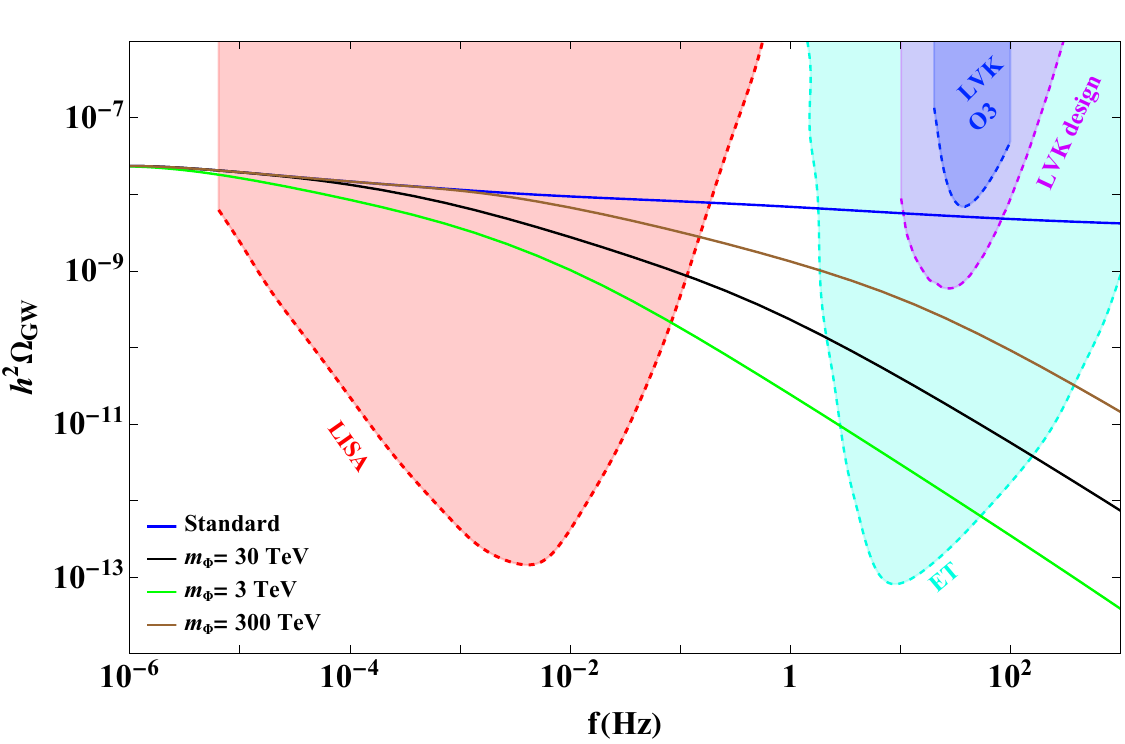}
    \caption{\it GW spectrum sourced by cosmic superstrings with $G\mu=2 \times 10^{-12}$ and $P=4 \times 10^{-3}$ in the large-scale interferometer band for the
    vanilla LVS scenario, discussed in Section~\ref{sec:vanilla-lvs}.
    A period of early matter domination induced by the volume modulus ends by gravitational decay.
    This induces a dip in the spectrum that will be robustly probed by the LISA mission.
    A reasonable estimate of the volume mass is 30 TeV (solid black) and is related to the dip frequency by Eq.~\eqref{eq:LISA-freq}.
    We also show the spectrum in ``Standard" cosmology, i.e: radiation domination with constant tension.}
    \label{fig:vanilla-LVS-small}
\end{figure}

Figure~\ref{fig:vanilla-LVS} illustrates the difference between the spectrum for standard cosmology and in vanilla LVS for the benchmark value $m_\Phi=30$ TeV, corresponding in LVS to the values in Eq.~\eqref{eq:benchmark}, together with $W_0\simeq1$ and modulo order one numbers.
The Figure also illustrates the mass-dependence of the frequency with the mass of the modulus.
A lighter volume mode corresponds to a lower-frequency signal (since the decay occurs later in time), although in vanilla scenarios these values might be in tension with achieving a successful Nucleosynthesis, as discussed in Sec.~\ref{ssc:mdmd}.
Ignoring this issue, a lighter volume mode can be achieved by choice of $W_0$ or by a stabilization mechanism leading to $m_\Phi \sim 1/\mathcal{V}^a$ with $a>3/2$.
A heavier volume, on the other hand, corresponds to a larger transition frequency.
Microscopically, this could be obtained for large values of $W_0$ (though note that $W_0$ is bounded by above for consistent description of the effective theory, see e.g.~\cite{Cicoli:2013swa}), or by a stabilization mechanism similar to the one discussed in Appendix~\ref{sec:scenario-highfreq}.
In addition, Figure~\ref{fig:vanilla-LVS-small} zooms into the interferometer band.

\subsection*{GW Detection prospects}
\label{subsec:detection_pros}

There are a multitude of presently operating and proposed GW experiments that could detect the SGWB studied in this Section\footnote{We have used the publicly available data from~\cite{Fu:2024rsm} to plot the experimental sensitivities.} (see~\cite{Caprini:2018mtu} for a review in sources and detection efforts of cosmological GW backgrounds and~\cite{Aggarwal:2020olq,Aggarwal:2025noe} for proposals at frequencies above the kHz).
In the plots, we show sensitivity curves for the following detectors:

\begin{itemize}
    \item \textbf{Ground based GW laser interferometers:} LIGO/VIRGO/KAGRA, including both latest bounds~\cite{KAGRA:2021kbb} and prospect sensitivities~\cite{LIGOScientific:2014pky,LIGOScientific:2016aoc,LIGOScientific:2019vic}, and the Einstein Telescope (ET)~\cite{Punturo:2010zz,Hild:2010id}.
    Other proposals include the Cosmic Explorer (CE)~\cite{Reitze:2019iox}.
    
    \item \textbf{Space based GW laser interferometers:}
    LISA~\cite{Bartolo:2016ami,Caprini:2019pxz,LISACosmologyWorkingGroup:2022jok} (see~\cite{Auclair:2019wcv} for prospects on cosmic strings).
    Other proposals include the Big-Bang Observatory (BBO)~\cite{Corbin:2005ny}, the Deci-Hertz Interferometer Gravitational-wave Observatory (DECIGO)~\cite{Kawamura:2006up}, Taiji~\cite{Ruan:2018tsw}, TianQin~\cite{TianQin:2015yph} and $\mu$-ARES~\cite{Sesana:2019vho}. 

    \item \textbf{Atomic interferometers:}
    AION~\cite{Badurina:2019hst,Badurina:2021rgt} (ground-based).
    Other (space-based) proposals include MAGIS~\cite{Graham:2017pmn} and AEDGE~\cite{Bertoldi:2019tck}.
    
    \item \textbf{Pulsar Timing Arrays (PTA):} 
    We use the data from the North American Nanohertz Observatory for Gravitational Waves (NANOGrav)~\cite{NANOGrav:2023gor} and European Pulsar Timing Array (EPTA)~\cite{EPTA:2023fyk} collaborations (the latter combined with data from the Indian Pulsar Timing Array~\cite{Joshi:2018ogr}).
    Other collaborations which have confirmed the detection of the SGWB are the Parkes Pulsar Timing Array (PPTA)~\cite{Reardon:2023gzh}  and the Chinese Pulsar Timing Array (CPTA)~\cite{Xu:2023wog}.
    We also include the sensitivity curves from the Square Kilometre Array (SKA)~\cite{Janssen:2014dka}.
\end{itemize} 

Let us now give reasonable bounds on the volume mass (and consequently on the transition frequency).
As discussed in Sec.~\ref{ssc:mdmd}, in scenarios with gravitational decay, successful nucleosynthesis requires the mass of the volume modulus to be $\mathcal{O}$ (TeV) or larger.
In the case of LARGE volume scenarios, it is difficult to make the mass of the volume mode larger while keeping the volume constant.\footnote{We will discuss such a possibility in Appendix~\ref{sec:scenario-highfreq}.}
In the presence of cosmic superstrings, a smaller volume (for a faster decay) implies a larger $G\mu$ which would be in tension with PTA observations.
The above arguments thus suggest that the mass of the volume mode lies in the order of magnitude of the tens of TeV, making the scenario predictive even without the requirement that the PTA signal is explained by cosmic superstrings.
The signal robustly lies in the LISA band.

\subsection{Short Volume Lifetime (SVL-LVS)}

Let us now consider the case where the tension varies.
We first provide a simple, unrealistic scenario where a background inspired in volume-modulus kination is followed by the hot Big Bang.
We obtain analytic expressions for the frequencies where the spectral index changes and for the maximum amplitude.
We then proceed to study the more realistic case where we solve numerically the dynamical system in Eqs.~\eqref{eq:sys}.
We can obtain an expression for the amplitude in terms of the number of e-folds of kination, tracker and matter domination.
We find the interesting result that the predicted signal is larger for a larger hierarchy between initial and late vacuum, because it allows for a longer period of kination with varying tension.
The standard LVS (even with decays given by Eq.~\eqref{eq:fasdtdecay}) does not allow for a large signal but other scenarios do.

\subsubsection{Toy example}

We now turn to a simple example which illustrates the basic features of the above analysis.
Consider a network formed at very early times (say, $\Lambda=10^{-4} M_P$) in a background inspired by volume modulus kination (i.e: $G\mu \sim 1/t$ and $a\sim t^{1/3}$).
In this simple example we assume that the period gives rise to standard radiation domination with constant tension, which is too strong an assumption to describe the scenarios of Sec~\ref{sec:stringy}.
In addition, we operate under the unphysical assumption that the network is already in a scaling regime.
Since, as we will see, the largest contribution to the spectrum arises from the earliest times, it would be interesting to compute the effects of the network formation or, if formed during inflation, the post-inflationary approach of the network to scaling~\cite{Guedes:2018afo,Cui:2019kkd,Gouttenoire:2019kij,Ferrer:2023uwz,Datta:2025yow}.

\begin{figure}[h!]
    \centering
    \includegraphics[width=1.0\linewidth]{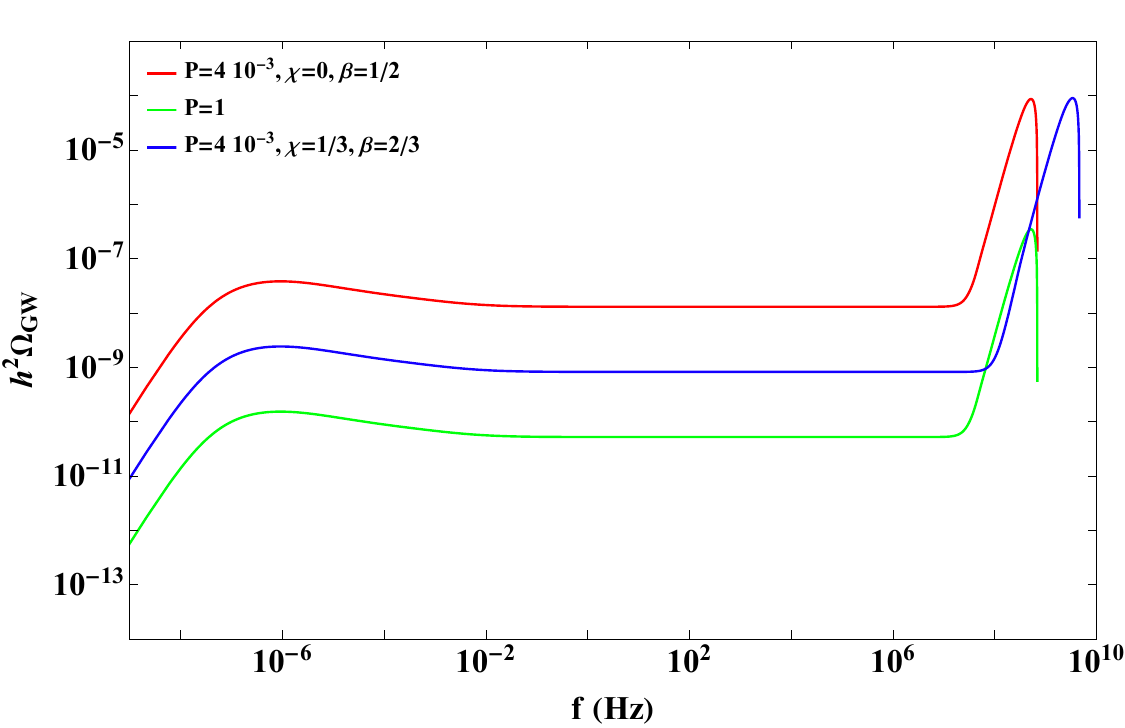}
    \caption{\it GW spectrum in an unrealistic scenario of volume modulus kination followed by the hot Big Bang, for different dependencies on the probability discussed in the literature, see the discussion around Eq.~\eqref{eq:p-dependence}.
    Generically, $P<1$ boosts the spectrum.
    In addition, a dependence in $P$ of the initial size of the loops (encoded in $\chi$) can affect the amplitude and the frequencies of features in the spectrum.
    Experimental limits are not showed in the plot because the configuration is unrealistic.}
    \label{fig:probsplots}
\end{figure}

Figure~\ref{fig:probsplots} shows the gravitational wave spectrum predicted by this simple scenario.
The amplitude reproduces the standard scenario at low frequencies but there is a sharp transition at high frequencies which yields a large amplitude.
The spectral indices agree with the considerations above, and the transition frequencies can be computed as follows:
\begin{itemize}
    \item \textbf{First turn.}
    The transition between the standard, flat spectrum (associated to loops created and sourcing during radiation domination) to the steep one (associated to loops created and sourcing during volume modulus kination) can be computed by comparing the relative energy densities.
    Evaluating Eq.~\eqref{eq:gw-standard} in the radiation period ($m=n=4$) and Eq.~\eqref{eq:gw-varying} in the kination period ($m=n=6$, $p=q=1$)\footnote{It is easy to show that the same spectral index is predicted for loops created during kination but sourcing during radiation ($m=6, \, n=4, \, p=1, \, q=0$) but the contribution is smaller.}, we find that the amplitudes are equal (in order of magnitude) at
    \begin{equation}\label{eq:f-transition-rad-kin}
        f=f_\Delta \lr{\frac{\alpha P^{\chi}}{\Gamma G\mu_\Delta}}^{3/8} \, ,
    \end{equation}
    which is the turning frequency.
    \item \textbf{Peak frequency.}
    We recall now that the largest contribution from loops forming during the kination era to the spectrum at a given frequency $f$ arises from the earliest times, namely from loops created with length $2/\tilde{f}$, at the time of formation.
    Moreover, loops created earlier contribute to higher frequencies, since they are formed with length $ t_i \sim a(t_i)/f$, which implies $f$ grows as a negative power of $t_i$.
    This implies that \textit{the peak of the spectrum carries information about the moment of formation of the network}.
    The peak frequency can thus be computed by imposing $t_i(t_f,f)=t_f$, giving 
    \begin{equation}\label{eq:peak-freq}
        f_{peak}=f_F=\frac{2}{\alpha P^{\chi}}\frac{1}{t_F z_F } \, .
    \end{equation}
    We emphasize here that the details of the spectrum at the peak will be sensitive to the formation process of the network.
    Our predictions should be understood as qualitative and a motivation for future research that can take into account this process.
\end{itemize}

Before concluding our toy example, we note that we can compute the maximum amplitude in terms of the parameters by evaluating Eq.~\eqref{eq:gw-varying} at the peak. In the particular case of volume-modulus kination:
\begin{equation}\label{eq:peak-amp}
    h^2 \Omega_{\rm{GW}}^{(\rm{peak})} \simeq 2A_\Delta \lr{\frac{f_F}{f_\Delta}}^4 \log [t_\Delta/t_F]\, .
\end{equation}

Figure~\ref{fig:probsplots} shows the spectrum in the toy model predicted by three different scenarios for probability-dependence.
For $P < 1$, the amplitude is boosted, although the factor depends on the equation of state of the background.
For instance, the radiation plateau can be shown to depend on $A_\Delta P^{3\chi/2}$ (as can be easily shown by integrating Eq.~\eqref{eq:gw-std-lf} with integration limit given by Eq.~\eqref{eq:t-max-std}).
On the other hand, the peak amplitude only depends on $P$ through $A_\Delta \sim P^{2(1-\beta)+\chi}$, which in our two benchmarks ($\beta=1/2$, $\chi=0$) and ($\beta=2/3$, $\chi=1/3$) give the same power of $P$, thus explaining that the peak amplitude is the same in these cases.
The frequencies are however different, and this can be explained by Eqs.~\eqref{eq:f-transition-rad-kin} and~\eqref{eq:peak-freq}.
According to Eq.~\eqref{eq:f-transition-rad-kin}, the flat-kination transition depends very weakly on $P$ as $f \sim P^{-5\chi/8}$, which explains that the $P=1$ and $\chi=0$ cases occur at a similar frequency, while the transition in the $\chi=1/3$ case occurs at a slightly larger frequency.
The peak frequency in Eq.~\eqref{eq:peak-freq}, however, depends on $P^\chi$ which explains the order of magnitude difference between the peak positions in the different scenarios.

\subsubsection{Realistic scenario}\label{sec-realistic}

\begin{figure}[t]
    \centering
    \includegraphics[width=1.0\linewidth]{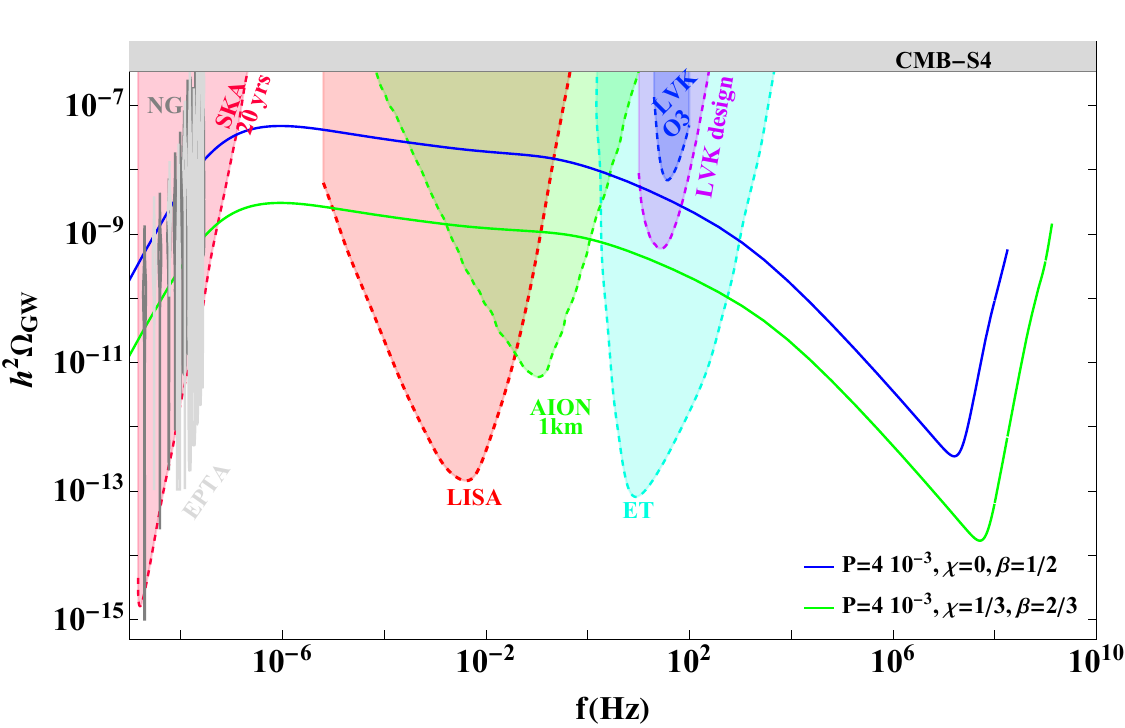}
    \caption{\it GW spectrum predicted by the SVL-LVS scenario as described in Sec.~\ref{sec-realistic} and App.~\ref{sec:scenario-highfreq}.
    The flat plateau characteristic of radiation domination is changed to a dip in the spectrum due to the early matter domination.
    This is preceded by a period of volume modulus kination, which induces a large high-frequency boost in the signal, with amplitude as in Eq.~\eqref{eq:kination-peak}.
    The plot shows two different probability dependencies quoted in the literature, see the discussion around Eq.~\eqref{eq:p-dependence}.
    We also show CMB-S4 projected bounds~\cite{CMB-S4:2020lpa}.}
    \label{fig:full-kination}
\end{figure}

Equation~\eqref{eq:peak-amp} computes the value of the GW spectrum sourced by volume-modulus kination at its high-frequency peak.
We notice, in particular, that the details of the posterior cosmology (including the other expected non-standard epochs such as trackers and matter domination) are included in the redshift factors.
We can thus use these equations to determine whether the high-frequency spectrum is observationally interesting/viable from the details of a given scenario; an example is shown in Fig.~\ref{fig:full-kination}, for a choice of potential: 
\begin{equation}
    10^{10}V=
    \left(C_1 e^{-\sqrt{2/3}\frac{\Phi}{M_P}}-e^{-2\sqrt{2/3}\frac{\Phi}{M_P}}  \right)^2
    \, , \qquad C_1^{3/2}=3\cdot 10^{-10}\, ,
\end{equation}
whose motivation is further described in the Appendix.
The peak amplitude is, in terms of number of e-folds:

\begin{equation}\label{eq:kination-peak}
    h^2 \Omega_{\rm{GW}}^{(\rm{peak})}\simeq 
    5\times 10^2 \times
    \frac{C_{\rm{eff}}}{30.6} \times
    \frac{\Gamma}{50} \times 
    \frac{0.1}{\alpha} \times
    \frac{4 \times 10^{-3}}{P^{2(1-\beta)+\chi}} \times 
    3N_{kin}\, \times
    \lr{G\mu_0}^2 \times
    \exp \lr{{2N_{kin}-N_m}}\, .
\end{equation}

This is the same dependence on the number of e-folds that one would get in a scenario with constant tension, with the important proviso that the tension is now evaluated at the earliest time (and so the amplitude is much larger than in a scenario with constant tension for same number of e-folds).
Of course, in scenarios with volume modulus kination $N_m$ tends to be large, whilst $N_{kin}$ grows very mildly with the parameters described in Sec.~\ref{sec:stringy} and Appendix~\ref{sec:scenario-highfreq}.
We can understand qualitatively the dependence of $N_{kin}$ and $N_m$ on the properties of the potential as follows:
\begin{itemize}
    \item $N_m$.
    The number of matter e-folds can be approximated noting that matter domination starts approximately at $t\sim 1/m_\Phi$ and the volume mode decays at $t_{reh}\sim 1/\Gamma$.
    Hence,
    \begin{equation}
        N_m \simeq \frac{2}{3}\ln \lr{\frac{m_\Phi}{\Gamma_\Phi}}\, .
    \end{equation}
    In the case where the fast decay channel is active, we have
    \begin{equation}
        N_m\simeq 12 +\frac{4}{3} \ln \lr{\frac{10^{10}}{\mathcal{V}}\times \frac{30 \text{ TeV}}{m_\Phi}}
    \end{equation}
    whilst for gravitational decay we recall $\Gamma^{(g)} =\Gamma^{(f)}/\mathcal{V}^2 $.
    In LARGE volume scenarios, this induces tens of additional e-folds of matter domination, resulting in an unobservable signal.

    \item $N_{kin}$.
    The number of e-folds of kination is somewhat harder to estimate.
    It is interesting, however, that the larger that the hierarchy between $V_{inf}$ and $V_{late}$ is, the larger is the field displacement and so the larger number of e-folds of kination.
    For illustrative purposes, inspired by~\cite{Cicoli:2024bwq}, we have considered a potential of the form $V=V_0e^{-\lambda \phi}$ with small $V_0=10^{-10}$ and $\lambda=4\sqrt{2/3}$ (see the Appendix~\ref{sec:scenario-highfreq} for details).

    The main obstruction to increasing $N_{kin}$ in this manner is that as $V_{late}$ gets smaller the easiest it is to overshoot, getting a phenomenologically unviable scenario.
    Thus this longer period where the field falls down its potential must be compensated by additional e-folds of tracker (which, as is evident from Eq.~\eqref{eq:kination-peak}, do not affect the amplitude).
    We have observed that $N_{kin}$ grows logarithmically with the hierarchy $V_{inf}/V_{late}$, reaching around 4-5 e-folds of kination in the scenario discussed in Appendix~\ref{sec:scenario-highfreq} but a more concrete analysis is left for future work.
    
\end{itemize}

We thus conclude that LARGE volume scenarios with hierarchically large separations between inflation (or other UV scale) and late-time scales featuring volume modulus kination source potentially interesting signals. Counterintuitively, it seems like a larger late-time volume (which leads to a longer period of early matter domination) yields a larger signal.
This is due to several factors on top of the usual phenomenon of amplitudes getting larger because the background redshifts faster than the GWs (see e.g.~\cite{Gouttenoire:2019kij}).
Firstly, the early-time emission is much stronger due to the enhancement of the tension at early times.
Second, the loop distribution is modified.
Together, these facts render an approximately constant contribution to the spectrum at all relevant frequency bins during kination, which is manifest in the logarithmic behaviour in Eq.~\eqref{eq:peak-amp}.
This is different to the constant tension case where at every frequency bin of relevance the largest contribution arises at the time $t_M(f)$ for given $f$, as discussed in Sec.~\ref{ssc:fdct}.

\section{Discussion}\label{sc:conc}

Let us now conclude and summarize our main findings.
After a review of cosmic strings in string theory and moduli dynamics in the early Universe in the context of LARGE volume scenarios, we have concluded that these models naturally accommodate the values of $G\mu$ and $P$ hinted at by PTA measurements.\footnote{We have taken as a proxy the values quoted in~\cite{Ellis:2023tsl} as best-fit, which took into account LVK constraints.
Our scenarios, however, predict a drop in the spectrum which invalidate these constraints.
Studying the best-fit of~\cite{Ellis:2023tsl} without the LVK constraints remains for future work, although we can see from Eq.~\eqref{eq:LISA-freq} that the drop in the spectrum will still lie in the LISA band for those values.}
At given $P$, larger values of the string tension would be excluded by PTA measurements, while smaller values would imply a volume modulus that is too light and hence in tension with BBN. Moreover, such models come with additional predictions for the GW spectrum across the whole frequency range. It is particularly interesting to observe how LVS scenarios where the volume mode decays gravitationally make a concrete prediction for a turning frequency which robustly lies within the LISA band, as shown in Figs. \ref{fig:vanilla-LVS} and \ref{fig:vanilla-LVS-small}. Notice how, in this context, the epoch of early matter domination needed to explain a non-detection by LVK is a natural prediction of the stringy cosmology, and does need to be added by hand (unlike, for example, \cite{Ellis:2023tsl}).

In order to provide quantitative estimates for the GW spectrum and its spectral index, we have generalized the approach in~\cite{Cui:2017ufi} to include the effects of varying tension, finding spectral indices that cannot be reproduced by cosmic strings with constant tension. In particular, we obtained analytic expressions for both spectral indices and peak frequency, Eqs. \eqref{eq:gw-varying} and \eqref{eq:peak-f} respectively. These results are model-independent and can be used for any situation where the tension of the strings varies with time as a power law.
Some possibilities have been discussed recently in~\cite{Revello:2024gwa,Brunelli:2025ems} and applying our results to those scenarios is a natural future direction.
Interestingly, for rapidly varying tension the GW spectrum at a given frequency is dominated by GW emission from the earliest times. This observation potentially provides a window into physics at very high energies, including the time of formation of the network.

Lastly, taking the PTA-suggested values for the intercommutation probability and string tension, we have distinguished two well-motivated scenarios and discussed in detail the corresponding predictions. In the vanilla LVS scenario, we have argued that there is a prediction for a tilt in the GW spectrum in the LISA band due to a period of early matter domination where the volume mode dominates the energy density (see Figures \ref{fig:vanilla-LVS} and \ref{fig:vanilla-LVS-small}).
The prediction for the tilt frequency in scenarios featuring gravitational decay is given in Eq.~\eqref{eq:LISA-freq} as a function of the parameter $\alpha$, the intercommutation probability and the mass of the volume mode.
As discussed in the last paragraph of Sec.~\ref{sec:vanilla-lvs}, the prediction is robust, even if the assumption that the PTA signal is explained by cosmic superstrings is lifted.
In the event of a positive detection of a GW background from cosmic superstrings, the LISA mission could provide information about microscopic, stringy parameters of our own universe, including the size of the extra dimensions and the mass of its associated mode.\footnote{Of course, this signal is degenerate with any other source of early matter domination ending slightly before BBN. However, having a hint of the mass of the volume mode and the size of the extra dimensions would suggest other complementary probes, such as collider signals.}

In addition, we have computed the peak amplitude of the spectrum sourced after a period with varying tension and matter domination in Eq.~\eqref{eq:kination-peak} (with peak frequency in Eq.~\eqref{eq:peak-freq}).
The result is the same dependence in number of e-folds of kination and matter as would be expected in a case with constant tension, provided one evaluates the dimensionless tension $G\mu_0$ at the time of emission.
The signal is thus potentially much larger (a ratio of $(G\mu_0/G\mu_{end})^2$) than in a case with constant tension.
This motivated us to study the SVL-LVS case, where we have argued that LARGE volume scenarios with volume-modulus kination (and the consequent varying tension) in the early Universe potentially provide large signals provided there is a large hierarchy between the UV and late-time scales, as shown in Figure \ref{fig:full-kination}.
This is another example of high-frequency GWs potentially providing access to the earliest times of the Universe, with larger signals sourced at earlier times~\cite{Aggarwal:2020olq,Aggarwal:2025noe}.\footnote{The paramount example is the Standard Model~\cite{Ghiglieri:2015nfa,Ghiglieri:2020mhm,Muia:2023wru} with a signal that grows with the reheating temperature, and the analogue in string theory~\cite{Frey:2024jqy} where the signal is larger than in the Standard Model and grows with the string scale.}
The main obstacles to claiming that this behavior is generic are the length of the matter-domination period and the overshoot problem. It would be very interesting to understand whether other scenarios, such as the ones discussed in~\cite{Brunelli:2025ems}, can improve the situation since the potentials and decay channels are different.
We emphasize here that the loops source the largest contribution to the GW spectrum as they are formed, in contrast with the constant tension case.

In the upcoming years Gravitational Wave astronomy aspires to achieve measurement precisions that will be orders of magnitude improved than the current GW detectors. It is conceivable that such measurements will also lead us closer to understanding fundamental aspects in the microscopic structure of our Universe.
In particular, the signals we discuss in this article represent a way to test cosmological consequences of certain String Theory constructions, and potentially probe the size and dynamics of the extra dimensions it predicts.

\medskip

\section*{Acknowledgements}

We thank Anastasios Avgoustidis, Joseph Conlon, Maurizio Gasperini, Antonio Iovino, Marek Lewicki, Fernando Quevedo, Gianmassimo Tasinato and Ivonne Zavala for discussions and comments on the manuscript.
The authors acknowledge feedback from the LISA Cosmology Working Group Consortium.
We are also grateful to the organizers of ``SUSY 23"; the workshop ``WISPs in String Cosmology"; and the workshop ``Ultra-high frequency gravitational waves: where to next?" where some of this work took place. The research of FR is funded through a junior postdoctoral fellowship of the Fonds Wetenschappelijk Onderzoek (FWO), project number 12A1Q25N, and was also partially supported by the Dutch Research Council (NWO) via a Start-Up grant and a Vici grant. The work of GV has been partially supported by STFC consolidated grant ST/P000681/1, ST/T000694/1. GV thanks the Institute for Theoretical Physics of Utrecht University for hospitality. This work is partly based upon work from the COST Action COSMIC WISPers CA21106, supported by COST (European Cooperation in Science and Technology).

\medskip

\newpage
\section*{Appendix}
\appendix

\section{Details on the benchmark scenario}\label{sec:scenario-highfreq}

In this appendix we discuss the scenario leading to the GW spectrum reported in Fig.~\ref{fig:full-kination}.
We emphasize that the behaviour is non-generic and the scenario we are studying is somewhat tuned.
We believe, however, that it illustrates certain interesting aspects of the physics of cosmic strings with varying tension.
This case, in which the overall volume mode drives the variation of the tension, only gives a meaningful signal in this sort of tuned scenario, where the period of matter domination is naturally large (and so one must aim to get as many e-folds of kination as possible).
However, other configurations where the string tension varies as a function of other moduli as those studied in~\cite{Brunelli:2025ems} could give less e-folds of matter domination, and we believe our discussion makes explicit the physics involved.
We have also taken as a proxy the PTA signal but there is nothing preventing from studying other late-time volume configurations, which may lead to interesting signals with less tuning.
These interesting avenues are left for future work.

The discussion is based on the potential reported in~\cite{Cicoli:2024bwq} (see~\cite{Villa:2025zmj} for a concise summary).
The idea is to consider the supersymmetric gravity sector of low-energy string vacua (i.e: closed string moduli, their axionic counterparts and their fermionic superpartners) and couple it to a sufficiently sequestered source of supersymmetry breaking, such as an antibrane in a warped throat.
This sort of system can be described in the nilpotent Goldstino formalism, started in~\cite{Komargodski:2009rz} (see~\cite{Kallosh:2015nia} and references therein for applications in string theory).
For our present purposes, the outcome is that the most general perturbative scalar potential for the volume modulus compatible with other symmetries generic in string vacua (see~\cite{Burgess:2020qsc}) is given by:

\begin{equation}\label{eq:scalar-pot-general}
    V=\frac{1}{U}\left[\left(f' W_X-3 g' W_0\right)^2 \right. 
-\left. f'' \left(f W_X^2  - 6 g W_X W_0 -9 h W_0^2\right)\right] \, ,
\end{equation}

where $f(\mathcal{V})$, $g(\mathcal{V})$ and $h(\mathcal{V})$ are functions of the volume modulus, which admit an expansion in inverse powers of $\mathcal{V}$ and that at tree level read $f(\mathcal{V})=\mathcal{V}^{2/3}$, $g(\mathcal{V})=0$ and $h(\mathcal{V})=1$.
$U$ is a function of $f$, $g$ and $h$ and at leading order $U= 3 f^2 h$.
Lastly, the contributions to the superpotential (which we take real) are $W_x$ and $W_0$ and do not depend on $\mathcal{V}$ (i.e: we are neglecting non-perturbative corrections to the superpotential).

It is well known that the first contribution to $f''$ arises at order $f''\sim 1/\mathcal{V}^3$ in the form of the BBHL~\cite{Becker:2002nn} correction and dominates at small $\mathcal{V}$, providing the $\lambda=\sqrt{27/2}$ behaviour characteristic of LVS models.
We neglect this contribution in our scenario (as could occur if the extra dimensions where a self-mirror CY manifold) and assume $f''=0$ at a sufficiently high order allowing for the behaviour $V \sim \mathcal{V}^{-8/3}$ at small $\tau$.
The potential is then a perfect square.
We assume on the grounds of effective field theory reasoning that quantum effects due to the antibrane generate a term of the form $g'=\tilde{A}/\mathcal{V}^a$ (note that functions in string vacua generically admit expansions in inverse powers of cycle volumes).
It is important to emphasize, however, that this correction is not the result of a stringy computation.
Assuming it, the minimum is located at a position
\begin{equation}
    \mathcal{V}^{2/3}=\lr{\frac{3\tilde{A}W_0}{W_x}}^{1/a}\, ,
\end{equation}
which can be exponentially large in the case where the Goldstino field is describing the effect of an antibrane in a warped throat.
Indeed, in that case it is well known that $W_x^2\sim \exp \lr{-8\pi K/3g_sM} $, with $K$ and $M$ a choice of integers (see~\cite{Aparicio:2015psl}).
We will take as a proxy $a=1$, which would occur for a logarithmic correction of the form $g(\mathcal{V})=\tilde{A}\log \mathcal{V}$, and so $\tilde{A}$ naturally carries powers of $g_s$.
In terms of the canonically normalized field $\Phi$, the perfect square contribution to the potential in Eq.~\eqref{eq:scalar-pot-general} is:\footnote{This contribution would then be uplifted to a de Sitter vacuum by a higher order correction to $f''$, which does not alter the discussion.}
\begin{equation}\label{eq:bLVS-pot}
    V=e^{-2\sqrt{2/3}\frac{\Phi}{M_P}}|W_x|^2/3-2\tilde{A}e^{-\sqrt{6}\frac{\Phi}{M_P}}W_xW_0+3\tilde{A}^2W_0^2e^{-4\sqrt{\frac{2}{3}}\frac{\Phi}{M_P}}\, .
\end{equation}
This particular choice of parameters corresponds, in the notation of Eq.~\eqref{eq:vexp},to 
\begin{equation*}
    V_0=3\tilde{A}^2W_0^2 \, , \qquad B=-2\tilde{A} W_x W_0/V_0 \, , \qquad A=W_x^2/3V_0
\end{equation*}
\begin{equation}
    \lambda_3 = 4\sqrt{2/3} \equiv \lambda\, , \qquad \lambda_1= \lambda/2\, , \qquad \lambda_2= 3\lambda/4 \, .
\end{equation}
At small volume, the last term dominates providing the $\lambda=4\sqrt{\frac{2}{3}}$ behaviour in the notation of Eq.~\eqref{eq:vexp}.
On a log-scale, this is a smaller slope than $\lambda=\sqrt{27/2}$ and so allows for more field range between the putative high-scale inflationary minimum, which we take at $\Lambda=10^{16}$ GeV, and the late-time vacuum.
For our plots, we choose parameters $W_x=3\tilde{A}W_0 (3/10^{10})^{2/3}$, ensuring a late-time volume $\mathcal{V}=10^{10}/3$, and $\sqrt{3}\tilde{A}W_0=10^{-5}$ which allows to have a large initial energy density $V(\Phi_0)=10^{16}$ GeV with a moderate hierarchy $G\mu \simeq 8.8 \times 10^{-4}$, although the initial volume $\mathcal{V}=7.5$ is unreasonably small ($\Phi=1.65 M_P$).
In addition, we take as initial conditions $\varepsilon_0=4.8 \times 10^{-6}$, $x^2-\varepsilon_0=0.99$ and $y^2=0.01$, in the notation of Sec.~\ref{ssc:kin}.
We stop the dynamical system evolution and patch to a period of early matter domination after the first oscillation around the minimum.
Early matter domination ends at a time $1/\Gamma_{\Phi \to hh}$, with
\begin{equation}
    \Gamma_{\Phi \to hh}=\lr{\frac{\mathcal{V}}{16\pi}}^2\frac{m_\Phi^2}{M_P^3}\, ,
\end{equation}
c.f. Eq.~\eqref{eq:fasdtdecay}, and the mass of the volume is computed from Eq.~\eqref{eq:bLVS-pot}, which for the choice of parameters indicated turns out to be $m_\Phi\simeq 8$ GeV.

\newpage
\bibliography{biblio}

\providecommand{\href}[2]{#2}\begingroup\raggedright\begin{thebibliography}{100}

\bibitem{LIGOScientific:2016aoc}
{\scshape LIGO Scientific, Virgo} collaboration, B.~P. Abbott et~al.,
  \emph{{Observation of Gravitational Waves from a Binary Black Hole Merger}},
  \href{https://doi.org/10.1103/PhysRevLett.116.061102}{\emph{Phys. Rev. Lett.}
  {\bfseries 116} (2016) 061102}
  [\href{https://arxiv.org/abs/1602.03837}{{\ttfamily 1602.03837}}].

\bibitem{Bailes:2021tot}
M.~Bailes et~al., \emph{{Gravitational-wave physics and astronomy in the 2020s
  and 2030s}}, \href{https://doi.org/10.1038/s42254-021-00303-8}{\emph{Nature
  Rev. Phys.} {\bfseries 3} (2021) 344}.

\bibitem{NANOGrav:2023gor}
{\scshape NANOGrav} collaboration, G.~Agazie et~al., \emph{{The NANOGrav 15 yr
  Data Set: Evidence for a Gravitational-wave Background}},
  \href{https://doi.org/10.3847/2041-8213/acdac6}{\emph{Astrophys. J. Lett.}
  {\bfseries 951} (2023) L8}
  [\href{https://arxiv.org/abs/2306.16213}{{\ttfamily 2306.16213}}].

\bibitem{EPTA:2023fyk}
{\scshape EPTA, InPTA:} collaboration, J.~Antoniadis et~al., \emph{{The second
  data release from the European Pulsar Timing Array - III. Search for
  gravitational wave signals}},
  \href{https://doi.org/10.1051/0004-6361/202346844}{\emph{Astron. Astrophys.}
  {\bfseries 678} (2023) A50}
  [\href{https://arxiv.org/abs/2306.16214}{{\ttfamily 2306.16214}}].

\bibitem{Reardon:2023gzh}
D.~J. Reardon et~al., \emph{{Search for an Isotropic Gravitational-wave
  Background with the Parkes Pulsar Timing Array}},
  \href{https://doi.org/10.3847/2041-8213/acdd02}{\emph{Astrophys. J. Lett.}
  {\bfseries 951} (2023) L6}
  [\href{https://arxiv.org/abs/2306.16215}{{\ttfamily 2306.16215}}].

\bibitem{Xu:2023wog}
H.~Xu et~al., \emph{{Searching for the Nano-Hertz Stochastic Gravitational Wave
  Background with the Chinese Pulsar Timing Array Data Release I}},
  \href{https://doi.org/10.1088/1674-4527/acdfa5}{\emph{Res. Astron.
  Astrophys.} {\bfseries 23} (2023) 075024}
  [\href{https://arxiv.org/abs/2306.16216}{{\ttfamily 2306.16216}}].

\bibitem{Ellis:2023tsl}
J.~Ellis, M.~Lewicki, C.~Lin and V.~Vaskonen, \emph{{Cosmic superstrings
  revisited in light of NANOGrav 15-year data}},
  \href{https://doi.org/10.1103/PhysRevD.108.103511}{\emph{Phys. Rev. D}
  {\bfseries 108} (2023) 103511}
  [\href{https://arxiv.org/abs/2306.17147}{{\ttfamily 2306.17147}}].

\bibitem{Datta:2024bqp}
S.~Datta and R.~Samanta, \emph{{Cosmic superstrings, metastable strings and
  ultralight primordial black holes: from NANOGrav to LIGO and beyond}},
  \href{https://doi.org/10.1007/JHEP02(2025)095}{\emph{JHEP} {\bfseries 02}
  (2025) 095} [\href{https://arxiv.org/abs/2409.03498}{{\ttfamily
  2409.03498}}].

\bibitem{Balasubramanian:2005zx}
V.~Balasubramanian, P.~Berglund, J.~P. Conlon and F.~Quevedo,
  \emph{{Systematics of moduli stabilisation in Calabi-Yau flux
  compactifications}},
  \href{https://doi.org/10.1088/1126-6708/2005/03/007}{\emph{JHEP} {\bfseries
  03} (2005) 007} [\href{https://arxiv.org/abs/hep-th/0502058}{{\ttfamily
  hep-th/0502058}}].

\bibitem{Conlon:2005ki}
J.~P. Conlon, F.~Quevedo and K.~Suruliz, \emph{{Large-volume flux
  compactifications: Moduli spectrum and D3/D7 soft supersymmetry breaking}},
  \href{https://doi.org/10.1088/1126-6708/2005/08/007}{\emph{JHEP} {\bfseries
  08} (2005) 007} [\href{https://arxiv.org/abs/hep-th/0505076}{{\ttfamily
  hep-th/0505076}}].

\bibitem{Conzinu:2024cwl}
P.~Conzinu, G.~Fanizza, M.~Gasperini, E.~Pavone, L.~Tedesco and G.~Veneziano,
  \emph{{Constraints on the Pre-Big Bang scenario from a cosmological
  interpretation of the NANOGrav data}},
  \href{https://doi.org/10.1088/1475-7516/2025/02/039}{\emph{JCAP} {\bfseries
  02} (2025) 039} [\href{https://arxiv.org/abs/2412.01734}{{\ttfamily
  2412.01734}}].

\bibitem{Gasperini:1992em}
M.~Gasperini and G.~Veneziano, \emph{{Pre - big bang in string cosmology}},
  \href{https://doi.org/10.1016/0927-6505(93)90017-8}{\emph{Astropart. Phys.}
  {\bfseries 1} (1993) 317}
  [\href{https://arxiv.org/abs/hep-th/9211021}{{\ttfamily hep-th/9211021}}].

\bibitem{Ellis:2023dgf}
J.~Ellis, M.~Fairbairn, G.~H\"utsi, J.~Raidal, J.~Urrutia, V.~Vaskonen et~al.,
  \emph{{Gravitational waves from supermassive black hole binaries in light of
  the NANOGrav 15-year data}},
  \href{https://doi.org/10.1103/PhysRevD.109.L021302}{\emph{Phys. Rev. D}
  {\bfseries 109} (2024) L021302}
  [\href{https://arxiv.org/abs/2306.17021}{{\ttfamily 2306.17021}}].

\bibitem{EPTA:2023xxk}
{\scshape EPTA, InPTA} collaboration, J.~Antoniadis et~al., \emph{{The second
  data release from the European Pulsar Timing Array - IV. Implications for
  massive black holes, dark matter, and the early Universe}},
  \href{https://doi.org/10.1051/0004-6361/202347433}{\emph{Astron. Astrophys.}
  {\bfseries 685} (2024) A94}
  [\href{https://arxiv.org/abs/2306.16227}{{\ttfamily 2306.16227}}].

\bibitem{NANOGrav:2023hvm}
{\scshape NANOGrav} collaboration, A.~Afzal et~al., \emph{{The NANOGrav 15 yr
  Data Set: Search for Signals from New Physics}},
  \href{https://doi.org/10.3847/2041-8213/acdc91}{\emph{Astrophys. J. Lett.}
  {\bfseries 951} (2023) L11}
  [\href{https://arxiv.org/abs/2306.16219}{{\ttfamily 2306.16219}}].

\bibitem{Figueroa:2023zhu}
D.~G. Figueroa, M.~Pieroni, A.~Ricciardone and P.~Simakachorn,
  \emph{{Cosmological Background Interpretation of Pulsar Timing Array Data}},
  \href{https://doi.org/10.1103/PhysRevLett.132.171002}{\emph{Phys. Rev. Lett.}
  {\bfseries 132} (2024) 171002}
  [\href{https://arxiv.org/abs/2307.02399}{{\ttfamily 2307.02399}}].

\bibitem{Ellis:2023oxs}
J.~Ellis, M.~Fairbairn, G.~Franciolini, G.~H\"utsi, A.~Iovino, M.~Lewicki
  et~al., \emph{{What is the source of the PTA GW signal?}},
  \href{https://doi.org/10.1103/PhysRevD.109.023522}{\emph{Phys. Rev. D}
  {\bfseries 109} (2024) 023522}
  [\href{https://arxiv.org/abs/2308.08546}{{\ttfamily 2308.08546}}].

\bibitem{Ratzinger:2020koh}
W.~Ratzinger and P.~Schwaller, \emph{{Whispers from the dark side: Confronting
  light new physics with NANOGrav data}},
  \href{https://doi.org/10.21468/SciPostPhys.10.2.047}{\emph{SciPost Phys.}
  {\bfseries 10} (2021) 047}
  [\href{https://arxiv.org/abs/2009.11875}{{\ttfamily 2009.11875}}].

\bibitem{Avgoustidis:2025svu}
A.~Avgoustidis, E.~J. Copeland, A.~Moss and J.~Raidal, \emph{{The stochastic
  gravitational wave background from cosmic superstrings}},
  \href{https://arxiv.org/abs/2503.10361}{{\ttfamily 2503.10361}}.

\bibitem{Ellis:2020ena}
J.~Ellis and M.~Lewicki, \emph{{Cosmic String Interpretation of NANOGrav Pulsar
  Timing Data}},
  \href{https://doi.org/10.1103/PhysRevLett.126.041304}{\emph{Phys. Rev. Lett.}
  {\bfseries 126} (2021) 041304}
  [\href{https://arxiv.org/abs/2009.06555}{{\ttfamily 2009.06555}}].

\bibitem{Blasi:2020mfx}
S.~Blasi, V.~Brdar and K.~Schmitz, \emph{{Has NANOGrav found first evidence for
  cosmic strings?}},
  \href{https://doi.org/10.1103/PhysRevLett.126.041305}{\emph{Phys. Rev. Lett.}
  {\bfseries 126} (2021) 041305}
  [\href{https://arxiv.org/abs/2009.06607}{{\ttfamily 2009.06607}}].

\bibitem{Blanco-Pillado:2021ygr}
J.~J. Blanco-Pillado, K.~D. Olum and J.~M. Wachter, \emph{{Comparison of cosmic
  string and superstring models to NANOGrav 12.5-year results}},
  \href{https://doi.org/10.1103/PhysRevD.103.103512}{\emph{Phys. Rev. D}
  {\bfseries 103} (2021) 103512}
  [\href{https://arxiv.org/abs/2102.08194}{{\ttfamily 2102.08194}}].

\bibitem{Albrecht:1984xv}
A.~Albrecht and N.~Turok, \emph{{Evolution of Cosmic Strings}},
  \href{https://doi.org/10.1103/PhysRevLett.54.1868}{\emph{Phys. Rev. Lett.}
  {\bfseries 54} (1985) 1868}.

\bibitem{Bennett:1987vf}
D.~P. Bennett and F.~R. Bouchet, \emph{{Evidence for a Scaling Solution in
  Cosmic String Evolution}},
  \href{https://doi.org/10.1103/PhysRevLett.60.257}{\emph{Phys. Rev. Lett.}
  {\bfseries 60} (1988) 257}.

\bibitem{Allen:1990tv}
B.~Allen and E.~P.~S. Shellard, \emph{{Cosmic string evolution: a numerical
  simulation}}, \href{https://doi.org/10.1103/PhysRevLett.64.119}{\emph{Phys.
  Rev. Lett.} {\bfseries 64} (1990) 119}.

\bibitem{KAGRA:2013rdx}
{\scshape KAGRA, LIGO Scientific, Virgo} collaboration, B.~P. Abbott et~al.,
  \emph{{Prospects for observing and localizing gravitational-wave transients
  with Advanced LIGO, Advanced Virgo and KAGRA}},
  \href{https://doi.org/10.1007/s41114-020-00026-9}{\emph{Living Rev. Rel.}
  {\bfseries 19} (2016) 1} [\href{https://arxiv.org/abs/1304.0670}{{\ttfamily
  1304.0670}}].

\bibitem{Sathyaprakash:2012jk}
B.~Sathyaprakash et~al., \emph{{Scientific Objectives of Einstein Telescope}},
  \href{https://doi.org/10.1088/0264-9381/29/12/124013}{\emph{Class. Quant.
  Grav.} {\bfseries 29} (2012) 124013}
  [\href{https://arxiv.org/abs/1206.0331}{{\ttfamily 1206.0331}}].

\bibitem{Audley:2017drz}
{\scshape LISA} collaboration, P.~Amaro-Seoane et~al., \emph{{Laser
  Interferometer Space Antenna}},
  \href{https://arxiv.org/abs/1702.00786}{{\ttfamily 1702.00786}}.

\bibitem{Blanco-Pillado:2024aca}
{\scshape LISA Cosmology Working Group} collaboration, J.~J. Blanco-Pillado,
  Y.~Cui, S.~Kuroyanagi, M.~Lewicki, G.~Nardini, M.~Pieroni et~al.,
  \emph{{Gravitational waves from cosmic strings in LISA: reconstruction
  pipeline and physics interpretation}},
  \href{https://arxiv.org/abs/2405.03740}{{\ttfamily 2405.03740}}.

\bibitem{Gouttenoire:2019kij}
Y.~Gouttenoire, G.~Servant and P.~Simakachorn, \emph{{Beyond the Standard
  Models with Cosmic Strings}},
  \href{https://doi.org/10.1088/1475-7516/2020/07/032}{\emph{JCAP} {\bfseries
  07} (2020) 032} [\href{https://arxiv.org/abs/1912.02569}{{\ttfamily
  1912.02569}}].

\bibitem{Conlon:2022pnx}
J.~P. Conlon and F.~Revello, \emph{{Catch-me-if-you-can: the overshoot problem
  and the weak/inflation hierarchy}},
  \href{https://doi.org/10.1007/JHEP11(2022)155}{\emph{JHEP} {\bfseries 11}
  (2022) 155} [\href{https://arxiv.org/abs/2207.00567}{{\ttfamily
  2207.00567}}].

\bibitem{Apers:2022cyl}
F.~Apers, J.~P. Conlon, M.~Mosny and F.~Revello, \emph{{Kination, meet Kasner:
  on the asymptotic cosmology of string compactifications}},
  \href{https://doi.org/10.1007/JHEP08(2023)156}{\emph{JHEP} {\bfseries 08}
  (2023) 156} [\href{https://arxiv.org/abs/2212.10293}{{\ttfamily
  2212.10293}}].

\bibitem{Revello:2023hro}
F.~Revello, \emph{{Attractive (s)axions: cosmological trackers at the boundary
  of moduli space}}, \href{https://doi.org/10.1007/JHEP05(2024)037}{\emph{JHEP}
  {\bfseries 05} (2024) 037}
  [\href{https://arxiv.org/abs/2311.12429}{{\ttfamily 2311.12429}}].

\bibitem{Apers:2024ffe}
F.~Apers, J.~P. Conlon, E.~J. Copeland, M.~Mosny and F.~Revello, \emph{{String
  theory and the first half of the universe}},
  \href{https://doi.org/10.1088/1475-7516/2024/08/018}{\emph{JCAP} {\bfseries
  08} (2024) 018} [\href{https://arxiv.org/abs/2401.04064}{{\ttfamily
  2401.04064}}].

\bibitem{Conlon:2024hgw}
J.~P. Conlon, \emph{{Out of the dark: WISPs in String Theory and the Early
  Universe}}, \href{https://doi.org/10.22323/1.454.0001}{\emph{PoS} {\bfseries
  COSMICWISPers} (2024) 001}
  [\href{https://arxiv.org/abs/2402.04725}{{\ttfamily 2402.04725}}].

\bibitem{Conlon:2024ene}
J.~P. Conlon, \emph{{String Theory and the Early Universe: Constraints and
  Opportunities}},  in \emph{{58th Rencontres de Moriond on Cosmology}}, 5,
  2024, \href{https://arxiv.org/abs/2405.19118}{{\ttfamily 2405.19118}}.

\bibitem{Cicoli:2023opf}
M.~Cicoli, J.~P. Conlon, A.~Maharana, S.~Parameswaran, F.~Quevedo and
  I.~Zavala, \emph{{String cosmology: From the early universe to today}},
  \href{https://doi.org/10.1016/j.physrep.2024.01.002}{\emph{Phys. Rept.}
  {\bfseries 1059} (2024) 1}
  [\href{https://arxiv.org/abs/2303.04819}{{\ttfamily 2303.04819}}].

\bibitem{Allahverdi:2020bys}
R.~Allahverdi et~al., \emph{{The First Three Seconds: a Review of Possible
  Expansion Histories of the Early Universe}},
  \href{https://arxiv.org/abs/2006.16182}{{\ttfamily 2006.16182}}.

\bibitem{Cicoli:2024bwq}
M.~Cicoli, C.~Hughes, A.~R. Kamal, F.~Marino, F.~Quevedo, M.~Ramos-Hamud
  et~al., \emph{{Back to the origins of brane-antibrane inflation}},
  \href{https://arxiv.org/abs/2410.00097}{{\ttfamily 2410.00097}}.

\bibitem{Aggarwal:2020olq}
N.~Aggarwal et~al., \emph{{Challenges and opportunities of gravitational-wave
  searches at MHz to GHz frequencies}},
  \href{https://doi.org/10.1007/s41114-021-00032-5}{\emph{Living Rev. Rel.}
  {\bfseries 24} (2021) 4} [\href{https://arxiv.org/abs/2011.12414}{{\ttfamily
  2011.12414}}].

\bibitem{Roshan:2024qnv}
R.~Roshan and G.~White, \emph{{Using gravitational waves to see the first
  second of the Universe}},  \href{https://arxiv.org/abs/2401.04388}{{\ttfamily
  2401.04388}}.

\bibitem{Aggarwal:2025noe}
N.~Aggarwal et~al., \emph{{Challenges and Opportunities of Gravitational Wave
  Searches above 10 kHz}},  \href{https://arxiv.org/abs/2501.11723}{{\ttfamily
  2501.11723}}.

\bibitem{Revello:2024gwa}
F.~Revello and G.~Villa, \emph{{Cosmic (super)strings with a time-varying
  tension}}, \href{https://doi.org/10.1088/1475-7516/2025/04/049}{\emph{JCAP}
  {\bfseries 04} (2025) 049}
  [\href{https://arxiv.org/abs/2411.04186}{{\ttfamily 2411.04186}}].

\bibitem{Reece:2023czb}
M.~Reece, \emph{{TASI Lectures: (No) Global Symmetries to Axion Physics}},
  \href{https://doi.org/10.22323/1.439.0008}{\emph{PoS} {\bfseries TASI2022}
  (2024) 008} [\href{https://arxiv.org/abs/2304.08512}{{\ttfamily
  2304.08512}}].

\bibitem{Witten:1985fp}
E.~Witten, \emph{{Cosmic Superstrings}},
  \href{https://doi.org/10.1016/0370-2693(85)90540-4}{\emph{Phys. Lett. B}
  {\bfseries 153} (1985) 243}.

\bibitem{Polchinski:1988cn}
J.~Polchinski, \emph{{Collision of Macroscopic Fundamental Strings}},
  \href{https://doi.org/10.1016/0370-2693(88)90942-2}{\emph{Phys. Lett. B}
  {\bfseries 209} (1988) 252}.

\bibitem{Copeland:2003bj}
E.~J. Copeland, R.~C. Myers and J.~Polchinski, \emph{{Cosmic F and D strings}},
  \href{https://doi.org/10.1088/1126-6708/2004/06/013}{\emph{JHEP} {\bfseries
  06} (2004) 013} [\href{https://arxiv.org/abs/hep-th/0312067}{{\ttfamily
  hep-th/0312067}}].

\bibitem{Damour:2004kw}
T.~Damour and A.~Vilenkin, \emph{{Gravitational radiation from cosmic
  (super)strings: Bursts, stochastic background, and observational windows}},
  \href{https://doi.org/10.1103/PhysRevD.71.063510}{\emph{Phys. Rev. D}
  {\bfseries 71} (2005) 063510}
  [\href{https://arxiv.org/abs/hep-th/0410222}{{\ttfamily hep-th/0410222}}].

\bibitem{Jackson:2004zg}
M.~G. Jackson, N.~T. Jones and J.~Polchinski, \emph{{Collisions of cosmic F and
  D-strings}}, \href{https://doi.org/10.1088/1126-6708/2005/10/013}{\emph{JHEP}
  {\bfseries 10} (2005) 013}
  [\href{https://arxiv.org/abs/hep-th/0405229}{{\ttfamily hep-th/0405229}}].

\bibitem{Copeland:2009ga}
E.~J. Copeland and T.~W.~B. Kibble, \emph{{Cosmic Strings and Superstrings}},
  \href{https://doi.org/10.1098/rspa.2009.0591}{\emph{Proc. Roy. Soc. Lond. A}
  {\bfseries 466} (2010) 623}
  [\href{https://arxiv.org/abs/0911.1345}{{\ttfamily 0911.1345}}].

\bibitem{Charnock:2016nzm}
T.~Charnock, A.~Avgoustidis, E.~J. Copeland and A.~Moss, \emph{{CMB constraints
  on cosmic strings and superstrings}},
  \href{https://doi.org/10.1103/PhysRevD.93.123503}{\emph{Phys. Rev. D}
  {\bfseries 93} (2016) 123503}
  [\href{https://arxiv.org/abs/1603.01275}{{\ttfamily 1603.01275}}].

\bibitem{EPTA:2023sfo}
{\scshape EPTA} collaboration, J.~Antoniadis et~al., \emph{{The second data
  release from the European Pulsar Timing Array - I. The dataset and timing
  analysis}}, \href{https://doi.org/10.1051/0004-6361/202346841}{\emph{Astron.
  Astrophys.} {\bfseries 678} (2023) A48}
  [\href{https://arxiv.org/abs/2306.16224}{{\ttfamily 2306.16224}}].

\bibitem{Marfatia:2023fvh}
D.~Marfatia and Y.-L. Zhou, \emph{{Gravitational waves from cosmic superstrings
  and gauge strings}},
  \href{https://doi.org/10.1007/JHEP07(2024)204}{\emph{JHEP} {\bfseries 07}
  (2024) 204} [\href{https://arxiv.org/abs/2312.10455}{{\ttfamily
  2312.10455}}].

\bibitem{Arkani-Hamed:1998jmv}
N.~Arkani-Hamed, S.~Dimopoulos and G.~R. Dvali, \emph{{The Hierarchy problem
  and new dimensions at a millimeter}},
  \href{https://doi.org/10.1016/S0370-2693(98)00466-3}{\emph{Phys. Lett. B}
  {\bfseries 429} (1998) 263}
  [\href{https://arxiv.org/abs/hep-ph/9803315}{{\ttfamily hep-ph/9803315}}].

\bibitem{Antoniadis:1998ig}
I.~Antoniadis, N.~Arkani-Hamed, S.~Dimopoulos and G.~R. Dvali, \emph{{New
  dimensions at a millimeter to a Fermi and superstrings at a TeV}},
  \href{https://doi.org/10.1016/S0370-2693(98)00860-0}{\emph{Phys. Lett. B}
  {\bfseries 436} (1998) 257}
  [\href{https://arxiv.org/abs/hep-ph/9804398}{{\ttfamily hep-ph/9804398}}].

\bibitem{Dasgupta:1999ss}
K.~Dasgupta, G.~Rajesh and S.~Sethi, \emph{{M theory, orientifolds and G -
  flux}}, \href{https://doi.org/10.1088/1126-6708/1999/08/023}{\emph{JHEP}
  {\bfseries 08} (1999) 023}
  [\href{https://arxiv.org/abs/hep-th/9908088}{{\ttfamily hep-th/9908088}}].

\bibitem{Giddings:2001yu}
S.~B. Giddings, S.~Kachru and J.~Polchinski, \emph{{Hierarchies from fluxes in
  string compactifications}},
  \href{https://doi.org/10.1103/PhysRevD.66.106006}{\emph{Phys. Rev. D}
  {\bfseries 66} (2002) 106006}
  [\href{https://arxiv.org/abs/hep-th/0105097}{{\ttfamily hep-th/0105097}}].

\bibitem{Giddings:2005ff}
S.~B. Giddings and A.~Maharana, \emph{{Dynamics of warped compactifications and
  the shape of the warped landscape}},
  \href{https://doi.org/10.1103/PhysRevD.73.126003}{\emph{Phys. Rev. D}
  {\bfseries 73} (2006) 126003}
  [\href{https://arxiv.org/abs/hep-th/0507158}{{\ttfamily hep-th/0507158}}].

\bibitem{Kachru:2003aw}
S.~Kachru, R.~Kallosh, A.~D. Linde and S.~P. Trivedi, \emph{{De Sitter vacua in
  string theory}},
  \href{https://doi.org/10.1103/PhysRevD.68.046005}{\emph{Phys. Rev. D}
  {\bfseries 68} (2003) 046005}
  [\href{https://arxiv.org/abs/hep-th/0301240}{{\ttfamily hep-th/0301240}}].

\bibitem{OCallaghan:2010rlo}
E.~O'Callaghan, S.~Chadburn, G.~Geshnizjani, R.~Gregory and I.~Zavala,
  \emph{{Effect of Extra Dimensions on Gravitational Waves from Cosmic
  Strings}}, \href{https://doi.org/10.1103/PhysRevLett.105.081602}{\emph{Phys.
  Rev. Lett.} {\bfseries 105} (2010) 081602}
  [\href{https://arxiv.org/abs/1003.4395}{{\ttfamily 1003.4395}}].

\bibitem{OCallaghan:2010mtk}
E.~O'Callaghan, S.~Chadburn, G.~Geshnizjani, R.~Gregory and I.~Zavala,
  \emph{{The effect of extra dimensions on gravity wave bursts from cosmic
  string cusps}},
  \href{https://doi.org/10.1088/1475-7516/2010/09/013}{\emph{JCAP} {\bfseries
  09} (2010) 013} [\href{https://arxiv.org/abs/1005.3220}{{\ttfamily
  1005.3220}}].

\bibitem{Avgoustidis:2007ju}
A.~Avgoustidis, \emph{{Cosmic String Dynamics and Evolution in Warped
  Spacetime}}, \href{https://doi.org/10.1103/PhysRevD.78.023501}{\emph{Phys.
  Rev. D} {\bfseries 78} (2008) 023501}
  [\href{https://arxiv.org/abs/0712.3224}{{\ttfamily 0712.3224}}].

\bibitem{Avgoustidis:2012vc}
A.~Avgoustidis, S.~Chadburn and R.~Gregory, \emph{{Cosmic superstring
  trajectories in warped compactifications}},
  \href{https://doi.org/10.1103/PhysRevD.86.063516}{\emph{Phys. Rev. D}
  {\bfseries 86} (2012) 063516}
  [\href{https://arxiv.org/abs/1204.0973}{{\ttfamily 1204.0973}}].

\bibitem{Dvali:1998pa}
G.~R. Dvali and S.~H.~H. Tye, \emph{{Brane inflation}},
  \href{https://doi.org/10.1016/S0370-2693(99)00132-X}{\emph{Phys. Lett. B}
  {\bfseries 450} (1999) 72}
  [\href{https://arxiv.org/abs/hep-ph/9812483}{{\ttfamily hep-ph/9812483}}].

\bibitem{Burgess:2001fx}
C.~P. Burgess, M.~Majumdar, D.~Nolte, F.~Quevedo, G.~Rajesh and R.-J. Zhang,
  \emph{{The Inflationary brane anti-brane universe}},
  \href{https://doi.org/10.1088/1126-6708/2001/07/047}{\emph{JHEP} {\bfseries
  07} (2001) 047} [\href{https://arxiv.org/abs/hep-th/0105204}{{\ttfamily
  hep-th/0105204}}].

\bibitem{Jones:2002cv}
N.~T. Jones, H.~Stoica and S.~H.~H. Tye, \emph{{Brane interaction as the origin
  of inflation}},
  \href{https://doi.org/10.1088/1126-6708/2002/07/051}{\emph{JHEP} {\bfseries
  07} (2002) 051} [\href{https://arxiv.org/abs/hep-th/0203163}{{\ttfamily
  hep-th/0203163}}].

\bibitem{Sarangi:2002yt}
S.~Sarangi and S.~H.~H. Tye, \emph{{Cosmic string production towards the end of
  brane inflation}},
  \href{https://doi.org/10.1016/S0370-2693(02)01824-5}{\emph{Phys. Lett. B}
  {\bfseries 536} (2002) 185}
  [\href{https://arxiv.org/abs/hep-th/0204074}{{\ttfamily hep-th/0204074}}].

\bibitem{Jones:2003da}
N.~T. Jones, H.~Stoica and S.~H.~H. Tye, \emph{{The Production, spectrum and
  evolution of cosmic strings in brane inflation}},
  \href{https://doi.org/10.1016/S0370-2693(03)00592-6}{\emph{Phys. Lett. B}
  {\bfseries 563} (2003) 6}
  [\href{https://arxiv.org/abs/hep-th/0303269}{{\ttfamily hep-th/0303269}}].

\bibitem{Pogosian:2003mz}
L.~Pogosian, S.~H.~H. Tye, I.~Wasserman and M.~Wyman, \emph{{Observational
  constraints on cosmic string production during brane inflation}},
  \href{https://doi.org/10.1103/PhysRevD.68.023506}{\emph{Phys. Rev. D}
  {\bfseries 68} (2003) 023506}
  [\href{https://arxiv.org/abs/hep-th/0304188}{{\ttfamily hep-th/0304188}}].

\bibitem{Kachru:2003sx}
S.~Kachru, R.~Kallosh, A.~D. Linde, J.~M. Maldacena, L.~P. McAllister and S.~P.
  Trivedi, \emph{{Towards inflation in string theory}},
  \href{https://doi.org/10.1088/1475-7516/2003/10/013}{\emph{JCAP} {\bfseries
  10} (2003) 013} [\href{https://arxiv.org/abs/hep-th/0308055}{{\ttfamily
  hep-th/0308055}}].

\bibitem{Burgess:2022nbx}
C.~P. Burgess and F.~Quevedo, \emph{{RG-induced modulus stabilization:
  perturbative de Sitter vacua and improved D3-$ \overline{\mathrm{D}3} $
  inflation}}, \href{https://doi.org/10.1007/JHEP06(2022)167}{\emph{JHEP}
  {\bfseries 06} (2022) 167}
  [\href{https://arxiv.org/abs/2202.05344}{{\ttfamily 2202.05344}}].

\bibitem{Frey:2023khe}
A.~R. Frey, R.~Mahanta, A.~Maharana, F.~Muia, F.~Quevedo and G.~Villa,
  \emph{{String thermodynamics in and out of equilibrium: Boltzmann equations
  and random walks}},
  \href{https://doi.org/10.1007/JHEP03(2024)112}{\emph{JHEP} {\bfseries 03}
  (2024) 112} [\href{https://arxiv.org/abs/2310.11494}{{\ttfamily
  2310.11494}}].

\bibitem{Frey:2024jqy}
A.~R. Frey, R.~Mahanta, A.~Maharana, F.~Quevedo and G.~Villa,
  \emph{{Gravitational waves from high temperature strings}},
  \href{https://doi.org/10.1007/JHEP12(2024)174}{\emph{JHEP} {\bfseries 12}
  (2024) 174} [\href{https://arxiv.org/abs/2408.13803}{{\ttfamily
  2408.13803}}].

\bibitem{Atick:1988si}
J.~J. Atick and E.~Witten, \emph{{The Hagedorn Transition and the Number of
  Degrees of Freedom of String Theory}},
  \href{https://doi.org/10.1016/0550-3213(88)90151-4}{\emph{Nucl. Phys. B}
  {\bfseries 310} (1988) 291}.

\bibitem{Barbon:2004dd}
J.~L.~F. Barbon and E.~Rabinovici, \emph{{Touring the Hagedorn ridge}},  in
  \emph{{From Fields to Strings: Circumnavigating Theoretical Physics: A
  Conference in Tribute to Ian Kogan}}, pp.~1973--2008, 8, 2004,
  \href{https://arxiv.org/abs/hep-th/0407236}{{\ttfamily hep-th/0407236}},
  \href{https://doi.org/10.1142/9789812775344_0048}{DOI}.

\bibitem{Horowitz:1996nw}
G.~T. Horowitz and J.~Polchinski, \emph{{A Correspondence principle for black
  holes and strings}},
  \href{https://doi.org/10.1103/PhysRevD.55.6189}{\emph{Phys. Rev. D}
  {\bfseries 55} (1997) 6189}
  [\href{https://arxiv.org/abs/hep-th/9612146}{{\ttfamily hep-th/9612146}}].

\bibitem{Horowitz:1997jc}
G.~T. Horowitz and J.~Polchinski, \emph{{Selfgravitating fundamental strings}},
  \href{https://doi.org/10.1103/PhysRevD.57.2557}{\emph{Phys. Rev. D}
  {\bfseries 57} (1998) 2557}
  [\href{https://arxiv.org/abs/hep-th/9707170}{{\ttfamily hep-th/9707170}}].

\bibitem{Abel:1999rq}
S.~A. Abel, J.~L.~F. Barbon, I.~I. Kogan and E.~Rabinovici, \emph{{String
  thermodynamics in D-brane backgrounds}},
  \href{https://doi.org/10.1088/1126-6708/1999/04/015}{\emph{JHEP} {\bfseries
  04} (1999) 015} [\href{https://arxiv.org/abs/hep-th/9902058}{{\ttfamily
  hep-th/9902058}}].

\bibitem{Conlon:2024uob}
J.~P. Conlon, E.~J. Copeland, E.~Hardy and N.~S. Gonz\'alez, \emph{{Percolating
  Cosmic String Networks from Kination}},
  \href{https://arxiv.org/abs/2406.12637}{{\ttfamily 2406.12637}}.

\bibitem{Brunelli:2025ems}
L.~Brunelli, M.~Cicoli and F.~G. Pedro, \emph{{Growth of Cosmic Strings beyond
  Kination}},  \href{https://arxiv.org/abs/2503.11293}{{\ttfamily 2503.11293}}.

\bibitem{Kibble:1984hp}
T.~W.~B. Kibble, \emph{{Evolution of a system of cosmic strings}},
  \href{https://doi.org/10.1016/0550-3213(85)90596-6}{\emph{Nucl. Phys. B}
  {\bfseries 252} (1985) 227}.

\bibitem{Bennett:1985qt}
D.~P. Bennett, \emph{{The evolution of cosmic strings}},
  \href{https://doi.org/10.1103/PhysRevD.33.872}{\emph{Phys. Rev. D} {\bfseries
  33} (1986) 872}.

\bibitem{Austin:1993rg}
D.~Austin, E.~J. Copeland and T.~W.~B. Kibble, \emph{{Evolution of cosmic
  string configurations}},
  \href{https://doi.org/10.1103/PhysRevD.48.5594}{\emph{Phys. Rev. D}
  {\bfseries 48} (1993) 5594}
  [\href{https://arxiv.org/abs/hep-ph/9307325}{{\ttfamily hep-ph/9307325}}].

\bibitem{Vilenkin:2000jqa}
A.~Vilenkin and E.~P.~S. Shellard, \emph{{Cosmic Strings and Other Topological
  Defects}}. Cambridge University Press, 7, 2000.

\bibitem{Martins:1996jp}
C.~J. A.~P. Martins and E.~P.~S. Shellard, \emph{{Quantitative string
  evolution}}, \href{https://doi.org/10.1103/PhysRevD.54.2535}{\emph{Phys. Rev.
  D} {\bfseries 54} (1996) 2535}
  [\href{https://arxiv.org/abs/hep-ph/9602271}{{\ttfamily hep-ph/9602271}}].

\bibitem{Martins:2000cs}
C.~J. A.~P. Martins and E.~P.~S. Shellard, \emph{{Extending the velocity
  dependent one scale string evolution model}},
  \href{https://doi.org/10.1103/PhysRevD.65.043514}{\emph{Phys. Rev. D}
  {\bfseries 65} (2002) 043514}
  [\href{https://arxiv.org/abs/hep-ph/0003298}{{\ttfamily hep-ph/0003298}}].

\bibitem{Emond:2021vts}
W.~T. Emond, S.~Ramazanov and R.~Samanta, \emph{{Gravitational waves from
  melting cosmic strings}},
  \href{https://doi.org/10.1088/1475-7516/2022/01/057}{\emph{JCAP} {\bfseries
  01} (2022) 057} [\href{https://arxiv.org/abs/2108.05377}{{\ttfamily
  2108.05377}}].

\bibitem{Avgoustidis:2005nv}
A.~Avgoustidis and E.~P.~S. Shellard, \emph{{Effect of reconnection probability
  on cosmic (super)string network density}},
  \href{https://doi.org/10.1103/PhysRevD.73.041301}{\emph{Phys. Rev. D}
  {\bfseries 73} (2006) 041301}
  [\href{https://arxiv.org/abs/astro-ph/0512582}{{\ttfamily
  astro-ph/0512582}}].

\bibitem{Sakellariadou:2004wq}
M.~Sakellariadou, \emph{{A Note on the evolution of cosmic string/superstring
  networks}}, \href{https://doi.org/10.1088/1475-7516/2005/04/003}{\emph{JCAP}
  {\bfseries 04} (2005) 003}
  [\href{https://arxiv.org/abs/hep-th/0410234}{{\ttfamily hep-th/0410234}}].

\bibitem{Avelino:2012qy}
P.~P. Avelino and L.~Sousa, \emph{{Scaling laws for weakly interacting cosmic
  (super)string and p-brane networks}},
  \href{https://doi.org/10.1103/PhysRevD.85.083525}{\emph{Phys. Rev. D}
  {\bfseries 85} (2012) 083525}
  [\href{https://arxiv.org/abs/1202.6298}{{\ttfamily 1202.6298}}].

\bibitem{Sousa:2016ggw}
L.~Sousa and P.~P. Avelino, \emph{{Probing Cosmic Superstrings with
  Gravitational Waves}},
  \href{https://doi.org/10.1103/PhysRevD.94.063529}{\emph{Phys. Rev. D}
  {\bfseries 94} (2016) 063529}
  [\href{https://arxiv.org/abs/1606.05585}{{\ttfamily 1606.05585}}].

\bibitem{Avgoustidis:2007aa}
A.~Avgoustidis and E.~P.~S. Shellard, \emph{{Velocity-Dependent Models for
  Non-Abelian/Entangled String Networks}},
  \href{https://doi.org/10.1103/PhysRevD.78.103510}{\emph{Phys. Rev. D}
  {\bfseries 78} (2008) 103510}
  [\href{https://arxiv.org/abs/0705.3395}{{\ttfamily 0705.3395}}].

\bibitem{Avgoustidis:2009ke}
A.~Avgoustidis and E.~J. Copeland, \emph{{The effect of kinematic constraints
  on multi-tension string network evolution}},
  \href{https://doi.org/10.1103/PhysRevD.81.063517}{\emph{Phys. Rev. D}
  {\bfseries 81} (2010) 063517}
  [\href{https://arxiv.org/abs/0912.4004}{{\ttfamily 0912.4004}}].

\bibitem{Pourtsidou:2010gu}
A.~Pourtsidou, A.~Avgoustidis, E.~J. Copeland, L.~Pogosian and D.~A. Steer,
  \emph{{Scaling configurations of cosmic superstring networks and their
  cosmological implications}},
  \href{https://doi.org/10.1103/PhysRevD.83.063525}{\emph{Phys. Rev. D}
  {\bfseries 83} (2011) 063525}
  [\href{https://arxiv.org/abs/1012.5014}{{\ttfamily 1012.5014}}].

\bibitem{Siemens:2006yp}
X.~Siemens, V.~Mandic and J.~Creighton, \emph{{Gravitational wave stochastic
  background from cosmic (super)strings}},
  \href{https://doi.org/10.1103/PhysRevLett.98.111101}{\emph{Phys. Rev. Lett.}
  {\bfseries 98} (2007) 111101}
  [\href{https://arxiv.org/abs/astro-ph/0610920}{{\ttfamily
  astro-ph/0610920}}].

\bibitem{Blanco-Pillado:2013qja}
J.~J. Blanco-Pillado, K.~D. Olum and B.~Shlaer, \emph{{The number of cosmic
  string loops}}, \href{https://doi.org/10.1103/PhysRevD.89.023512}{\emph{Phys.
  Rev. D} {\bfseries 89} (2014) 023512}
  [\href{https://arxiv.org/abs/1309.6637}{{\ttfamily 1309.6637}}].

\bibitem{Cui:2018rwi}
Y.~Cui, M.~Lewicki, D.~E. Morrissey and J.~D. Wells, \emph{{Probing the pre-BBN
  universe with gravitational waves from cosmic strings}},
  \href{https://doi.org/10.1007/JHEP01(2019)081}{\emph{JHEP} {\bfseries 01}
  (2019) 081} [\href{https://arxiv.org/abs/1808.08968}{{\ttfamily
  1808.08968}}].

\bibitem{Vilenkin:1981bx}
A.~Vilenkin, \emph{{Gravitational radiation from cosmic strings}},
  \href{https://doi.org/10.1016/0370-2693(81)91144-8}{\emph{Phys. Lett. B}
  {\bfseries 107} (1981) 47}.

\bibitem{Vachaspati:1984gt}
T.~Vachaspati and A.~Vilenkin, \emph{{Gravitational Radiation from Cosmic
  Strings}}, \href{https://doi.org/10.1103/PhysRevD.31.3052}{\emph{Phys. Rev.
  D} {\bfseries 31} (1985) 3052}.

\bibitem{Turok:1984cn}
N.~Turok, \emph{{Grand Unified Strings and Galaxy Formation}},
  \href{https://doi.org/10.1016/0550-3213(84)90407-3}{\emph{Nucl. Phys. B}
  {\bfseries 242} (1984) 520}.

\bibitem{Burden:1985md}
C.~J. Burden, \emph{{Gravitational Radiation From a Particular Class of Cosmic
  Strings}}, \href{https://doi.org/10.1016/0370-2693(85)90326-0}{\emph{Phys.
  Lett. B} {\bfseries 164} (1985) 277}.

\bibitem{Olum:1999sg}
K.~D. Olum and J.~J. Blanco-Pillado, \emph{{Radiation from cosmic string
  standing waves}},
  \href{https://doi.org/10.1103/PhysRevLett.84.4288}{\emph{Phys. Rev. Lett.}
  {\bfseries 84} (2000) 4288}
  [\href{https://arxiv.org/abs/astro-ph/9910354}{{\ttfamily
  astro-ph/9910354}}].

\bibitem{Moore:2001px}
J.~N. Moore, E.~P.~S. Shellard and C.~J. A.~P. Martins, \emph{{On the evolution
  of Abelian-Higgs string networks}},
  \href{https://doi.org/10.1103/PhysRevD.65.023503}{\emph{Phys. Rev. D}
  {\bfseries 65} (2002) 023503}
  [\href{https://arxiv.org/abs/hep-ph/0107171}{{\ttfamily hep-ph/0107171}}].

\bibitem{Blanco-Pillado:2017oxo}
J.~J. Blanco-Pillado and K.~D. Olum, \emph{{Stochastic gravitational wave
  background from smoothed cosmic string loops}},
  \href{https://doi.org/10.1103/PhysRevD.96.104046}{\emph{Phys. Rev. D}
  {\bfseries 96} (2017) 104046}
  [\href{https://arxiv.org/abs/1709.02693}{{\ttfamily 1709.02693}}].

\bibitem{Apers:2024dtn}
F.~Apers, J.~P. Conlon and M.~Mosny, \emph{{A Note on 4d Kination and
  Higher-Dimensional Uplifts}},
  \href{https://arxiv.org/abs/2409.08049}{{\ttfamily 2409.08049}}.

\bibitem{Conlon:2006gv}
J.~P. Conlon, \emph{{Moduli Stabilisation and Applications in IIB String
  Theory}}, \href{https://doi.org/10.1002/prop.200610334}{\emph{Fortsch. Phys.}
  {\bfseries 55} (2007) 287}
  [\href{https://arxiv.org/abs/hep-th/0611039}{{\ttfamily hep-th/0611039}}].

\bibitem{Conlon:2005jm}
J.~P. Conlon and F.~Quevedo, \emph{{Kahler moduli inflation}},
  \href{https://doi.org/10.1088/1126-6708/2006/01/146}{\emph{JHEP} {\bfseries
  01} (2006) 146} [\href{https://arxiv.org/abs/hep-th/0509012}{{\ttfamily
  hep-th/0509012}}].

\bibitem{Conlon:2006tq}
J.~P. Conlon, \emph{{The QCD axion and moduli stabilisation}},
  \href{https://doi.org/10.1088/1126-6708/2006/05/078}{\emph{JHEP} {\bfseries
  05} (2006) 078} [\href{https://arxiv.org/abs/hep-th/0602233}{{\ttfamily
  hep-th/0602233}}].

\bibitem{Conlon:2007gk}
J.~P. Conlon and F.~Quevedo, \emph{{Astrophysical and cosmological implications
  of large volume string compactifications}},
  \href{https://doi.org/10.1088/1475-7516/2007/08/019}{\emph{JCAP} {\bfseries
  08} (2007) 019} [\href{https://arxiv.org/abs/0705.3460}{{\ttfamily
  0705.3460}}].

\bibitem{Blumenhagen:2009gk}
R.~Blumenhagen, J.~P. Conlon, S.~Krippendorf, S.~Moster and F.~Quevedo,
  \emph{{SUSY Breaking in Local String/F-Theory Models}},
  \href{https://doi.org/10.1088/1126-6708/2009/09/007}{\emph{JHEP} {\bfseries
  09} (2009) 007} [\href{https://arxiv.org/abs/0906.3297}{{\ttfamily
  0906.3297}}].

\bibitem{Cicoli:2012sz}
M.~Cicoli, M.~Goodsell and A.~Ringwald, \emph{{The type IIB string axiverse and
  its low-energy phenomenology}},
  \href{https://doi.org/10.1007/JHEP10(2012)146}{\emph{JHEP} {\bfseries 10}
  (2012) 146} [\href{https://arxiv.org/abs/1206.0819}{{\ttfamily 1206.0819}}].

\bibitem{Cicoli:2012aq}
M.~Cicoli, J.~P. Conlon and F.~Quevedo, \emph{{Dark radiation in LARGE volume
  models}}, \href{https://doi.org/10.1103/PhysRevD.87.043520}{\emph{Phys. Rev.
  D} {\bfseries 87} (2013) 043520}
  [\href{https://arxiv.org/abs/1208.3562}{{\ttfamily 1208.3562}}].

\bibitem{Hebecker:2014gka}
A.~Hebecker, P.~Mangat, F.~Rompineve and L.~T. Witkowski, \emph{{Dark Radiation
  predictions from general Large Volume Scenarios}},
  \href{https://doi.org/10.1007/JHEP09(2014)140}{\emph{JHEP} {\bfseries 09}
  (2014) 140} [\href{https://arxiv.org/abs/1403.6810}{{\ttfamily 1403.6810}}].

\bibitem{Becker:2002nn}
K.~Becker, M.~Becker, M.~Haack and J.~Louis, \emph{{Supersymmetry breaking and
  alpha-prime corrections to flux induced potentials}},
  \href{https://doi.org/10.1088/1126-6708/2002/06/060}{\emph{JHEP} {\bfseries
  06} (2002) 060} [\href{https://arxiv.org/abs/hep-th/0204254}{{\ttfamily
  hep-th/0204254}}].

\bibitem{Bonetti:2016dqh}
F.~Bonetti and M.~Weissenbacher, \emph{{The Euler characteristic correction to
  the K\"ahler potential \textemdash{} revisited}},
  \href{https://doi.org/10.1007/JHEP01(2017)003}{\emph{JHEP} {\bfseries 01}
  (2017) 003} [\href{https://arxiv.org/abs/1608.01300}{{\ttfamily
  1608.01300}}].

\bibitem{Burgess:2020qsc}
C.~P. Burgess, M.~Cicoli, D.~Ciupke, S.~Krippendorf and F.~Quevedo, \emph{{UV
  Shadows in EFTs: Accidental Symmetries, Robustness and No-Scale
  Supergravity}}, \href{https://doi.org/10.1002/prop.202000076}{\emph{Fortsch.
  Phys.} {\bfseries 68} (2020) 2000076}
  [\href{https://arxiv.org/abs/2006.06694}{{\ttfamily 2006.06694}}].

\bibitem{Cicoli:2021rub}
M.~Cicoli, F.~Quevedo, R.~Savelli, A.~Schachner and R.~Valandro,
  \emph{{Systematics of the \ensuremath{\alpha}' expansion in F-theory}},
  \href{https://doi.org/10.1007/JHEP08(2021)099}{\emph{JHEP} {\bfseries 08}
  (2021) 099} [\href{https://arxiv.org/abs/2106.04592}{{\ttfamily
  2106.04592}}].

\bibitem{Weissenbacher:2019mef}
M.~Weissenbacher, \emph{{F-theory vacua and $\alpha'$-corrections}},
  \href{https://doi.org/10.1007/JHEP04(2020)032}{\emph{JHEP} {\bfseries 04}
  (2020) 032} [\href{https://arxiv.org/abs/1901.04758}{{\ttfamily
  1901.04758}}].

\bibitem{Weissenbacher:2020cyf}
M.~Weissenbacher, \emph{{On $\alpha'$-effects from $D$-branes in $4d \;
  \mathcal{N} = 1$}},
  \href{https://doi.org/10.1007/JHEP11(2020)076}{\emph{JHEP} {\bfseries 11}
  (2020) 076} [\href{https://arxiv.org/abs/2006.15552}{{\ttfamily
  2006.15552}}].

\bibitem{Klaewer:2020lfg}
D.~Klaewer, S.-J. Lee, T.~Weigand and M.~Wiesner, \emph{{Quantum corrections in
  4d $N$ = 1 infinite distance limits and the weak gravity conjecture}},
  \href{https://doi.org/10.1007/JHEP03(2021)252}{\emph{JHEP} {\bfseries 03}
  (2021) 252} [\href{https://arxiv.org/abs/2011.00024}{{\ttfamily
  2011.00024}}].

\bibitem{Antoniadis:1997eg}
I.~Antoniadis, S.~Ferrara, R.~Minasian and K.~S. Narain, \emph{{R**4 couplings
  in M and type II theories on Calabi-Yau spaces}},
  \href{https://doi.org/10.1016/S0550-3213(97)00572-5}{\emph{Nucl. Phys. B}
  {\bfseries 507} (1997) 571}
  [\href{https://arxiv.org/abs/hep-th/9707013}{{\ttfamily hep-th/9707013}}].

\bibitem{Minasian:2015bxa}
R.~Minasian, T.~G. Pugh and R.~Savelli, \emph{{F-theory at order $\alpha'^3$}},
  \href{https://doi.org/10.1007/JHEP10(2015)050}{\emph{JHEP} {\bfseries 10}
  (2015) 050} [\href{https://arxiv.org/abs/1506.06756}{{\ttfamily
  1506.06756}}].

\bibitem{Antoniadis:2018hqy}
I.~Antoniadis, Y.~Chen and G.~K. Leontaris, \emph{{Perturbative moduli
  stabilisation in type IIB/F-theory framework}},
  \href{https://doi.org/10.1140/epjc/s10052-018-6248-4}{\emph{Eur. Phys. J. C}
  {\bfseries 78} (2018) 766}
  [\href{https://arxiv.org/abs/1803.08941}{{\ttfamily 1803.08941}}].

\bibitem{Antoniadis:2019rkh}
I.~Antoniadis, Y.~Chen and G.~K. Leontaris, \emph{{Logarithmic loop
  corrections, moduli stabilisation and de Sitter vacua in string theory}},
  \href{https://doi.org/10.1007/JHEP01(2020)149}{\emph{JHEP} {\bfseries 01}
  (2020) 149} [\href{https://arxiv.org/abs/1909.10525}{{\ttfamily
  1909.10525}}].

\bibitem{Leontaris:2022rzj}
G.~K. Leontaris and P.~Shukla, \emph{{Stabilising all K\"ahler moduli in
  perturbative LVS}},
  \href{https://doi.org/10.1007/JHEP07(2022)047}{\emph{JHEP} {\bfseries 07}
  (2022) 047} [\href{https://arxiv.org/abs/2203.03362}{{\ttfamily
  2203.03362}}].

\bibitem{Planck:2018vyg}
{\scshape Planck} collaboration, N.~Aghanim et~al., \emph{{Planck 2018 results.
  VI. Cosmological parameters}},
  \href{https://doi.org/10.1051/0004-6361/201833910}{\emph{Astron. Astrophys.}
  {\bfseries 641} (2020) A6}
  [\href{https://arxiv.org/abs/1807.06209}{{\ttfamily 1807.06209}}].

\bibitem{Conlon:2008cj}
J.~P. Conlon, R.~Kallosh, A.~D. Linde and F.~Quevedo, \emph{{Volume Modulus
  Inflation and the Gravitino Mass Problem}},
  \href{https://doi.org/10.1088/1475-7516/2008/09/011}{\emph{JCAP} {\bfseries
  09} (2008) 011} [\href{https://arxiv.org/abs/0806.0809}{{\ttfamily
  0806.0809}}].

\bibitem{Linde:1993cn}
A.~D. Linde, \emph{{Hybrid inflation}},
  \href{https://doi.org/10.1103/PhysRevD.49.748}{\emph{Phys. Rev. D} {\bfseries
  49} (1994) 748} [\href{https://arxiv.org/abs/astro-ph/9307002}{{\ttfamily
  astro-ph/9307002}}].

\bibitem{Chen:2024roo}
C.~Chen, K.~Dimopoulos, C.~Er\"oncel and A.~Ghoshal, \emph{{Enhanced primordial
  gravitational waves from a stiff postinflationary era due to an oscillating
  inflaton}}, \href{https://doi.org/10.1103/PhysRevD.110.063554}{\emph{Phys.
  Rev. D} {\bfseries 110} (2024) 063554}
  [\href{https://arxiv.org/abs/2405.01679}{{\ttfamily 2405.01679}}].

\bibitem{Andriot:2024jsh}
D.~Andriot, S.~Parameswaran, D.~Tsimpis, T.~Wrase and I.~Zavala,
  \emph{{Exponential Quintessence: curved, steep and stringy?}},
  \href{https://arxiv.org/abs/2405.09323}{{\ttfamily 2405.09323}}.

\bibitem{Andriot:2024sif}
D.~Andriot, \emph{{Quintessence: an analytical study, with theoretical and
  observational applications}},
  \href{https://arxiv.org/abs/2410.17182}{{\ttfamily 2410.17182}}.

\bibitem{Wetterich:1987fm}
C.~Wetterich, \emph{{Cosmology and the Fate of Dilatation Symmetry}},
  \href{https://doi.org/10.1016/0550-3213(88)90193-9}{\emph{Nucl. Phys. B}
  {\bfseries 302} (1988) 668}
  [\href{https://arxiv.org/abs/1711.03844}{{\ttfamily 1711.03844}}].

\bibitem{Copeland:1997et}
E.~J. Copeland, A.~R. Liddle and D.~Wands, \emph{{Exponential potentials and
  cosmological scaling solutions}},
  \href{https://doi.org/10.1103/PhysRevD.57.4686}{\emph{Phys. Rev. D}
  {\bfseries 57} (1998) 4686}
  [\href{https://arxiv.org/abs/gr-qc/9711068}{{\ttfamily gr-qc/9711068}}].

\bibitem{Ferreira:1997hj}
P.~G. Ferreira and M.~Joyce, \emph{{Cosmology with a primordial scaling
  field}}, \href{https://doi.org/10.1103/PhysRevD.58.023503}{\emph{Phys. Rev.
  D} {\bfseries 58} (1998) 023503}
  [\href{https://arxiv.org/abs/astro-ph/9711102}{{\ttfamily
  astro-ph/9711102}}].

\bibitem{Bhaumik:2022pil}
N.~Bhaumik, A.~Ghoshal and M.~Lewicki, \emph{{Doubly peaked induced stochastic
  gravitational wave background: testing baryogenesis from primordial black
  holes}}, \href{https://doi.org/10.1007/JHEP07(2022)130}{\emph{JHEP}
  {\bfseries 07} (2022) 130}
  [\href{https://arxiv.org/abs/2205.06260}{{\ttfamily 2205.06260}}].

\bibitem{Bhaumik:2022zdd}
N.~Bhaumik, A.~Ghoshal, R.~K. Jain and M.~Lewicki, \emph{{Distinct signatures
  of spinning PBH domination and evaporation: doubly peaked gravitational
  waves, dark relics and CMB complementarity}},
  \href{https://doi.org/10.1007/JHEP05(2023)169}{\emph{JHEP} {\bfseries 05}
  (2023) 169} [\href{https://arxiv.org/abs/2212.00775}{{\ttfamily
  2212.00775}}].

\bibitem{Brustein:1992nk}
R.~Brustein and P.~J. Steinhardt, \emph{{Challenges for superstring
  cosmology}}, \href{https://doi.org/10.1016/0370-2693(93)90384-T}{\emph{Phys.
  Lett. B} {\bfseries 302} (1993) 196}
  [\href{https://arxiv.org/abs/hep-th/9212049}{{\ttfamily hep-th/9212049}}].

\bibitem{Barreiro:1998aj}
T.~Barreiro, B.~de~Carlos and E.~J. Copeland, \emph{{Stabilizing the dilaton in
  superstring cosmology}},
  \href{https://doi.org/10.1103/PhysRevD.58.083513}{\emph{Phys. Rev. D}
  {\bfseries 58} (1998) 083513}
  [\href{https://arxiv.org/abs/hep-th/9805005}{{\ttfamily hep-th/9805005}}].

\bibitem{Huey:2000jx}
G.~Huey, P.~J. Steinhardt, B.~A. Ovrut and D.~Waldram, \emph{{A Cosmological
  mechanism for stabilizing moduli}},
  \href{https://doi.org/10.1016/S0370-2693(00)00152-0}{\emph{Phys. Lett. B}
  {\bfseries 476} (2000) 379}
  [\href{https://arxiv.org/abs/hep-th/0001112}{{\ttfamily hep-th/0001112}}].

\bibitem{Brustein:2004jp}
R.~Brustein, S.~P. de~Alwis and P.~Martens, \emph{{Cosmological stabilization
  of moduli with steep potentials}},
  \href{https://doi.org/10.1103/PhysRevD.70.126012}{\emph{Phys. Rev. D}
  {\bfseries 70} (2004) 126012}
  [\href{https://arxiv.org/abs/hep-th/0408160}{{\ttfamily hep-th/0408160}}].

\bibitem{Barreiro:2005ua}
T.~Barreiro, B.~de~Carlos, E.~Copeland and N.~J. Nunes, \emph{{Moduli evolution
  in the presence of flux compactifications}},
  \href{https://doi.org/10.1103/PhysRevD.72.106004}{\emph{Phys. Rev. D}
  {\bfseries 72} (2005) 106004}
  [\href{https://arxiv.org/abs/hep-ph/0506045}{{\ttfamily hep-ph/0506045}}].

\bibitem{Coughlan:1983ci}
G.~D. Coughlan, W.~Fischler, E.~W. Kolb, S.~Raby and G.~G. Ross,
  \emph{{Cosmological Problems for the Polonyi Potential}},
  \href{https://doi.org/10.1016/0370-2693(83)91091-2}{\emph{Phys. Lett. B}
  {\bfseries 131} (1983) 59}.

\bibitem{Banks:1993en}
T.~Banks, D.~B. Kaplan and A.~E. Nelson, \emph{{Cosmological implications of
  dynamical supersymmetry breaking}},
  \href{https://doi.org/10.1103/PhysRevD.49.779}{\emph{Phys. Rev. D} {\bfseries
  49} (1994) 779} [\href{https://arxiv.org/abs/hep-ph/9308292}{{\ttfamily
  hep-ph/9308292}}].

\bibitem{deCarlos:1993wie}
B.~de~Carlos, J.~A. Casas, F.~Quevedo and E.~Roulet, \emph{{Model independent
  properties and cosmological implications of the dilaton and moduli sectors of
  4-d strings}},
  \href{https://doi.org/10.1016/0370-2693(93)91538-X}{\emph{Phys. Lett. B}
  {\bfseries 318} (1993) 447}
  [\href{https://arxiv.org/abs/hep-ph/9308325}{{\ttfamily hep-ph/9308325}}].

\bibitem{Higaki:2012ar}
T.~Higaki and F.~Takahashi, \emph{{Dark Radiation and Dark Matter in Large
  Volume Compactifications}},
  \href{https://doi.org/10.1007/JHEP11(2012)125}{\emph{JHEP} {\bfseries 11}
  (2012) 125} [\href{https://arxiv.org/abs/1208.3563}{{\ttfamily 1208.3563}}].

\bibitem{Leedom:2024qgr}
J.~M. Leedom, M.~Putti, N.~Righi and A.~Westphal, \emph{{Preheating Axions in
  String Cosmology}},  \href{https://arxiv.org/abs/2411.18496}{{\ttfamily
  2411.18496}}.

\bibitem{Cicoli:2022fzy}
M.~Cicoli, A.~Hebecker, J.~Jaeckel and M.~Wittner, \emph{{Axions in string
  theory \textemdash{} slaying the Hydra of dark radiation}},
  \href{https://doi.org/10.1007/JHEP09(2022)198}{\emph{JHEP} {\bfseries 09}
  (2022) 198} [\href{https://arxiv.org/abs/2203.08833}{{\ttfamily
  2203.08833}}].

\bibitem{Aparicio:2014wxa}
L.~Aparicio, M.~Cicoli, S.~Krippendorf, A.~Maharana, F.~Muia and F.~Quevedo,
  \emph{{Sequestered de Sitter String Scenarios: Soft-terms}},
  \href{https://doi.org/10.1007/JHEP11(2014)071}{\emph{JHEP} {\bfseries 11}
  (2014) 071} [\href{https://arxiv.org/abs/1409.1931}{{\ttfamily 1409.1931}}].

\bibitem{Reece:2015qbf}
M.~Reece and W.~Xue, \emph{{SUSY\textquoteright{}s Ladder: reframing
  sequestering at Large Volume}},
  \href{https://doi.org/10.1007/JHEP04(2016)045}{\emph{JHEP} {\bfseries 04}
  (2016) 045} [\href{https://arxiv.org/abs/1512.04941}{{\ttfamily
  1512.04941}}].

\bibitem{Co:2021lkc}
R.~T. Co, D.~Dunsky, N.~Fernandez, A.~Ghalsasi, L.~J. Hall, K.~Harigaya et~al.,
  \emph{{Gravitational wave and CMB probes of axion kination}},
  \href{https://doi.org/10.1007/JHEP09(2022)116}{\emph{JHEP} {\bfseries 09}
  (2022) 116} [\href{https://arxiv.org/abs/2108.09299}{{\ttfamily
  2108.09299}}].

\bibitem{Gouttenoire:2021wzu}
Y.~Gouttenoire, G.~Servant and P.~Simakachorn, \emph{{Revealing the Primordial
  Irreducible Inflationary Gravitational-Wave Background with a Spinning
  Peccei-Quinn Axion}},  \href{https://arxiv.org/abs/2108.10328}{{\ttfamily
  2108.10328}}.

\bibitem{Gouttenoire:2021jhk}
Y.~Gouttenoire, G.~Servant and P.~Simakachorn, \emph{{Kination cosmology from
  scalar fields and gravitational-wave signatures}},
  \href{https://arxiv.org/abs/2111.01150}{{\ttfamily 2111.01150}}.

\bibitem{Chowdhury:2022gdc}
D.~Chowdhury, G.~Tasinato and I.~Zavala, \emph{{The rise of the primordial
  tensor spectrum from an early scalar-tensor epoch}},
  \href{https://doi.org/10.1088/1475-7516/2022/08/010}{\emph{JCAP} {\bfseries
  08} (2022) 010} [\href{https://arxiv.org/abs/2204.10218}{{\ttfamily
  2204.10218}}].

\bibitem{Chowdhury:2023opo}
D.~Chowdhury, G.~Tasinato and I.~Zavala, \emph{{Dark energy, D-branes and
  pulsar timing arrays}},
  \href{https://doi.org/10.1088/1475-7516/2023/11/090}{\emph{JCAP} {\bfseries
  11} (2023) 090} [\href{https://arxiv.org/abs/2307.01188}{{\ttfamily
  2307.01188}}].

\bibitem{Sousa:2024ytl}
L.~Sousa, \emph{{Cosmic strings and gravitational waves}},
  \href{https://doi.org/10.1007/s10714-024-03293-x}{\emph{Gen. Rel. Grav.}
  {\bfseries 56} (2024) 105}.

\bibitem{Ghoshal:2023sfa}
A.~Ghoshal, Y.~Gouttenoire, L.~Heurtier and P.~Simakachorn, \emph{{Primordial
  black hole archaeology with gravitational waves from cosmic strings}},
  \href{https://doi.org/10.1007/JHEP08(2023)196}{\emph{JHEP} {\bfseries 08}
  (2023) 196} [\href{https://arxiv.org/abs/2304.04793}{{\ttfamily
  2304.04793}}].

\bibitem{Sousa:2013aaa}
L.~Sousa and P.~P. Avelino, \emph{{Stochastic Gravitational Wave Background
  generated by Cosmic String Networks: Velocity-Dependent One-Scale model
  versus Scale-Invariant Evolution}},
  \href{https://doi.org/10.1103/PhysRevD.88.023516}{\emph{Phys. Rev. D}
  {\bfseries 88} (2013) 023516}
  [\href{https://arxiv.org/abs/1304.2445}{{\ttfamily 1304.2445}}].

\bibitem{Ebelt:2023clh}
J.~Ebelt, S.~Krippendorf and A.~Schachner, \emph{{W0\_sample =
  np.random.normal(0,1)?}},
  \href{https://doi.org/10.1016/j.physletb.2024.138786}{\emph{Phys. Lett. B}
  {\bfseries 855} (2024) 138786}
  [\href{https://arxiv.org/abs/2307.15749}{{\ttfamily 2307.15749}}].

\bibitem{Gukov:1999ya}
S.~Gukov, C.~Vafa and E.~Witten, \emph{{CFT's from Calabi-Yau four folds}},
  \href{https://doi.org/10.1016/S0550-3213(00)00373-4}{\emph{Nucl. Phys. B}
  {\bfseries 584} (2000) 69}
  [\href{https://arxiv.org/abs/hep-th/9906070}{{\ttfamily hep-th/9906070}}].

\bibitem{Cicoli:2013swa}
M.~Cicoli, J.~P. Conlon, A.~Maharana and F.~Quevedo, \emph{{A Note on the
  Magnitude of the Flux Superpotential}},
  \href{https://doi.org/10.1007/JHEP01(2014)027}{\emph{JHEP} {\bfseries 01}
  (2014) 027} [\href{https://arxiv.org/abs/1310.6694}{{\ttfamily 1310.6694}}].

\bibitem{Fu:2024rsm}
B.~Fu, A.~Ghoshal, S.~F. King and M.~H. Rahat, \emph{{Type-I two-Higgs-doublet
  model and gravitational waves from domain walls bounded by strings}},
  \href{https://doi.org/10.1007/JHEP08(2024)237}{\emph{JHEP} {\bfseries 08}
  (2024) 237} [\href{https://arxiv.org/abs/2404.16931}{{\ttfamily
  2404.16931}}].

\bibitem{Caprini:2018mtu}
C.~Caprini and D.~G. Figueroa, \emph{{Cosmological Backgrounds of Gravitational
  Waves}}, \href{https://doi.org/10.1088/1361-6382/aac608}{\emph{Class. Quant.
  Grav.} {\bfseries 35} (2018) 163001}
  [\href{https://arxiv.org/abs/1801.04268}{{\ttfamily 1801.04268}}].

\bibitem{KAGRA:2021kbb}
{\scshape KAGRA, Virgo, LIGO Scientific} collaboration, R.~Abbott et~al.,
  \emph{{Upper limits on the isotropic gravitational-wave background from
  Advanced LIGO and Advanced Virgo\textquoteright{}s third observing run}},
  \href{https://doi.org/10.1103/PhysRevD.104.022004}{\emph{Phys. Rev. D}
  {\bfseries 104} (2021) 022004}
  [\href{https://arxiv.org/abs/2101.12130}{{\ttfamily 2101.12130}}].

\bibitem{LIGOScientific:2014pky}
{\scshape LIGO Scientific} collaboration, J.~Aasi et~al., \emph{{Advanced
  LIGO}}, \href{https://doi.org/10.1088/0264-9381/32/7/074001}{\emph{Class.
  Quant. Grav.} {\bfseries 32} (2015) 074001}
  [\href{https://arxiv.org/abs/1411.4547}{{\ttfamily 1411.4547}}].

\bibitem{LIGOScientific:2019vic}
{\scshape LIGO Scientific, Virgo} collaboration, B.~P. Abbott et~al.,
  \emph{{Search for the isotropic stochastic background using data from
  Advanced LIGO\textquoteright{}s second observing run}},
  \href{https://doi.org/10.1103/PhysRevD.100.061101}{\emph{Phys. Rev. D}
  {\bfseries 100} (2019) 061101}
  [\href{https://arxiv.org/abs/1903.02886}{{\ttfamily 1903.02886}}].

\bibitem{Punturo:2010zz}
M.~Punturo et~al., \emph{{The Einstein Telescope: A third-generation
  gravitational wave observatory}},
  \href{https://doi.org/10.1088/0264-9381/27/19/194002}{\emph{Class. Quant.
  Grav.} {\bfseries 27} (2010) 194002}.

\bibitem{Hild:2010id}
S.~Hild et~al., \emph{{Sensitivity Studies for Third-Generation Gravitational
  Wave Observatories}},
  \href{https://doi.org/10.1088/0264-9381/28/9/094013}{\emph{Class. Quant.
  Grav.} {\bfseries 28} (2011) 094013}
  [\href{https://arxiv.org/abs/1012.0908}{{\ttfamily 1012.0908}}].

\bibitem{Reitze:2019iox}
D.~Reitze et~al., \emph{{Cosmic Explorer: The U.S. Contribution to
  Gravitational-Wave Astronomy beyond LIGO}}, {\emph{Bull. Am. Astron. Soc.}
  {\bfseries 51} (2019) 035}
  [\href{https://arxiv.org/abs/1907.04833}{{\ttfamily 1907.04833}}].

\bibitem{Bartolo:2016ami}
N.~Bartolo et~al., \emph{{Science with the space-based interferometer LISA. IV:
  Probing inflation with gravitational waves}},
  \href{https://doi.org/10.1088/1475-7516/2016/12/026}{\emph{JCAP} {\bfseries
  12} (2016) 026} [\href{https://arxiv.org/abs/1610.06481}{{\ttfamily
  1610.06481}}].

\bibitem{Caprini:2019pxz}
C.~Caprini, D.~G. Figueroa, R.~Flauger, G.~Nardini, M.~Peloso, M.~Pieroni
  et~al., \emph{{Reconstructing the spectral shape of a stochastic
  gravitational wave background with LISA}},
  \href{https://doi.org/10.1088/1475-7516/2019/11/017}{\emph{JCAP} {\bfseries
  11} (2019) 017} [\href{https://arxiv.org/abs/1906.09244}{{\ttfamily
  1906.09244}}].

\bibitem{LISACosmologyWorkingGroup:2022jok}
{\scshape LISA Cosmology Working Group} collaboration, P.~Auclair et~al.,
  \emph{{Cosmology with the Laser Interferometer Space Antenna}},
  \href{https://doi.org/10.1007/s41114-023-00045-2}{\emph{Living Rev. Rel.}
  {\bfseries 26} (2023) 5} [\href{https://arxiv.org/abs/2204.05434}{{\ttfamily
  2204.05434}}].

\bibitem{Auclair:2019wcv}
P.~Auclair et~al., \emph{{Probing the gravitational wave background from cosmic
  strings with LISA}},
  \href{https://doi.org/10.1088/1475-7516/2020/04/034}{\emph{JCAP} {\bfseries
  04} (2020) 034} [\href{https://arxiv.org/abs/1909.00819}{{\ttfamily
  1909.00819}}].

\bibitem{Corbin:2005ny}
V.~Corbin and N.~J. Cornish, \emph{{Detecting the cosmic gravitational wave
  background with the big bang observer}},
  \href{https://doi.org/10.1088/0264-9381/23/7/014}{\emph{Class. Quant. Grav.}
  {\bfseries 23} (2006) 2435}
  [\href{https://arxiv.org/abs/gr-qc/0512039}{{\ttfamily gr-qc/0512039}}].

\bibitem{Kawamura:2006up}
S.~Kawamura et~al., \emph{{The Japanese space gravitational wave antenna
  DECIGO}}, \href{https://doi.org/10.1088/0264-9381/23/8/S17}{\emph{Class.
  Quant. Grav.} {\bfseries 23} (2006) S125}.

\bibitem{Ruan:2018tsw}
W.-H. Ruan, Z.-K. Guo, R.-G. Cai and Y.-Z. Zhang, \emph{{Taiji program:
  Gravitational-wave sources}},
  \href{https://doi.org/10.1142/S0217751X2050075X}{\emph{Int. J. Mod. Phys. A}
  {\bfseries 35} (2020) 2050075}
  [\href{https://arxiv.org/abs/1807.09495}{{\ttfamily 1807.09495}}].

\bibitem{TianQin:2015yph}
{\scshape TianQin} collaboration, J.~Luo et~al., \emph{{TianQin: a space-borne
  gravitational wave detector}},
  \href{https://doi.org/10.1088/0264-9381/33/3/035010}{\emph{Class. Quant.
  Grav.} {\bfseries 33} (2016) 035010}
  [\href{https://arxiv.org/abs/1512.02076}{{\ttfamily 1512.02076}}].

\bibitem{Sesana:2019vho}
A.~Sesana et~al., \emph{{Unveiling the gravitational universe at $\mu$-Hz
  frequencies}}, \href{https://doi.org/10.1007/s10686-021-09709-9}{\emph{Exper.
  Astron.} {\bfseries 51} (2021) 1333}
  [\href{https://arxiv.org/abs/1908.11391}{{\ttfamily 1908.11391}}].

\bibitem{Badurina:2019hst}
L.~Badurina et~al., \emph{{AION: An Atom Interferometer Observatory and
  Network}}, \href{https://doi.org/10.1088/1475-7516/2020/05/011}{\emph{JCAP}
  {\bfseries 05} (2020) 011}
  [\href{https://arxiv.org/abs/1911.11755}{{\ttfamily 1911.11755}}].

\bibitem{Badurina:2021rgt}
L.~Badurina, O.~Buchmueller, J.~Ellis, M.~Lewicki, C.~McCabe and V.~Vaskonen,
  \emph{{Prospective sensitivities of atom interferometers to gravitational
  waves and ultralight dark matter}},
  \href{https://doi.org/10.1098/rsta.2021.0060}{\emph{Phil. Trans. A. Math.
  Phys. Eng. Sci.} {\bfseries 380} (2021) 20210060}
  [\href{https://arxiv.org/abs/2108.02468}{{\ttfamily 2108.02468}}].

\bibitem{Graham:2017pmn}
{\scshape MAGIS} collaboration, P.~W. Graham, J.~M. Hogan, M.~A. Kasevich,
  S.~Rajendran and R.~W. Romani, \emph{{Mid-band gravitational wave detection
  with precision atomic sensors}},
  \href{https://arxiv.org/abs/1711.02225}{{\ttfamily 1711.02225}}.

\bibitem{Bertoldi:2019tck}
{\scshape AEDGE} collaboration, Y.~A. El-Neaj et~al., \emph{{AEDGE: Atomic
  Experiment for Dark Matter and Gravity Exploration in Space}},
  \href{https://doi.org/10.1140/epjqt/s40507-020-0080-0}{\emph{EPJ Quant.
  Technol.} {\bfseries 7} (2020) 6}
  [\href{https://arxiv.org/abs/1908.00802}{{\ttfamily 1908.00802}}].

\bibitem{Joshi:2018ogr}
B.~C. Joshi et~al., \emph{{Precision pulsar timing with the ORT and the GMRT
  and its applications in pulsar astrophysics}}, .

\bibitem{Janssen:2014dka}
G.~Janssen et~al., \emph{{Gravitational wave astronomy with the SKA}},
  \href{https://doi.org/10.22323/1.215.0037}{\emph{PoS} {\bfseries AASKA14}
  (2015) 037} [\href{https://arxiv.org/abs/1501.00127}{{\ttfamily
  1501.00127}}].

\bibitem{Guedes:2018afo}
G.~S.~F. Guedes, P.~P. Avelino and L.~Sousa, \emph{{Signature of inflation in
  the stochastic gravitational wave background generated by cosmic string
  networks}}, \href{https://doi.org/10.1103/PhysRevD.98.123505}{\emph{Phys.
  Rev. D} {\bfseries 98} (2018) 123505}
  [\href{https://arxiv.org/abs/1809.10802}{{\ttfamily 1809.10802}}].

\bibitem{Cui:2019kkd}
Y.~Cui, M.~Lewicki and D.~E. Morrissey, \emph{{Gravitational Wave Bursts as
  Harbingers of Cosmic Strings Diluted by Inflation}},
  \href{https://doi.org/10.1103/PhysRevLett.125.211302}{\emph{Phys. Rev. Lett.}
  {\bfseries 125} (2020) 211302}
  [\href{https://arxiv.org/abs/1912.08832}{{\ttfamily 1912.08832}}].

\bibitem{Ferrer:2023uwz}
F.~Ferrer, A.~Ghoshal and M.~Lewicki, \emph{{Imprints of a supercooled phase
  transition in the gravitational wave spectrum from a cosmic string network}},
  \href{https://doi.org/10.1007/JHEP09(2023)036}{\emph{JHEP} {\bfseries 09}
  (2023) 036} [\href{https://arxiv.org/abs/2304.02636}{{\ttfamily
  2304.02636}}].

\bibitem{Datta:2025yow}
S.~Datta, A.~Ghosal, A.~Ghoshal and G.~White, \emph{{Complementarity between
  Cosmic String Gravitational Waves and long lived particle searches in
  laboratory}},  \href{https://arxiv.org/abs/2501.03326}{{\ttfamily
  2501.03326}}.

\bibitem{CMB-S4:2020lpa}
{\scshape CMB-S4} collaboration, K.~Abazajian et~al., \emph{{CMB-S4:
  Forecasting Constraints on Primordial Gravitational Waves}},
  \href{https://doi.org/10.3847/1538-4357/ac1596}{\emph{Astrophys. J.}
  {\bfseries 926} (2022) 54}
  [\href{https://arxiv.org/abs/2008.12619}{{\ttfamily 2008.12619}}].

\bibitem{Cui:2017ufi}
Y.~Cui, M.~Lewicki, D.~E. Morrissey and J.~D. Wells, \emph{{Cosmic Archaeology
  with Gravitational Waves from Cosmic Strings}},
  \href{https://doi.org/10.1103/PhysRevD.97.123505}{\emph{Phys. Rev. D}
  {\bfseries 97} (2018) 123505}
  [\href{https://arxiv.org/abs/1711.03104}{{\ttfamily 1711.03104}}].

\bibitem{Ghiglieri:2015nfa}
J.~Ghiglieri and M.~Laine, \emph{{Gravitational wave background from Standard
  Model physics: Qualitative features}},
  \href{https://doi.org/10.1088/1475-7516/2015/07/022}{\emph{JCAP} {\bfseries
  07} (2015) 022} [\href{https://arxiv.org/abs/1504.02569}{{\ttfamily
  1504.02569}}].

\bibitem{Ghiglieri:2020mhm}
J.~Ghiglieri, G.~Jackson, M.~Laine and Y.~Zhu, \emph{{Gravitational wave
  background from Standard Model physics: Complete leading order}},
  \href{https://doi.org/10.1007/JHEP07(2020)092}{\emph{JHEP} {\bfseries 07}
  (2020) 092} [\href{https://arxiv.org/abs/2004.11392}{{\ttfamily
  2004.11392}}].

\bibitem{Muia:2023wru}
F.~Muia, F.~Quevedo, A.~Schachner and G.~Villa, \emph{{Testing BSM physics with
  gravitational waves}},
  \href{https://doi.org/10.1088/1475-7516/2023/09/006}{\emph{JCAP} {\bfseries
  09} (2023) 006} [\href{https://arxiv.org/abs/2303.01548}{{\ttfamily
  2303.01548}}].

\bibitem{Villa:2025zmj}
G.~Villa, \emph{{Remarks on brane-antibrane inflation}},  in \emph{{2nd General
  Meeting of COST Action COSMIC WISPers (CA21106)}}, 1, 2025,
  \href{https://arxiv.org/abs/2501.09074}{{\ttfamily 2501.09074}}.

\bibitem{Komargodski:2009rz}
Z.~Komargodski and N.~Seiberg, \emph{{From Linear SUSY to Constrained
  Superfields}},
  \href{https://doi.org/10.1088/1126-6708/2009/09/066}{\emph{JHEP} {\bfseries
  09} (2009) 066} [\href{https://arxiv.org/abs/0907.2441}{{\ttfamily
  0907.2441}}].

\bibitem{Kallosh:2015nia}
R.~Kallosh, F.~Quevedo and A.~M. Uranga, \emph{{String Theory Realizations of
  the Nilpotent Goldstino}},
  \href{https://doi.org/10.1007/JHEP12(2015)039}{\emph{JHEP} {\bfseries 12}
  (2015) 039} [\href{https://arxiv.org/abs/1507.07556}{{\ttfamily
  1507.07556}}].

\bibitem{Aparicio:2015psl}
L.~Aparicio, F.~Quevedo and R.~Valandro, \emph{{Moduli Stabilisation with
  Nilpotent Goldstino: Vacuum Structure and SUSY Breaking}},
  \href{https://doi.org/10.1007/JHEP03(2016)036}{\emph{JHEP} {\bfseries 03}
  (2016) 036} [\href{https://arxiv.org/abs/1511.08105}{{\ttfamily
  1511.08105}}].

\end{thebibliography}\endgroup
\bibliographystyle{JHEP}

\end{document}